\documentclass[10pt,letterpaper]{article}
\usepackage[top=0.85in,left=2.75in,footskip=0.75in]{geometry}

\usepackage{amsmath,amssymb}

\usepackage{changepage}
\usepackage{algorithm}
\usepackage{algorithmic}
\usepackage[utf8x]{inputenc}
\usepackage[spaces,hyphens]{url}
\usepackage{textcomp,marvosym}

\usepackage{cite}
\usepackage{amsmath,amssymb,amsfonts,amsthm,bbm,mathrsfs,verbatim,bm} 
 \usepackage{graphicx}
\usepackage{nameref,hyperref}

\usepackage{microtype}
\DisableLigatures[f]{encoding = *, family = * }

\usepackage[table]{xcolor}

\usepackage{array}

\newcolumntype{+}{!{\vrule width 2pt}}

\newlength\savedwidth

\newcommand\thickhline{\noalign{\global\savedwidth\arrayrulewidth\global\arrayrulewidth 2pt}%
\hline
\noalign{\global\arrayrulewidth\savedwidth}}

\raggedright
\usepackage[aboveskip=1pt,labelfont=bf,labelsep=period,justification=raggedright,singlelinecheck=off]{caption}
\usepackage{subcaption}
 
\usepackage[english]{babel}
\makeatletter
\renewcommand{\@biblabel}[1]{\quad#1.}
\makeatother

\usepackage{lastpage,fancyhdr,graphicx}
\usepackage{epstopdf}
\usepackage[numbers, square, comma, sort&compress]{natbib}
\usepackage{csquotes}
\fancyheadoffset[L]{2.25in}
\fancyfootoffset[L]{2.25in}
\begin{document}
\vspace*{0.2in}

\begin{flushleft}
{\Large
\textbf\newline{Can a latent Hawkes process be used for epidemiological modelling?} 
}
\newline
\\
Stamatina Lamprinakou\textsuperscript{1*},
Axel Gandy\textsuperscript{1*},
Emma McCoy\textsuperscript{1}
\\
\bigskip
\textbf{1} Department of Mathematics, Imperial College London, London, United Kingdom
\\
\bigskip

%
%





* s.lamprinakou18@imperial.ac.uk, a.gandy@imperial.ac.uk

\end{flushleft}
\section*{Abstract}
Understanding the spread of COVID-19 has been the subject of numerous studies, highlighting the significance of reliable epidemic models. Here, we introduce a novel epidemic model using a latent Hawkes process with temporal covariates for modelling the infections. Unlike other models, we model the reported cases via a probability distribution driven by the underlying Hawkes process. Modelling the infections via a Hawkes process allows us to estimate by whom an infected individual was infected. We propose a Kernel Density Particle Filter (KDPF) for inference of both latent cases and reproduction number and for predicting the new cases in the near future. The computational effort is proportional to the number of infections making it possible to use particle filter type algorithms, such as the KDPF. We demonstrate the performance of the proposed algorithm on synthetic data sets and COVID-19 reported cases in various local authorities in the UK, and benchmark our model to alternative approaches.



\section*{Introduction}
The novel coronavirus disease (COVID-19) has been declared a Global Health Emergency of International Concern with over 557 million cases and 6.36 million deaths as of 3 August 2022 according to the World Health Organization. In the absence of vaccines, countries initially followed mitigation strategies or countermeasures to prevent the rapid spread of COVID-19, such as social distancing, quarantine, mask wearing, and lock-downs.

A large number of studies have been carried out to understand the spread of COVID-19, forecast new cases and when the peak of the pandemic will occur, and investigate  \enquote {what-if-scenarios}. For example, Ferguson et al.~\cite{ferguson2020report} presented the results of epidemiological modelling looking at a  variety  of nonpharmaceutical interventions. Several compartmental models \cite{Zou2020.05.24.20111989,chen2020time,wangping2020extended,roques2020using} using ordinary differential equations (ODE) have been proposed for modelling the spread of COVID-19. Various models using Hawkes processes \cite{garetto2021time,Kresin2020ComparisonOT,escobar2020hawkes,CHIANG2021,browning2021simple,koyama2021estimating,bertozzi2020challenges}, widely used to model contagion patterns, have been introduced as an alternative to ODE models. Others have used a Poisson autoregression model of the daily new observed cases \cite{agosto2020poisson} and a Bayesian model linking the infection cycle to observed deaths\cite{flaxman2020estimating}.

We introduce a novel epidemic model using a latent Hawkes process \cite{laub2015hawkes} with temporal covariates for modelling the infections. Unlike other Hawkes models, the Hawkes process is used as a latent, i.e. for modelling the actual unobserved infection cases. Observations, such as reported infection cases, are then modelled as random quantities driven by the latent Hawkes process. Other models that use the latent processes in epidemiological models (e.g.~\cite{flaxman2020estimating}) usually have time-aggregated counts of infections as latent process, i.e. the latent process works on a discrete scale. We propose using a Kernel Density Particle Filter (KDPF) \citep{sheinson2014comparison, liu2001combined} for inference of both latent cases and reproduction number and for predicting the new cases in the near future. It is feasible to employ particle filter type algorithms, like the KDPF, because the computational effort is linear to the number of infections. Modelling the infections via a Hawkes process allows us to estimate by whom an infected individual was infected. We demonstrate the performance of the proposed algorithm on synthetic data and COVID-19 reported cases in various local authorities in the UK. The methods \cite{koyama2021estimating,cori2013new} provide similar estimates of reproduction number to the proposed algorithm. The ability of our model to estimate individual latent cases and reveal epidemic dynamics provides an important advantage over other models.

\section*{Related work} \label{S1}
The Hawkes process is a well known self-exciting process in which the intensity function depends on all previous events assuming infinite population that allow for parametric or non-parametric estimation of the reproduction number (that is, the expected number of infections triggered per infected individual). Hawkes processes have been widely used in numerous applications such as social media, criminology and earthquake modelling. In this section, we present the application of the Hawkes processes in the modelling of COVID-19.

 First, we briefly review basic compartmental models and their connection with Hawkes process and COVID. The Susceptible-Infected-Recovered (SIR) and Susceptible-Exposed-Infected-Recovered (SEIR) models are the two basic compartmental epidemic models for modelling the spread of infectious disease \cite{jones2007notes,Zou2020.05.24.20111989}. The SIR model defines three classes of individuals: those susceptible to infection (S), those currently infected (I) and those recovered (R). The SEIR model involves an additional compartment (E) that models the exposed individuals without having obvious symptoms. For many diseases, including COVID-19, there is an incubation period during which exposed individuals to the virus may not be as contagious as the infectious individuals. A variant of the SEIR model called SuEIR was introduced by Zou et al. \cite{Zou2020.05.24.20111989} for modelling and forecasting the spread of COVID. The SuEIR compared to SEIR has an additional compartment (u) that models the unreported cases. Estimates based on compartmental models can be unreliable, as they are highly sensitive to initial conditions and parameters such as transmission and recovery rates~\cite{escobar2020hawkes}. 

A stochastic formulation of SIR called Stochastic SIR \citep{allen2008introduction} is a point process having events that are either the recovery times or the infection times of individuals with exponentially distributed recovery times. Rizoiu et al. \cite{rizoiu2018sir} introduced the SIR-Hawkes process (also known as HawkesN), which is a generalization of the Hawkes process concerning finite population. They showed that the conditional intensity of the SIR-Hawkes process with no background events and exponential infectious period distribution is identical to the expected conditional intensity of Stochastic SIR with respect to the recovery period distribution. The Hawkes process with gamma infectious period distribution can approximate stage compartment models if the average waiting times in the compartments follow an independent exponential distribution \cite{lloyd2001destabilization, CHIANG2021}. 

Kresin et al. \cite{Kresin2020ComparisonOT} claim that although the SEIR model is mostly used for COVID modelling compared to the Hawkes process, a Hawkes model offers more accurate forecasts. Specifically, they suggest a SEIR-Hawkes model in which the intensity of newly exposed cases is a function of infection times and size of the population. Chiang et al. \cite{CHIANG2021} introduced a Hawkes process model of COVID-19 that estimates the intensity of cases and the reproduction number. The reported cases are modelled via a Hawkes process. The reproduction number is estimated via a Poisson regression with spatial-temporal covariates including mobility indices and demographic features. Based on the branching nature of the Hawkes process, Escobar \cite{escobar2020hawkes} derived a simple expression for the intensities of reported and unreported COVID-19 cases. The key to this model is that at the beginning of a generation the infectious will either (1) be registered, (2) not be registered but continue being contagious, or (3) recover with fixed probabilities. However, we believe that the probability of remaining contagious and not being registered infectious should be a decreasing function of time and not fixed.

Garetto et al \cite{garetto2021time} proposed a modulated marked Hawkes process for modelling the spread of COVID-19 under the impact of countermeasures. Each mark corresponds to a different class of infectious individuals with specific kernel functions. Three classes of infectious are considered: symptomatic, asymptomatic and superspreader, for obtaining the average intensity function and the average total number of points up to a specific time. Symptomatic people are those who will develop evident symptoms and by extension they will be quarantined. Asymptomatic people are those who will not develop strong enough symptoms to be quarantined. Superspreaders are individuals who exert a high infection rate but do not get quarantined. The model estimates the reproduction number taking into account the amount of recourses employed by the health service to discover the infected population, the countermeasures, as well as the stages that all infectious go through: random incubation time, presymptomatic period, random disease period and random residual phase.

 Koyama et al. \cite{koyama2021estimating} developed a discrete-time Hawkes model for estimating the temporally changing reproduction number, and hence detecting the change points via assuming a negative binomial distribution for the daily cases. Further analysis in \cite{browning2021simple,triambak2021random} examined the daily death data to avoid the issues raised from the reported cases. Browning et al. \cite{browning2021simple} modelled the reported daily deaths using a discrete-time Hawkes process, where the cases are assumed Poisson distributed. They considered one fixed change point that breaks the period of analysis into two phases: the initial period where the virus is spreading rapidly and the period after the introduction of preventative measures. The model provides accurate predictions for short-time intervals. 
 
All the aforementioned stochastic Hawkes models use the Hawkes process for modelling either the reported infections or the newly exposed cases. Herein, we provide a novel epidemic model for the infections using a latent Hawkes process with temporal covariates and, in turn, the reported cases using a probability distribution driven by the underlying Hawkes process. Working on a continuous scale offers the inference of individual latent cases and reveals unobserved transmission paths of the epidemic. We apply particle methods for inferring the latent cases and the reproduction number and predicting observed cases over short time horizons. The simulation analysis shows that the estimated reproduction number and the intensity of latent cases depict the epidemic's development and capture the trajectory of cases.

\section*{Methods}
\subsection*{Model}
We introduce a novel epidemic model using a latent Hawkes process of infections that then trigger a process of reported infection cases.

We focus on an infinite homogeneous population and restrict our attention to an epidemic process over a horizon $[T_0, T)$, $T_0<T$, in which we assume immunity to re-infection. This immunity is a reasonable assumption over the time scales we consider. We break the horizon $[T_0,T)$ into $k$ subintervals $\mathcal{T}_j=[T_{j-1},T_j)$ for $j=1,..,k$ with $T_k=T$. We assume that the epidemic is triggered by a set of infectious individuals at the beginning of the process, the times of their infections denoted by a finite set $\mathcal{H}_0\subseteq (-\infty,T_0)$.

The epidemic process is seen as a counting process $N(t)$ with a set of jump times $\mathcal{T}^N=\{t_0<t_1<t_2<...\}$ and intensity given by 
\begin{equation*}
\lambda^N(t)= \sum\limits_{t_i\in h_t^0}R(t) h(t-t_i)
\end{equation*} for $t>0$ with $h_t^0=\{t_i|t_i<t\}\cup \mathcal{H}_0$ being the set of all infection events prior to time $t$. The kernel $h(t-t_i)$ represents the relative infectiousness at time $t$ of an infection at time $t_i$. We assume that the transition kernel $h$ is a probability density function with non-negative real-valued support: $ h:[0,\infty)\rightarrow [0,\infty)$ and  $\int\limits_{0}^{\infty}h(s)ds=1$. The process $R(t)$ represents the instantaneous reproduction number that is the average number of newly infected people that each infected individual would infect if the conditions, such as interventions and control measures for restriction of epidemic, remained as they were at time $t$~\cite{cori2013new}.  

It is natural to see the reported infections as a counting process $M(t)$ with a set of jump times $\mathcal{T}^M=\{\tau_1<\tau_2<...<\tau_m\}$ and intensity of observed cases at time $\tau$ as a function of the times of infection up to time $\tau$, namely 
\begin{equation} \label{EqIntOC}
\lambda^M(\tau)=\sum\limits_{t_i\in h_\tau^0} \beta g(\tau-t_i)
\end{equation} for $\tau>0$, where $\beta$ is the expected number of observed cases per infected individual at time $\tau$ (also known as ascertainment rate). The transition kernel $g(\tau -t_i)$ represents the relative delay between the infection at time $t_i$ and the time at $\tau$ the infection is detected. 
Similar to the transition kernel of latent cases $h$, we specify the transition kernel of observed cases $g$ to be a probability density function with non-negative real-valued support.

$M(t)$ is usually only observable in daily or weekly aggregates. We will use $\mathcal{T}_n$ as aggregation intervals and let $Y_n$ be the number of reported cases in $\mathcal{T}_n$. We model $Y_{n}$ via a distribution $G$ having mean $\mu_{n}$ equal to the expected observed cases in $\mathcal{T}_n$ given by 
\begin{equation*}
\mu_{n}=\beta\sum\limits_{t_w\in h_{T_n}^0\ }\int\limits_{\max(t_w, T_{n-1})}^{T_n}{g(s-t_w)}ds.
\end{equation*} The usual options of $G$ are Negative Binomial (NB)~\cite{koyama2021estimating,stocks2020model} and Poisson distribution \citep{browning2021simple, cori2013new}. We model the reproduction number $R(t)$ as a stepwise function having as many weights as the number of subintervals, that is, 
\begin{equation*}
R(t)=\prod_{n=1}^k R_{n}^{\mathbbm{1}(t\in \mathcal{T}_n)},
\end{equation*} where $\{R_n\}$ is assumed to be a Markov process. Usually, a random walk on a logarithmic scale \citep{storvik2022sequential} or a normal scale \citep{koyama2021estimating} is imposed as a prior on the weights $\{R_n\}$.

The model is described by the equations: 
\begin{align}
\label{Mds1a}
   \lambda^{N}(t)&=R(t)\sum\limits_{t_i\in h_t^0}h(t-t_i),\ t\in [T_0,T)\ ;\\
 Y_{n} &\sim G \text{ with mean } E(Y_n)=\mu_n,\ n=1,..,k\ ;\\
 R(t)&=\prod\limits_{n=1}^{k}R_{n}^{\mathbbm{1}\{t\in\mathcal{T}_n\}}\ ; \\
 \{R_n\}_{n=1}^k & \text{ is a Markov process}\ ; \\
\mu_{n}&=\beta\sum\limits_{t_w\in [0,T_n)}\int\limits_{\max(t_w,T_{n-1})}^{T_n}{g(s-t_w)}ds,\ n=1,..,k\ . \label{Mds1b}
\end{align}

\subsection*{Inference algorithm}
 
 Given a set of observed infections $\{Y_1,..,Y_k\}$, we seek to infer the counting process $N(t)$ and the reproduction number $R(t)$.

The proposed epidemic model described by the equations (\ref{Mds1a})-(\ref{Mds1b}) is seen as a state-space model with a latent state process $\{X_n:1\ \leq n\leq k\}$ and an observed process $\{Y_n:\  1\leq n \leq k\}$. Each hidden state $X_n$  consists of the reproduction number's weight $R_{n}$ associated to $\mathcal{T}_n$ and the set of latent cases $S_n^N$ falling into $\mathcal{T}_n$. The time-constant parameters are the parameters associated with the distribution $G$ and the prior imposed on the weights $\{R_n\}_{n=1}^k$. We apply a KDPF \citep{sheinson2014comparison, liu2001combined} for inferring the counting process $N(t)$, the weights $\{R_{n}\}_{n=1}^k$, and the time-constant parameters.

We focus on illustrating the performance of our model on COVID-19. As the COVID-19 reported cases are subject to erroneous observation and for the data we observe that the sample variance is larger than the sample mean, we model the observed cases $Y_{n}$ via a negative binomial distribution (NB) with mean $\mu_n$ and dispersion $v>0$. We use the following form of the negative binomial distribution
\begin{equation*}
P(Y_n|\mu_n,v)=\frac{\gamma(Y_n+v^{-1})}{Y_n!\gamma(v^{-1})}\left(\frac{1}{1+v\mu_n}\right)^{\frac{1}{v}}\left(\frac{v\mu_n}{v\mu_n+1}\right)^{Y_n}
\end{equation*}
with mean $E(Y_n)=\mu_n$ and variance $var(Y_n)=\mu_n(1+v\mu_n)$. Before we discuss the KDPF, we define the transition kernels of the observed and latent cases and the prior on weights $\{R_n\}_{n=1}^k$ for COVID-19. We also suggest a simple method to initialize $\mathcal{H}_0$.

\begin{figure}[!h] 
  \begin{subfigure}{6cm}
    \centering\includegraphics[width=6cm]{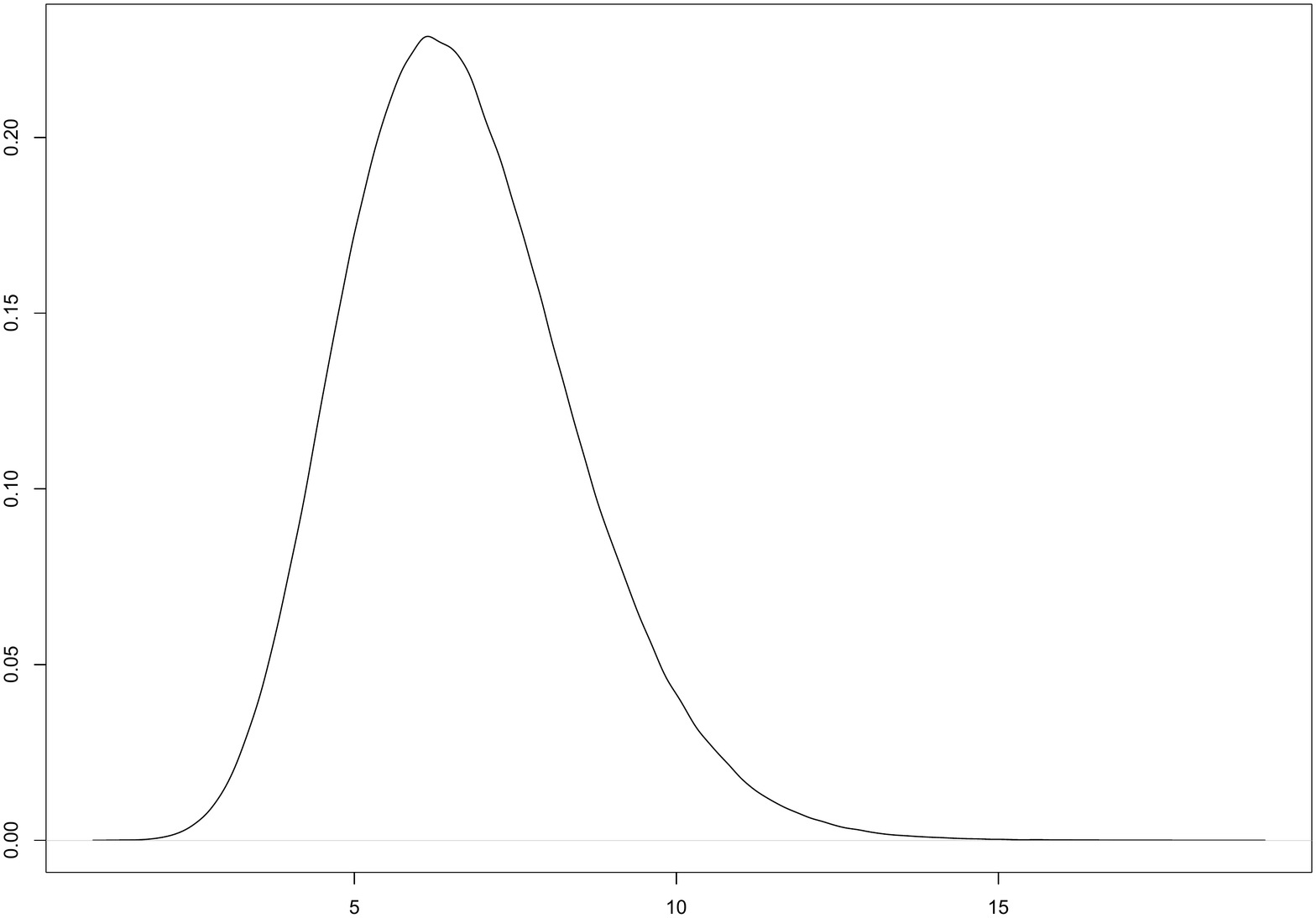}
  \end{subfigure}
  \begin{subfigure}{6cm}
    \centering\includegraphics[width=6cm]{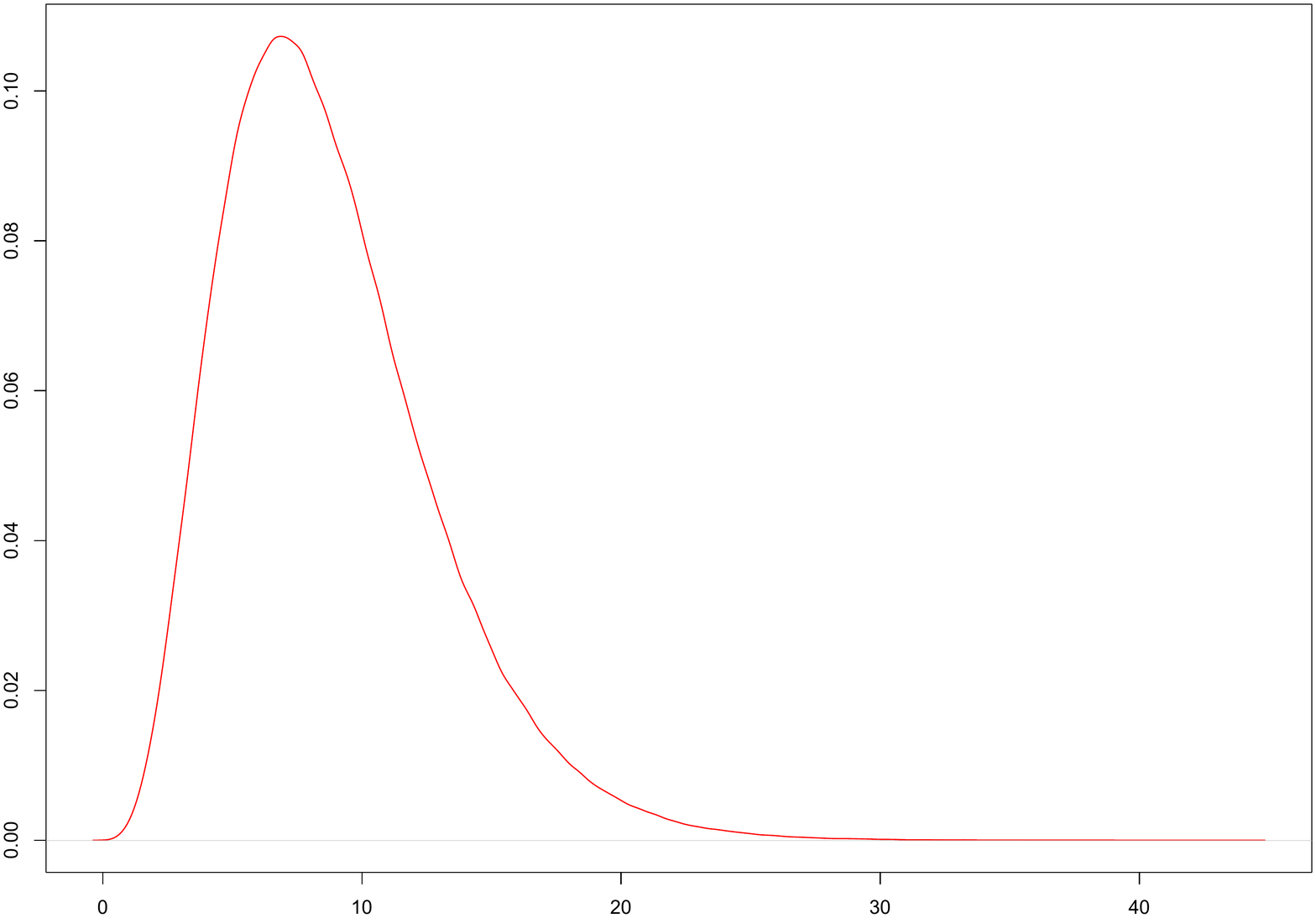}
  \end{subfigure}
  
  \caption{\bf The generation interval (GI) (black curve) and the period between observed and actual infection times (red curve).}
  \label{TK_10ma}
\end{figure}

\paragraph*{Transition Kernels} 
The dynamics of the latent and observed cases are determined by the generation interval (GI) and incubation period (IP) \citep{fine2003interval}. The generation interval is the time interval between the time of infection of the infector (the primary case) and that of the infectee (the secondary case generated by the primary case). The incubation period is the time interval between the infection and the onset of symptoms in a specific case. Zhao et al.~\cite{zhao2021estimating} assume that the GI and IP follow a gamma distribution. They infer that the mean and SD of GI are equal at 6.7 days and 1.8 days and those of IP at 6.8 and 4.1 days by using a maximum likelihood estimation approach and contact tracing data of COVID-19 cases. We follow the same assumption for the GI (namely, the transition kernel of latent cases is a gamma density with a mean at 6.7 days and  SD of 1.8 days). We model the time interval between the observed time and actual time of infection as a gamma density with a mean at 8.8 days and SD of 4.1 days  (that is, the transition kernel of observed cases is a gamma density having mean equal at 8.8 days and  SD of 4.1 days). For the transition kernel of the observed events, we adopt the values inferred by Zhao et al.~\cite{zhao2021estimating}  for IP with a slightly increased mean to consider the necessary time for conducting a test against COVID-19. Figure \ref{TK_10ma} illustrates the transition kernels. 

\paragraph*{Set of infectious at the beginning of the process, $\mathcal{H}_0$} We adopt a heuristic approach to initialize $\mathcal{H}_0$.  The transition kernel of latent cases illustrated in Figure \ref{TK_10ma} shows that a latent case at $t_w$ can influence the latent intensity at $t$ if $t_w$ has occurred at most 21 days before $t$. Otherwise, the influence of $t_w$ is negligible. Therefore, as the history of the process, we consider the latent cases of 21 days/3 weeks before the beginning of the process. The mode of the transition kernel of the observed cases equal to 6.216 demonstrates that an event is most likely to be observed 7 days after the actual infection time. Considering the observed cases are daily, we initialize the history of latent case, $\mathcal{H}_{0}$ by uniformly spreading on the day $-i$ the number of cases occurred on the day $(-i+7)$ times $1/\beta$. In simulation analysis, we propose initialization of $\mathcal{H}_0$ when we deal with weekly reported cases.

\paragraph*{Imposed prior on weights \begin{math} \mathbf{\{R_n\}_{n=1}^k} \end{math} }  A geometric random walk (RW) is imposed as prior on weights \begin{math}\{R_{n}\}_{n=1}^k \end{math} :
\begin{align*}
\log R_{n}&=\log R_{n-1}+\log\epsilon_n, \ \ \epsilon_n \sim\mbox{gamma}(d,d)\\
R_{1}&\sim \mbox{Uniform}(\alpha,b).
\end{align*} We impose a gamma prior on the noise of RW $\epsilon_n$ with equal shape and rate at $d$. This induces that the weight $R_{n}$ is gamma distributed with a mean equal to $R_{n-1}$ and standard deviation $R_{n-1}/ \sqrt{d}$. The stronger fluctuations in the observed data, the more flexible modelling we need. Smaller values of $d$  have higher standard deviation and lead to a wider range of possible values of $R_{n}$ increasing the flexibility of the model.

\subsubsection*{Kernel Density Particle Filter}
 We apply a KDPF (Algorithm \ref{APAlga})
 for inferring the counting process $N(t)$, the weights $\{R_{n}\}_{n=1}^k$, and the time-constant parameters. The time-constant parameters for modelling COVID-19 infections are the shape $d$ of the noise $\epsilon_n$ and the dispersion parameter $v$.

The KDPF builds on the auxiliary particle filter (APF) \citep{pitt1999filtering, kantas2015particle, doucet2009tutorial} by adding small random perturbations to all the parameter particles to reduce the sample degeneracy by modelling the time-constant parameters as random quantities and their posterior via a mixture of normal distributions.  We assume independence among the time-constant parameters, and, following Sheinson et al.~\cite{sheinson2014comparison}, we use logarithms for the time-constant parameters, as they have positive support: 
\begin{equation*} p(\log d_{n+1}, \log v_{n+1}|Y_{1:(n+1)})=p(\log d_{n+1}|Y_{1:{(n+1)}})p(\log v_{n+1}|Y_{1:(n+1)}).\end{equation*}
The posteriors $p(\log d_{n+1}|Y_{1:(n+1)})$ and $p(\log v_{n+1}|Y_{1:(n+1)})$ are smoothly approximated via a mixture of normal distributions weighted by the sample weights $w_{jn}$ given by 

\begin{align*}
    p(\log d_{n+1}|Y_{1:(n+1)}) &\approx  \sum\limits_{j=1}^N\omega_{jn}\mathcal{N}( \log d_{n+1} |m_{j,dn}^{(L)},h^2 V_{nd}^{(L)}), \\
      p(\log v_{n+1}|Y_{1:(n+1)}) &\approx  \sum\limits_{j=1}^N\omega_{jn}\mathcal{N}( \log v_{n+1} |m_{j,vn}^{(L)},h^2 V_{nv}^{(L)}),
\end{align*} where $\mathcal{N}(\mu,\sigma^2)$ is a Gaussian pdf with mean $\mu$ and variance $\sigma^2$. The KDPF uses a tuning parameter $\Delta \in (0,1]$ and two quantities as a function of that parameter: $h^2=1-((3\Delta-1)/(2/\Delta))^2$ and $a^2=1-h^2$. The parameter $\Delta$ is typically taken to be between 0.95 and 0.99 for reducing the chance of degeneracy \citep{sheinson2014comparison,west1993approximating}. 

The mean values and the variances of the posteriors of time-constant parameters are defined as follows\citep{west1993approximating,sheinson2014comparison}:
\begin{align*}
m_{j,dn}^{(L)}&=a\log d_{jn} + (1-a)\bar{d}_{Ln}, \hspace{0.5cm} \bar{d}_{Ln}=\sum\limits_{j=1}^Nw_{jn}\log d_{jn}, \\
m_{j,vn}^{(L)}&=a\log v_{jn} + (1-a)\bar{v}_{Ln}, \hspace{0.5cm} \bar{v}_{Ln}=\sum\limits_{j=1}^Nw_{jn}\log v_{jn}, \\
V_{nv}^{(L)}&=\frac{V_1}{V_1^2-V_2}\sum\limits_{j=1}^N\omega_{jn}(\log v_{jn}-\bar{v}_{Ln})^2, \\
V_{nd}^{(L)}&=\frac{V_1}{V_1^2-V_2}\sum\limits_{j=1}^N \omega_{jn}(\log d_{jn}-\bar{d}_{Ln})^2,  \\
\end{align*}
with $V_1=\sum\limits_{j=1}^Nw_{jn}$ and $V_2=\sum\limits_{j=1}^Nw_{jn}^2$.

Following Sheinson et al.~\cite{sheinson2014comparison}, we define the initial densities of parameters $d$ and $v$ to be log-normal:
\begin{align*}
\log d &\sim \mathcal{N}(\mu_d,\sigma^2_d), \  \mu_d=\frac{\log(d_{max}) + \log(d_{min})}{2}, \  \sigma_d=\frac{\log(d_{max}) - \log(d_{min})}{8},\\
\log v &\sim \mathcal{N}(\mu_v,\sigma^2_v), \  \mu_v=\frac{\log(v_{max}) + \log(v_{min})}{2}, \  \sigma_v=\frac{\log(v_{max}) - \log(v_{min})}{8},\\
\end{align*}
considering that $d_{min}\leq d\leq d_{max}$ and $v_{min}\leq v \leq v_{max}$ . The transition densities of the time-constant parameters are given by 
\begin{align*}
p(\log d_{n+1}| \log d_n) &\sim \mathcal{N}(\log d_{n+1}| a\log d_n + (1-a) \bar{d}_{Ln},h^2V_{nd}^{(L)}), \\
p(\log v_{n+1}| \log v_n) &\sim \mathcal{N}(\log v_{n+1}| a\log v_n + (1-a) \bar{v}_{Ln},h^2V_{nv}^{(L)}).
\end{align*}
 
The initial density of the hidden process is given by 

\begin{equation*}
f(x_1|\mathcal{H}_0)=\mbox{U}\left(R_{1};\alpha, b\right)P(S_1^N|R_{1},\mathcal{H}_0), 
\end{equation*} while the transition density is given by \begin{align*}f(x_n|x_{1:(n-1)},\mathcal{H}_0,d,v)&=P(S_n^N|R_{n},S_{1:(n-1)}^N,\mathcal{H}_0)P(R_{n}|R_{n-1},d).\end{align*} $U(R_1;\alpha,b)$ denotes that $R_1$ is uniformly distributed within the interval $[\alpha,b]$.

\paragraph*{Sampling the hidden latent cases}
We sample the latent cases $S_n^N$ falling into the subinterval $\mathcal{T}_n$ by applying Algorithm \ref{MdsAlg1a}, which is a simulation procedure based on the branching structure of the Hawkes process \citep{laub2015hawkes}. The proposed algorithm is a superposition of Poisson processes in the interval $\mathcal{T}_n$; the descendants of each latent event at $t_i$ form an inhomogeneous Poisson process with intensity \begin{equation*}\lambda_i(t)=R_nh(t-t_i)\end{equation*}
for $t>t_i$ and $t\in[T_{n-1},T_n)$. This induces that: 
\begin{itemize}
\item{The number of events $n_i$ triggered by an event at $t_i$ in the interval $\mathcal{T}_n$ is Poisson distributed with parameter  
\begin{equation*}\lambda=R_n\int\limits_{\max(t_i, T_{n-1})}^{T_n}h(s-t_i)ds.\end{equation*}}
\item{ The arrival times of the $n_i$ descendants are  $t_i+E_i$ with $E_i$ being iid random variables with pdf the truncated distribution $h(t)$ in $[\max(t_i,T_{n-1}),T_n)$.}
\end{itemize}

 \begin{algorithm}[!h] 
\caption{Sample $S_n^N|S_{1:(n-1)}^N,\ \mathcal{H}_0,R_{n}$} 
\label{MdsAlg1a}
\begin{algorithmic}[1]
\STATE{Input: $S_{1:(n-1)}^N$,\ $\mathcal{H}_0$,\ $R_{n}$ }\\
\STATE{Initialize an empty queue: $Q_t$.}\\
\STATE{$Q_t=\mathcal{H}_0 \cup \{S_{v}^N\}_{v=n-\eta}^{n-1}$ with $n-\eta \geq 1$ and $\eta$ being the number of former subintervals we consider (the value of $\eta$ is determined by the transition kernel of latent cases)}.
\WHILE{$Q_t$ is not empty }

\STATE{Remove the first element $t_i$ from $Q_t$.}\\
\STATE{Draw the number of events $n_{i}$ triggered by an event at $t_i$ from a Poisson distribution with parameter  $\lambda=R_{n}\int\limits_{\max(t_i,T_{n-1})}^{T_n}h(s-t_i)ds$ that is the average number of offsprings generated by an event at $t_i$ in $\mathcal{T}_n$.}
\STATE{Generate $n_{i}$ events from the truncated distribution $h(t)$ in $[\max(t_i,T_{n-1}),T_n)$, and add the new elements to the back of queue $Q_t$. }\\
\ENDWHILE
\STATE{Return $Q_t$.}
\end{algorithmic}
\end{algorithm}

\paragraph*{Who infected whom} The Hawkes process is an excellent option for modelling the evolution of an epidemic due to its mutually exciting nature, making it feasible to estimate by whom an infected individual was infected. 
Bertozzi et al.~\cite{bertozzi2020challenges} describe how we can infer the primary infection $i$ that triggered a secondary infection $j$ using a self-exciting branching process. The parent of each infection $j$ is assumed to be sampled from a multinomial distribution parameterized by $\pi_j$, where $\pi_j=\{\pi_{ji}\}_{i\in h_j}$ with 
\begin{equation*}
    \pi_{ji}=\frac{h(t_j-t_i)}{\sum_{t_w\in h_{t_j}^0}h(t_j-t_w)}
\end{equation*} being the probability of secondary infection $j$ having been caused by primary infection $i$, 
$h_j=\{i: t_i\in h_{t_j}^0, t_i \in \cup_{v=j-\eta}^j\mathcal{T}_v, t_j\in \mathcal{T}_j\}$ and $\eta$ the number of former subintervals that influence the latent cases falling into $\mathcal{T}_j$ determined by the transmission kernel of latent cases ($\eta= 21$ days for COVID-19). Alternatively, by recording the parent of each latent infection at step 7 of Algorithm \ref{MdsAlg1a}, the proposed model can show the branching structure of the process. This approach increases the computational complexity of the algorithm, as more memory units will be required.

\paragraph*{Complexity} The computational costs of propagation (step 12 of KDPF) and finding weights (step 14 of KDPF) at state (interval) $j$ are equal at $O\left(N\sum\limits_{v=j-\eta}^j|S_v^N|\right)$. The computational cost of finding auxiliary weights (step 9 of KDPF) at state (interval) $j$ is the combined costs of propagation and finding weights.  Hence, the computational cost of the algorithm over all states (intervals) is $O\left(N \left(\eta+1\right)|\mathcal{T}^N|\right)$. The effectiveness of Algorithm \ref{MdsAlg1a} allows for a computationally realistic cost of Algorithm \ref{APAlga}. $N$ is the number of particles, $S_j^N$ the set of latent cases falling into subinterval $\mathcal{T}_j$, $\mathcal{T}^N=\cup_{j}S_j^N$ and $\eta$ the number of former subintervals that influence the latent cases falling into $\mathcal{T}_j$ determined by the transition kernel of latent cases. The $O$-notation denotes the asymptotic upper bound \citep{cormen2022introduction}. The algorithm is easily parallelized over $N$.
 
\paragraph*{Fixed-lag smoothing densities} As the resampling step leads to path degeneracy, it is difficult to obtain a good approximation of the smoothing density $p(x_{1:T}|y_{1:T})$ for large $T$ via SMC. Therefore, we use SMC to sample from the fixed-lag smoothing densities with lag $L$. Resampling results in replicating samples, and in the long run results in a lack of diversity called particle degeneracy \citep{endo2019introduction}. We apply the multinomial resampling step when the Effective Sample Size (ESS) is less than the $80\%$ of the number of particles. We reduce the frequency of resampling to balance the loss of information due to degeneracy with the loss of information due to the additional variability caused by resampling \citep{sheinson2014comparison}.

\begin{algorithm}[!h] 
\algsetup{linenosize=\tiny}
\tiny 
\caption{\bf Kernel density particle filter} 
\label{APAlga}
\begin{algorithmic}[1]
\STATE{Initialize the parameters $\{\theta_{j1}\}_{j=1}^N$, $\theta_{j1}=(d_{j1},v_{j1})$ with $d_{min}\leq d \leq d_{max}$ and $v_{min}\leq v \leq v_{max}$:\\
\hspace*{0.5em} \textbf{for} $j$ in $1:N$ \textbf{do} \\
\hspace*{1.5em}$\log d_{j1}=\mathcal{N}(\mu_d, \sigma_d^2)$ with $\mu_d=\frac{\log d_{max} +\log d_{min}}{2}$ and $\sigma_d=\frac{\log d_{max} -\log d_{min}}{8}$  \\
\hspace*{1.5em}$\log v_{j1}=\mathcal{N}(\mu_v, \sigma_v^2)$ with $\mu_v=\frac{\log v_{max} +\log v_{min}}{2}$ and $\sigma_v=\frac{\log v_{max} -\log v_{min}}{8}$\\
\hspace*{0.5em}\textbf{end for} \\
}
\STATE{Sample $N$ particles $\{X_{j1}\}_{j=1}^N$, $X_{j1}=\left(R_{j1},\ S_{j1}^N\right)$: \\
\hspace*{0.5em} \textbf{for} $j$ in $1:N$ \textbf{do} \\
\hspace*{1.5em}$R_{j1}\sim \mbox{Uniform}(\alpha,b)$ \\
\hspace*{1.5em}$S_{j1}^N \sim P\left(S_1^N|R_{j1},\mathcal{H}_0\right)$\\
\hspace*{0.5em}\textbf{end for}
}

\STATE{Find the weights, $\tilde{w}_1=\{\tilde{w}_{j1}\}_{j=1}^N$:\\
\hspace*{0.5em} \textbf{for} $j$ in $1:N$ \textbf{do} \\
\hspace*{1.5em}$\tilde{w}_{j1}=P\left(Y_1|S_{j1}^N,\beta,\mathcal{H}_0,v_{j1}\right)$ \\
\hspace*{0.5em}\textbf{end for}\\
}

\STATE{Normalize the weights, $w_1=\{w_{j1}\}_{j=1}^N$:\\
\hspace*{0.5em} \textbf{for} $j$ in $1:N$ \textbf{do} \\
\hspace*{1.5em}$w_{j1}=\frac{\tilde{w}_{j1}}{\sum\limits_{j=1}^N\tilde{w}_{j1}}$ \\
\hspace*{0.5em}\textbf{end for}
}

\FOR{$n=1,..,k$}
\STATE{
\hspace*{0.5em} \textbf{for} $j$ in $1:N$ \textbf{do} \\
\hspace*{1.5em}$m_{j,dn}^{(L)}=a\log d_{jn} + (1-a)\bar{d}_{Ln}, \hspace{0.5cm} \bar{d}_{Ln}=\sum\limits_{j=1}^Nw_{jn}\log d_{jn}$ \\
\hspace*{1.5em}$m_{j,vn}^{(L)}=a\log v_{jn} + (1-a)\bar{v}_{Ln}, \hspace{0.5cm} \bar{v}_{Ln}=\sum\limits_{j=1}^Nw_{jn}\log v_{jn}$ \\
\hspace*{1.5em}$m_{j,dn}=a d_{jn} + (1-a)\bar{d}_{n}, \hspace{0.5cm} \bar{d}_{n}=\sum\limits_{j=1}^Nw_{jn} d_{jn}$ \\
\hspace*{1.5em}$m_{j,vn}=a  v_{jn} + (1-a)\bar{v}_{n}, \hspace{0.5cm} \bar{v}_{n}=\sum\limits_{j=1}^N w_{jn} v_{jn}$ \\
\hspace*{0.5em}\textbf{end for}

}
\STATE{ For each particle $j$, we calculate an estimate of $X_{j,n+1}$ called $\tilde{X}_{j,n+1}$ by drawing a sample from $P(X_{n+1}|X_n,\mathcal{H}_0)$: \\
\hspace*{0.5em} \textbf{for} $j$ in $1:N$ \textbf{do} \\
\hspace*{1.5em}$\tilde{R}_{j,n+1}\sim P\left(R_{n+1}|R_{jn},m_{j,dn}\right)$  \\
\hspace*{1.5em}$\tilde{S}_{j,n+1}^N\sim P\left(S_{n+1}^N|S_{j,1:n}^N,\tilde{R}_{j,n+1},\mathcal{H}_0\right)$\\
\hspace*{0.5em}\textbf{end for}
}
\STATE{Find the auxiliary weights, $\tilde{g}_{n+1}=\{\tilde{g}_{j,n+1}\}_{j=1}^N$:\\
\hspace*{0.5em} \textbf{for} $j$ in $1:N$ \textbf{do} \\
\hspace*{1.5em}$\tilde{g}_{j,n+1}=g_{jn}w_{jn}P\left(Y_{n+1}|S_{j,1:n}^N,\tilde{S}_{j,n+1}^N,\beta,\mathcal{H}_0, m_{j,vn}\right)$\\  
\hspace*{0.5em}\textbf{end for}\\
}
\STATE{Normalize the auxiliary weights, $g_{n+1}=\{g_{j,n+1}\}_{j=1}^N$:\\
\hspace*{0.5em} \textbf{for} $j$ in $1:N$ \textbf{do} \\
\hspace*{1.5em}$g_{j,n+1}=\frac{\tilde{g}_{j,n+1}}{\sum\limits_{j=1}^N\tilde{g}_{j,n+1}}$ \\
\hspace*{0.5em}\textbf{end for}}

\STATE{
\textbf{if}\Big($ESS(g_{n+1})=1/\sum\limits_{j=1}^Ng_{j,n+1}^2<0.8N$\Big) \textbf{then}
resample and form $N$ equally weighted particles, $\Bar{X}_{n}=\{\Bar{X}_{n}^i\}_{i=1}^N$:\\
\hspace{0.5em}\textbf{for} $j$ in $1:N$ \textbf{do} \\
\hspace*{1em}(i) sample index $i_j$ from a multinomial distribution with probabilities $g_{n+1}$\\
\hspace*{1em}(ii) $\bar{X}_{n}^j=X_{i_j,n}$\\
\hspace*{1em}(iii) $g_{j,n+1}=1$\\
\hspace{0.5em}\textbf{end for}\\
\textbf{end if}
}

\STATE{Regenerate the fixed parameters: \\\hspace*{0.5em} \textbf{for} $j$ in $1:N$ \textbf{do} \\
\hspace*{1.5em}$ \log v_{j,n+1}\sim \mathcal{N}(m_{i_j,vn}^{(L)},h^2V_{nv}^{(L)})$ \\
\hspace*{1.5em}$\log d_{j,n+1}\sim \mathcal{N}(m_{i_j,dn}^{(L)},h^2V_{nd}^{(L)})$\\
\hspace*{0.5em}\textbf{end for}\\
where $V^{(L)}_{nv}$ is the weighted variance of $\{\log v_{jn}\}_{n=1}^N$ and $V^{(L)}_{nd}$ the weighted variance of $\{\log d_{jn}\}_{n=1}^N$.
}
\STATE{Using $\bar{X}_{n}$ propagate: \\
\hspace*{0.5em} \textbf{for} $j$ in $1:N$ \textbf{do} \\
\hspace*{1.5em}$R_{j,n+1}\sim P(R_{n+1}|R_{jn},d_{j,n+1})$ \\
\hspace*{1.5em}$S_{j,n+1}^N \sim P\left(S_{n+1}^N|S_{j,1:n}^N,R_{j,n+1},\mathcal{H}_0\right)$\\
\hspace{1.5em}Set $X_{j,n+1}=(\bar{X}_{n}^j,\left(R_{j,n+1},S_{j,n+1}^N)\right)$\\
\hspace*{0.5em}\textbf{end for}
}

\STATE{Find the weights, $\tilde{w}_{n+1}=\{\tilde{w}_{j,n+1}\}_{j=1}^N$:\\
\hspace*{0.5em} \textbf{for} $j$ in $1:N$ \textbf{do} \\
\hspace*{1.5em}$\tilde{w}_{j,n+1}=\frac{P\left(Y_{n+1}|S_{j,1:n+1}^N,\beta,\mathcal{H}_0,v_{j,n+1}\right)}{P\left(Y_{n+1}|S_{j,1:n}^N,\tilde{S}_{i_j,n+1}^N,\beta,\mathcal{H}_0,m_{i_j,vn}\right)}$ \\
\hspace*{0.5em}\textbf{end for}
}

\STATE{Normalize the weights, $w_{n+1}=\{w_{j,n+1}\}_{j=1}^N$:\\
\hspace*{0.5em} \textbf{for} $j$ in $1:N$ \textbf{do} \\
\hspace*{1.5em}$w_{j,n+1}=\frac{\tilde{w}_{j,n+1}}{\sum\limits_{j=1}^N\tilde{w}_{j,n+1}}$ \\
\hspace*{0.5em}\textbf{end for}
}

\STATE{ To draw a sample from $P\left(X_{1:n+1}|Y_{1:n+1}\right)$. We do resampling with weights $\{w_{j,n+1}\}_{j=1}^N\}$ if resampling was performed at step 10. Otherwise, we do resampling with weights $k_{j,n+1}\propto \tilde{w}_{j,n+1}g_{j,n+1}$.
}

\ENDFOR

\end{algorithmic}
\end{algorithm}

\section*{Results}
\subsection*{Simulation Analysis}\label{SSimulation}

We carried out a simulation study on synthetic data to illustrate the performance of the KDPF (Algorithm \ref{APAlga}) for inferring the intensity of latent cases, the reproduction number and the time-constant parameters.

\begin{figure}[!h] 
    \centering\includegraphics[width=8cm]{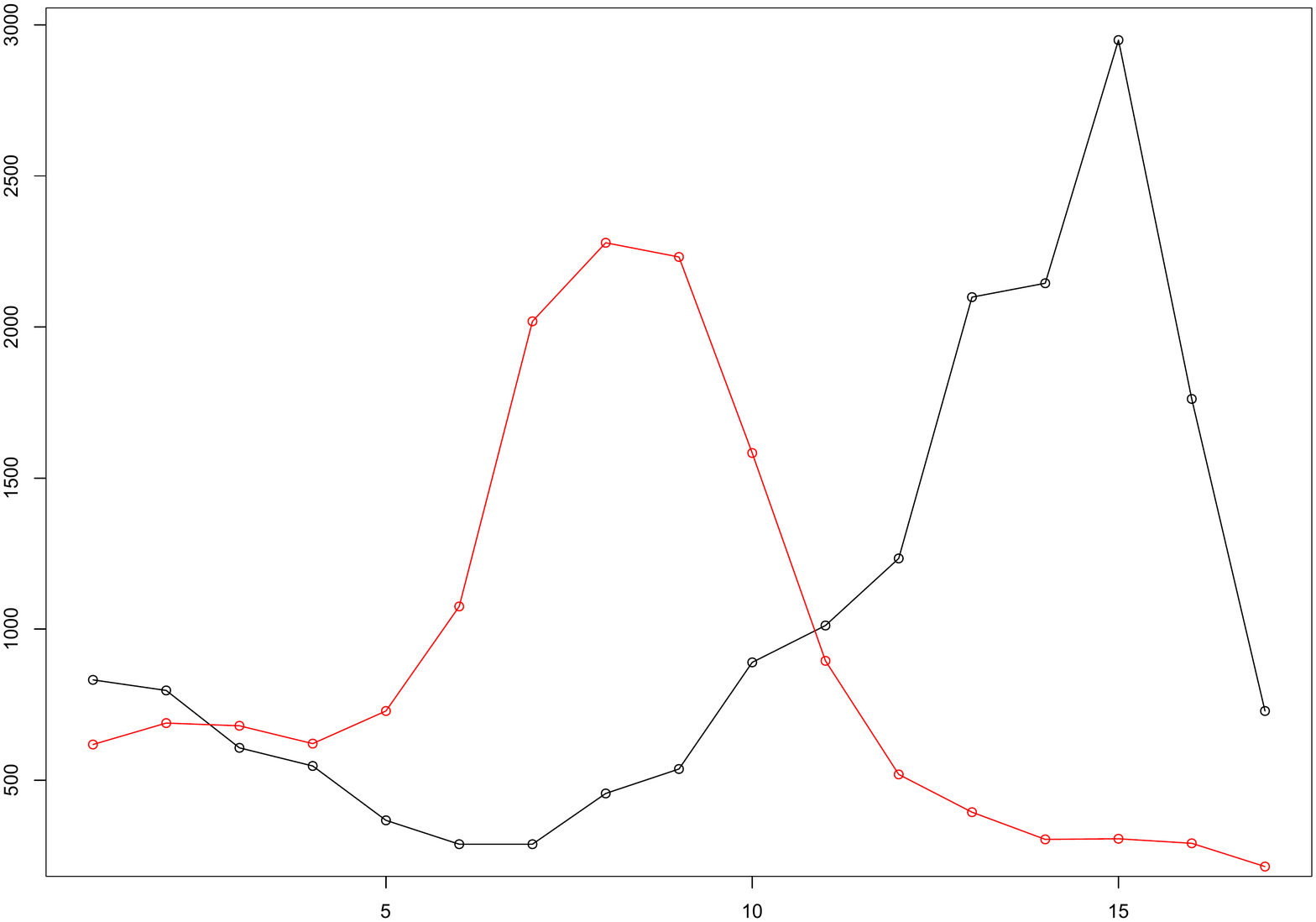}
  \caption{\bf {Weekly Observed Data} (Scenario A (black line); Scenario B (red line)) .}
  \label{FigWOD}
\end{figure}

Two different scenarios illustrated in Figures 1 and 2 were simulated as follows:
\begin{itemize}
\item{\textbf{Scenario A}: The process is triggered by 1745 infectious and the times of their infections, $\mathcal{H}_0$ , are uniformly allocated in 3 weeks ($[0,21)$) with a day being the time unit. We generate weekly latent and observed cases according to the model equations \ref{Mds1a}-\ref{Mds1b} for weeks 1-20 ($[21,161)$) given $\mathcal{H}_0$, $v=0.014$, $d=14.44$, $\beta=0.5$ and $R_{1}=1.79$. We are interested in inferring the latent cases in weeks $4-19$ with $\mathcal{H}_0$ being the set of times of latent infections in weeks 1-3. Using the generated observed cases in weeks 2-4, we estimate the latent infections in weeks 1-3 as follows: The latent cases in the week $i$ are equal to the number of observed events in the week $(i + 1)$ times $1/\beta$, and are spread uniformly in $[(i-1)\times7 + 21, i\times7 +21)$ for $1\leq i \leq 3$. We assume $\alpha=1$, $b=2$, $d_{min}=10$, $d_{max}=20$, $v_{min}=0.0001$ and $v_{max}=0.5$. The ground truth is characterized by $\mathcal{H}_0$ consisting of 4855 seeds, while the estimated seeds are 4228. The observed cases in weeks 4-20 are 17540 (see Figure \ref{FigWOD}).  
}
\item{\textbf{Scenario B}: The process is triggered by 1176 seeds and generated as described above. We assume $d=15.28$, $v=0.001$, $\beta=0.5$, $R_1=1.51$, $d_{min}=10$, $d_{max}=20$, $v_{min}=0.001$, $v_{max}=0.5$, $\alpha=1$ and $b=2$. The observed cases in weeks 4-20 are 15448 (see Figure \ref{FigWOD}). }
\end{itemize}

We deal with 16 hidden states $\{X_n\}_{n=1}^{16}$. Each state $X_n$ is associated with the latent cases falling during the week $\mathcal{T}_n$ and the parameter $R_n$ associated with that week. We infer the latent intensity $\lambda^N(t)$ and the weights $\{R_{n}\}_{n=1}^{16}$ as well as the weekly latent cases via the  particle sample derived by drawing samples from the smoothing density with lag equal to 4. 

\begin{figure}[!h] 
  \begin{subfigure}{6cm}
    \centering\includegraphics[width=6cm]{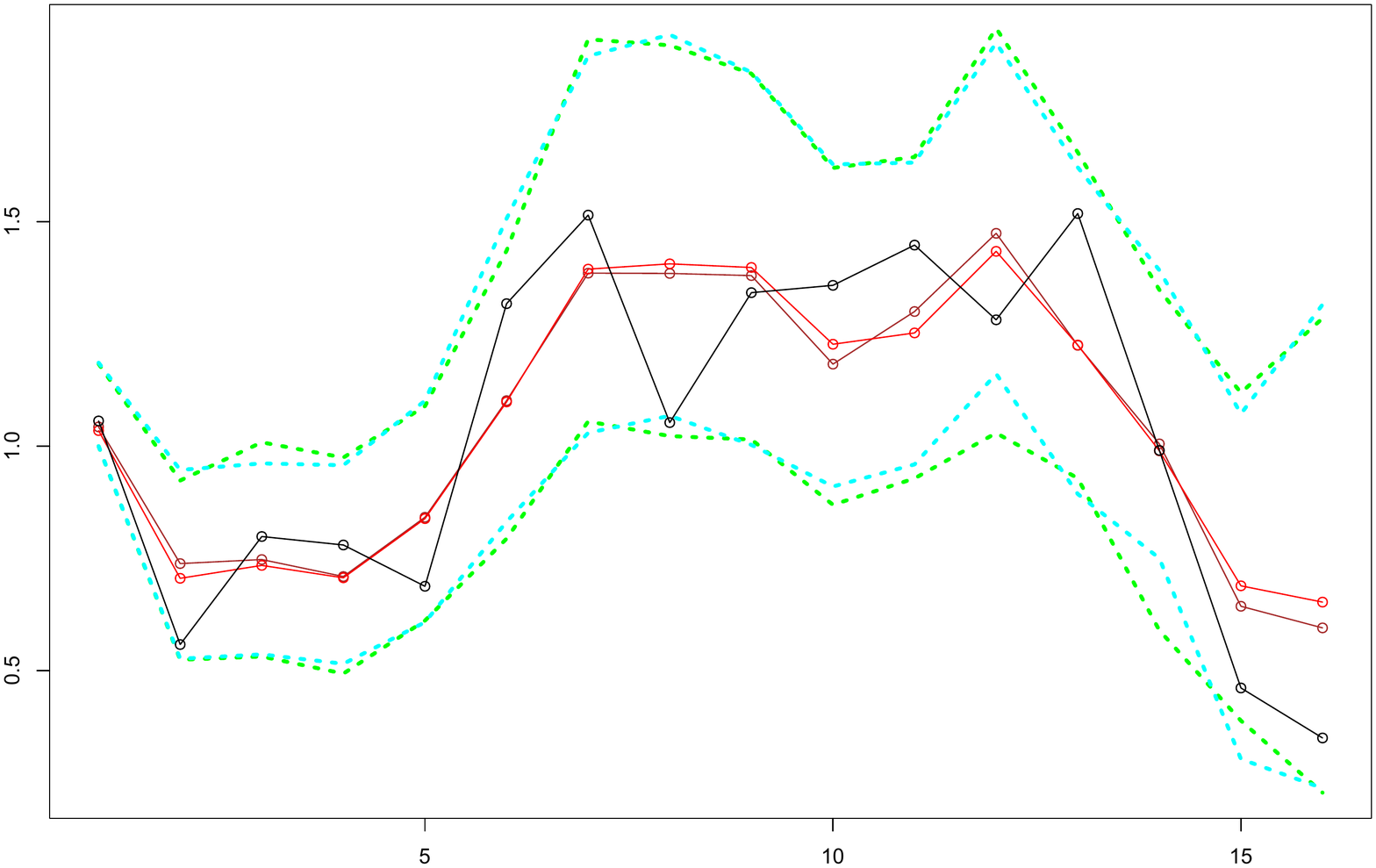}
 \caption{$\{R_{n}\}_{n=1}^{16}$}
 \end{subfigure}
  \begin{subfigure}{6cm}
    \centering\includegraphics[width=6cm]{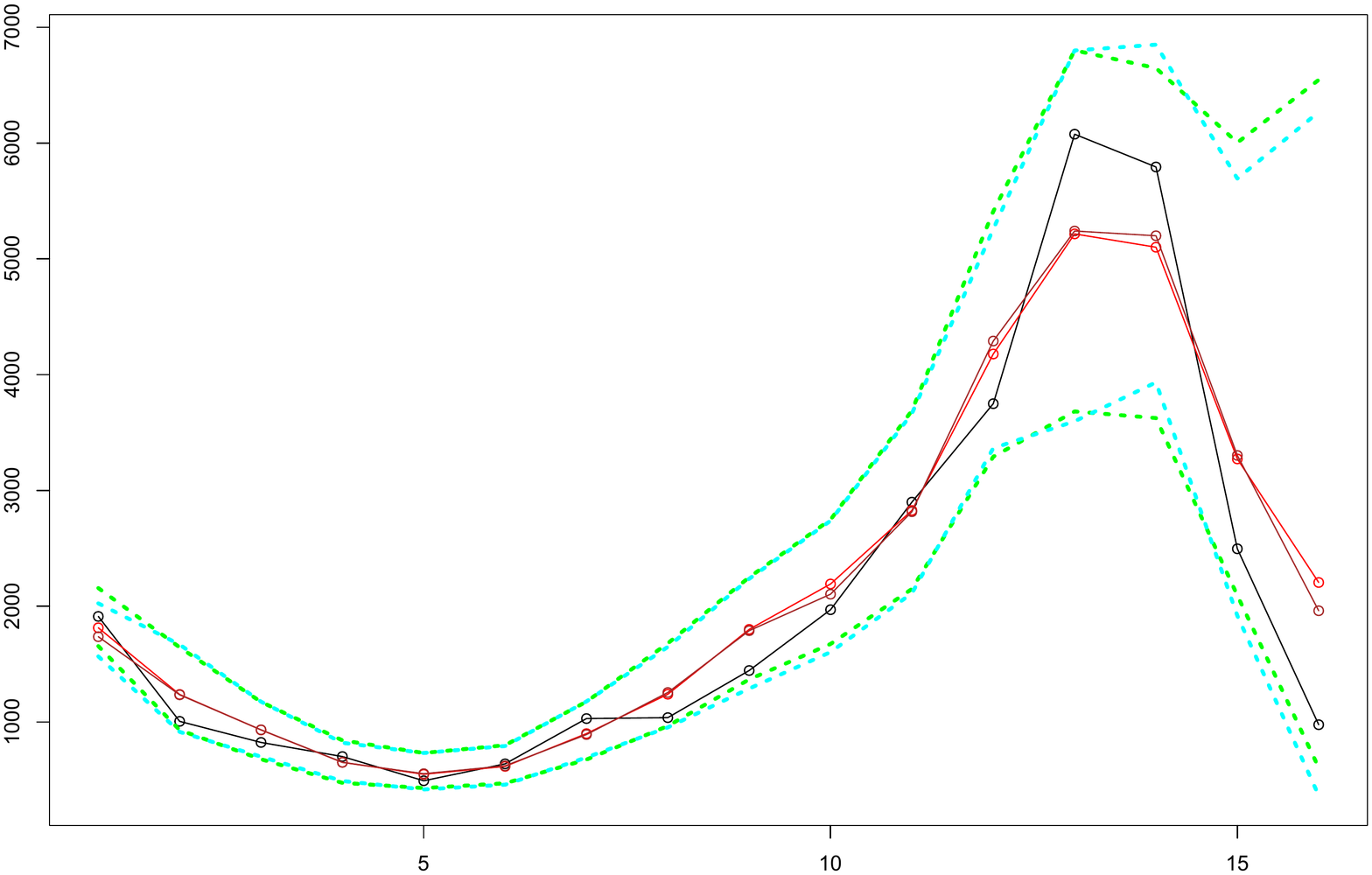}
 \caption{Weekly latent cases.}
 \end{subfigure}
 \begin{subfigure}{6cm}
    \centering\includegraphics[width=6cm]{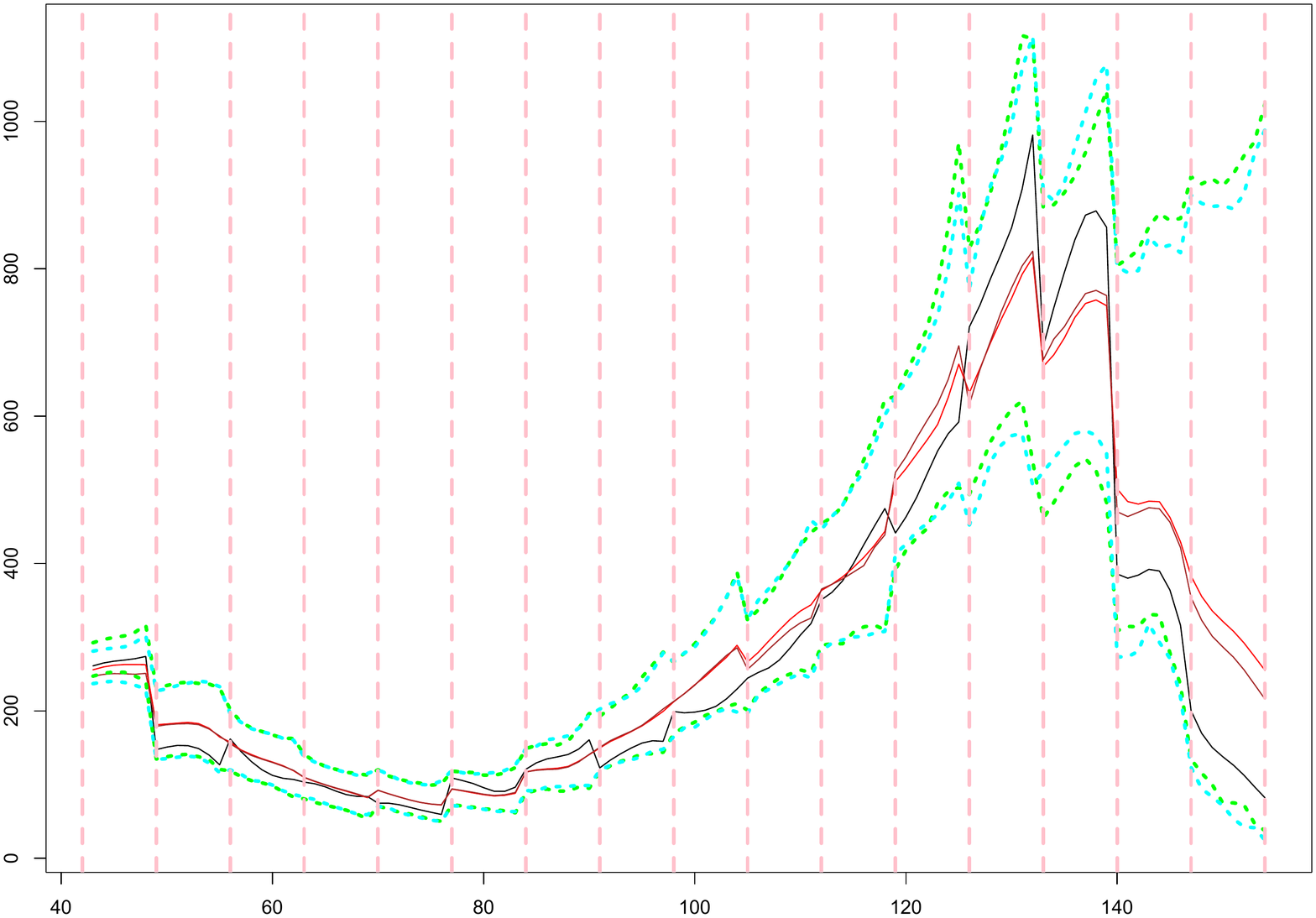}
  \caption{Intensity of latent cases.}
 \end{subfigure}
  \caption{{\bf The true (black line) and the estimated weighs $\{R_{n}\}_{n=1}^{16}$, weekly latent cases and latent intensity (with estimated seeds (posterior median (brown line) ; 99\% CI (cyan line)), and true seeds (posterior median (red line) ; 99\% CI (green line))).} The vertical dotted lines show the beginning of each week in the period we examine.}
  \label{FigSc1a}
\end{figure}

\begin{figure}[!h] 
  \begin{subfigure}{6cm}
    \centering\includegraphics[width=6cm]{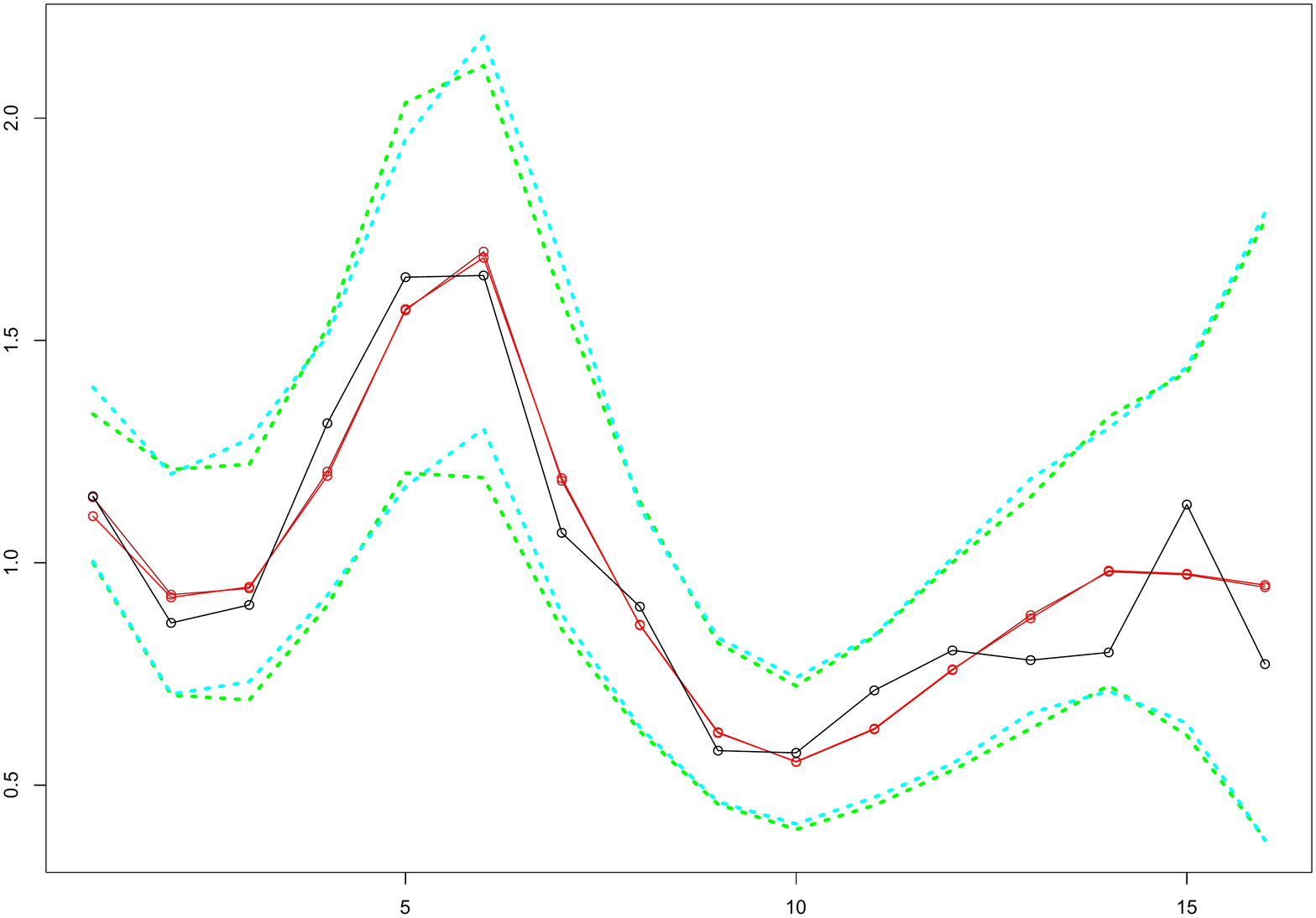}
 \caption{$\{R_{n}\}_{n=1}^{16}$}
 \end{subfigure}
  \begin{subfigure}{6cm}
    \centering\includegraphics[width=6cm]{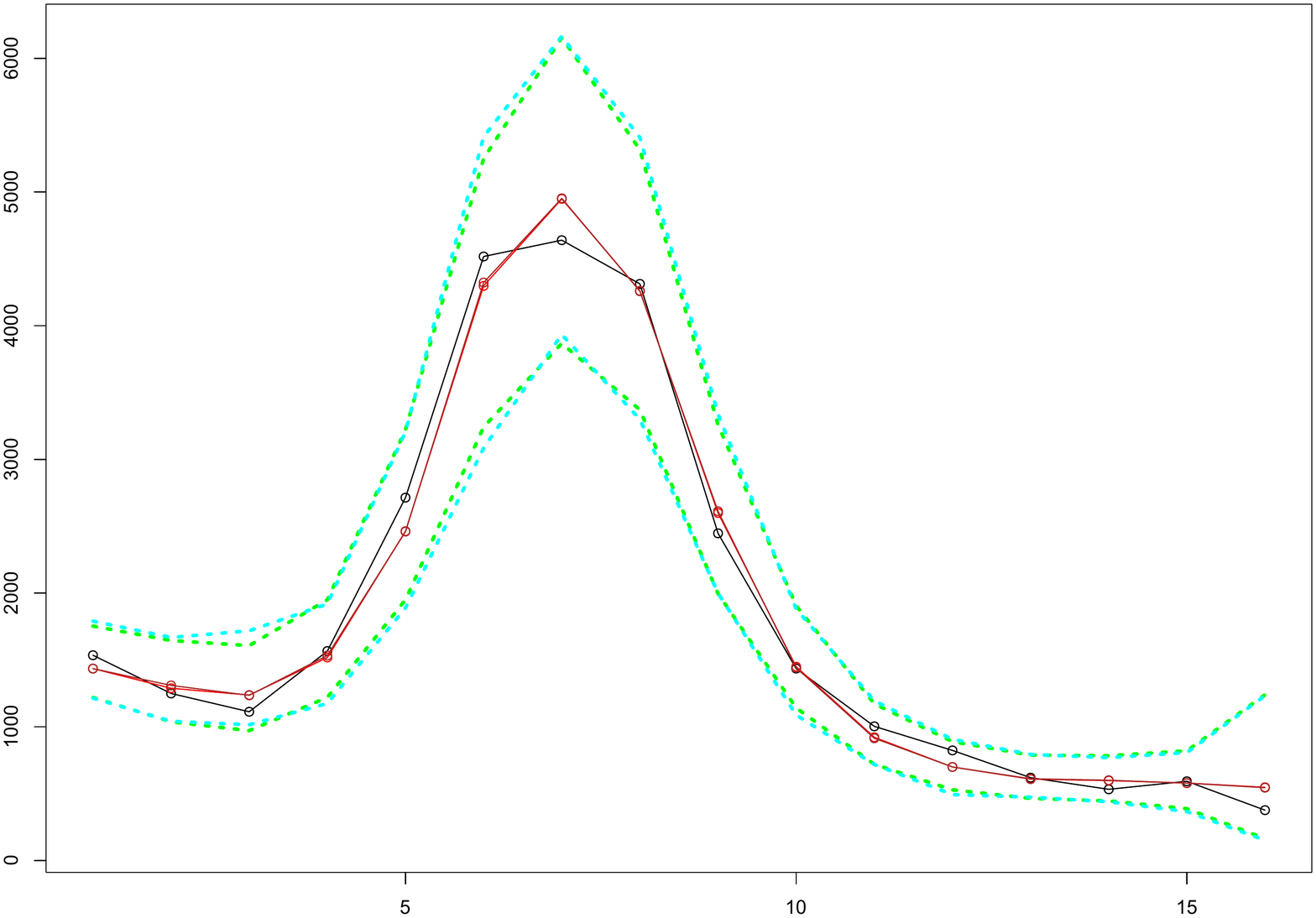}
 \caption{Weekly latent cases.}
 \end{subfigure}
 \begin{subfigure}{6cm}
    \centering\includegraphics[width=6cm]{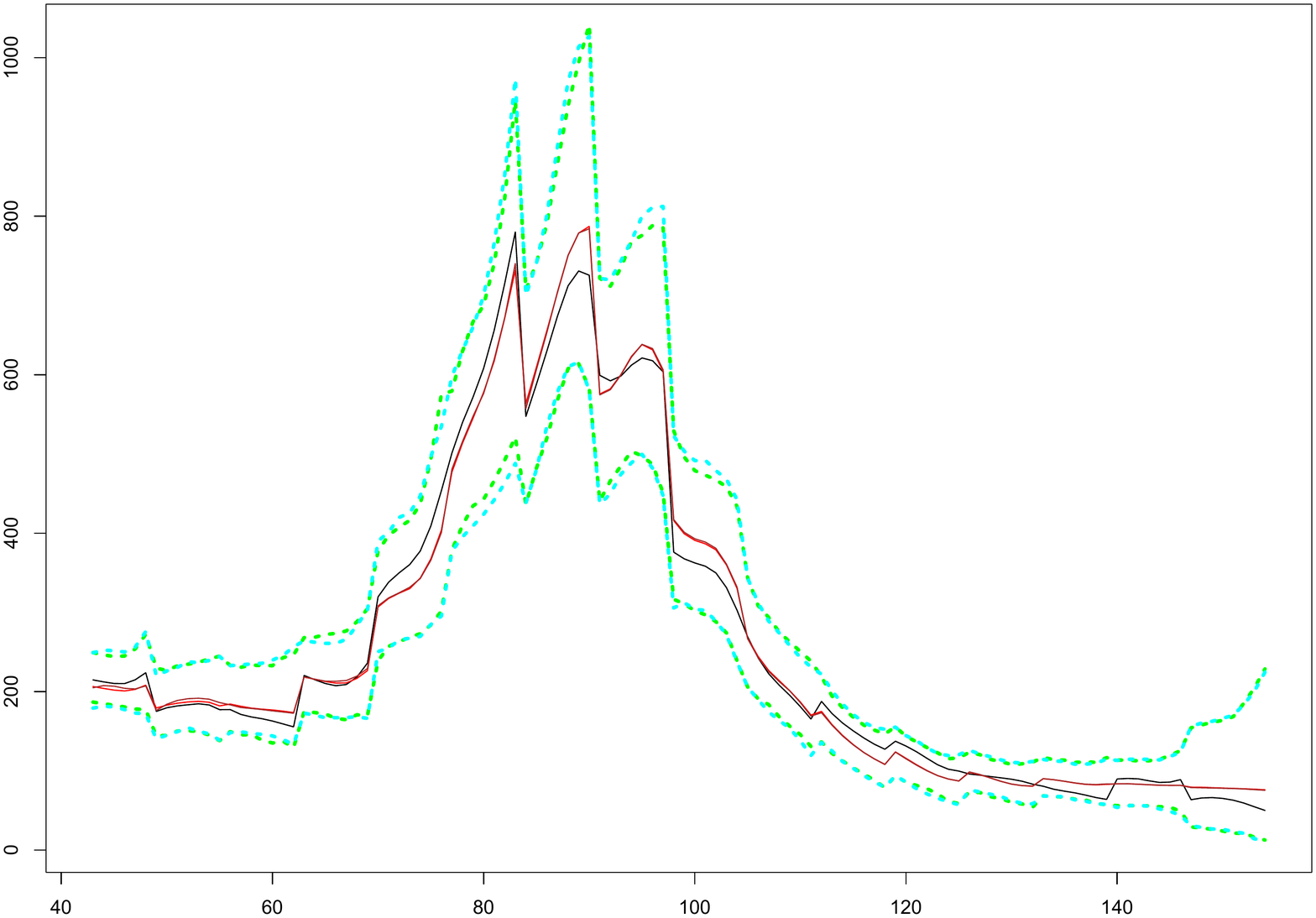}
  \caption{Intensity of latent cases.}
 \end{subfigure}
  \caption{{\bf The true (black line) and the estimated weighs $\{R_{n}\}_{n=1}^{16}$, weekly latent cases and latent intensity (with estimated seeds (posterior median (brown line) ; 99\% CI (cyan line)), and true seeds (posterior median (red line) ; 99\% CI (green line))).} The vertical dotted lines show the beginning of each week in the period we examine.}
  \label{FigSc2a}
\end{figure}

Figures \ref{FigSc1a} - \ref{FigSc2a} illustrate the estimated latent intensity, the estimated weekly hidden cases and the estimated weights of the reproduction number for both scenarios using 40000 particles. We note that the 99\% Credible Intervals (CIs) of the time-constant parameters include the actual values of the parameters. The simulation analysis shows that the KDPF approaches well the ground truth.   
 
To confirm the convergence of posterior estimates of weights and weekly hidden cases concerning the number of particles ($N$), we find the associated Monte Carlo Standard Errors (MCSEs). The MCSEs of posterior means of weights and weekly latent cases are given by
\begin{equation*}
    MCSE(R)=\frac{1}{16}\sum_{i=1}^{16}\left(\frac{\mathrm{var}(R_{i})}{N} \right)^{1/2}
\end{equation*} and
\begin{equation*}
    MCSE(Y)=\frac{1}{16}\sum_{i=1}^{16}\left(\frac{\mathrm{var}(Y_{i})}{N} \right)^{1/2}
\end{equation*}
where $\mathrm{var}(z)$ is the variance of $z$ and $Y_{i}$ the aggregate latent cases in $i_{th}$ week. The MCSE verifies the convergence of posterior estimates concerning the number of particles (see Tables \ref{tableSc1a} - \ref{tableSc2a}).

\begin{table}[!ht]
\centering
\caption{\bf MCSEs of posterior means of weights and  weekly hidden cases in Scenario A.}
\begin{tabular}{ |l|l|l|l|}
\hline
 \multicolumn{4}{|l|}{\bf Convergence of the posterior estimates}  \\
 \thickhline
  $MCSE$ & $N=20000$ & $N=30000$ & $N=40000$\\
 \hline
 $R$& 0.000626& 0.000662& 0.000668\\ \hline
 $Y$&1.617828  &  1.554482 & 1.529521 \\  
 \hline
\end{tabular}

\label{tableSc1a}
\end{table}

\begin{table}[!ht]
\centering
\caption{\bf MCSEs of posterior means of weights and  weekly hidden cases in Scenario B.}
\begin{tabular}{|l|l|l|l|}
 \hline
 \multicolumn{4}{|l|}{\bf Convergence of the posterior estimates}  \\
\thickhline
$MCSE$ & $N=20000$ & $N=30000$ & $N=40000$\\
 \hline
 $R$ & 0.000637 & 0.000655 & 0.000629\\ \hline
 $Y$ & 0.997832 & 0.993364  & 0.978517 \\
 \hline
\end{tabular}

\label{tableSc2a}
\end{table}

\begin{figure}[!h] 
  \begin{subfigure}{6cm}
    \centering\includegraphics[width=6cm]{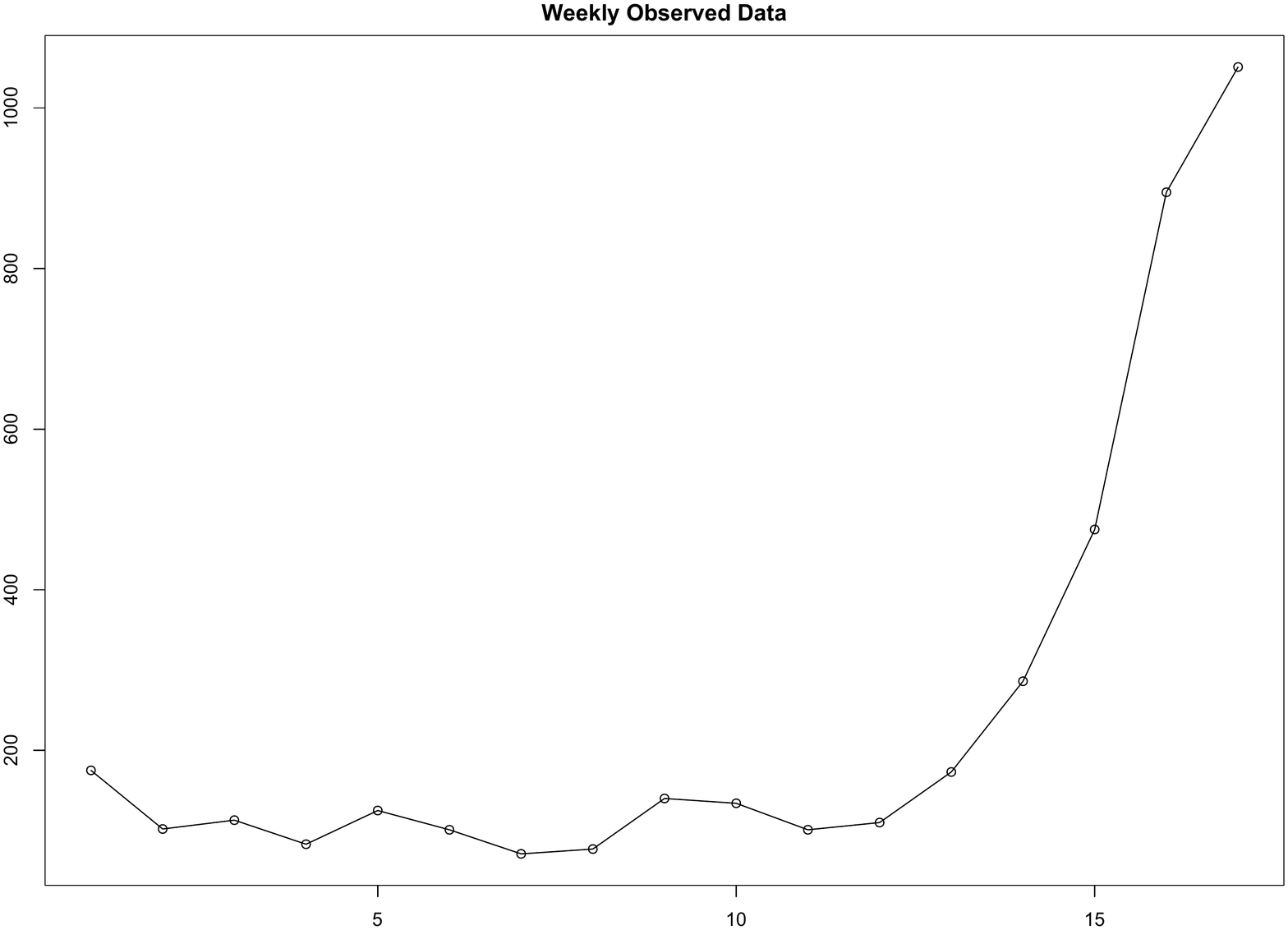}
    \caption{The weekly observed cases.} 
  \end{subfigure}
  \begin{subfigure}{6cm}
    \centering\includegraphics[width=6cm]{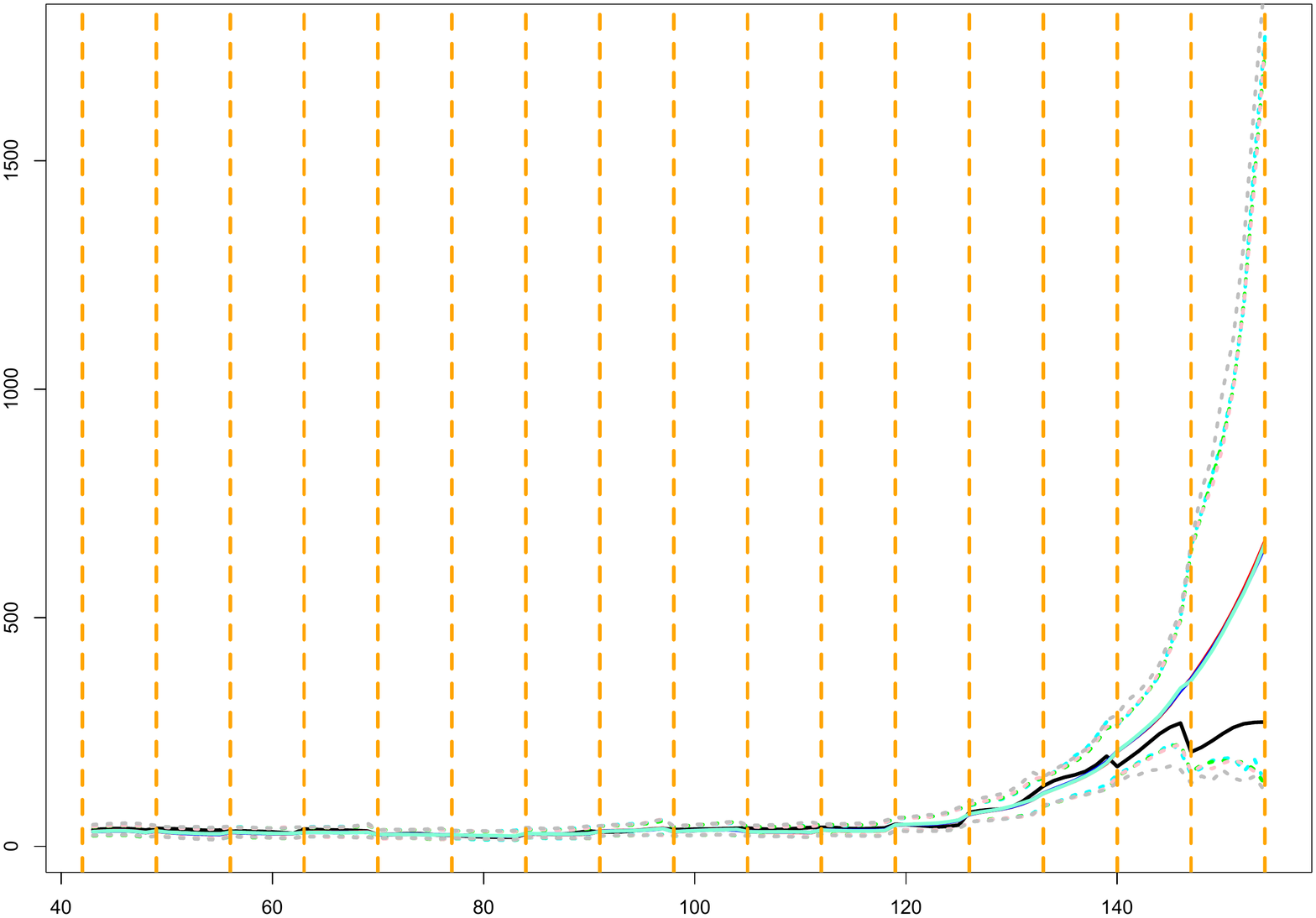}
   \caption{The estimated intensity of latent cases.}
  \end{subfigure}
  
  \begin{subfigure}{6cm}
    \centering\includegraphics[width=6cm]{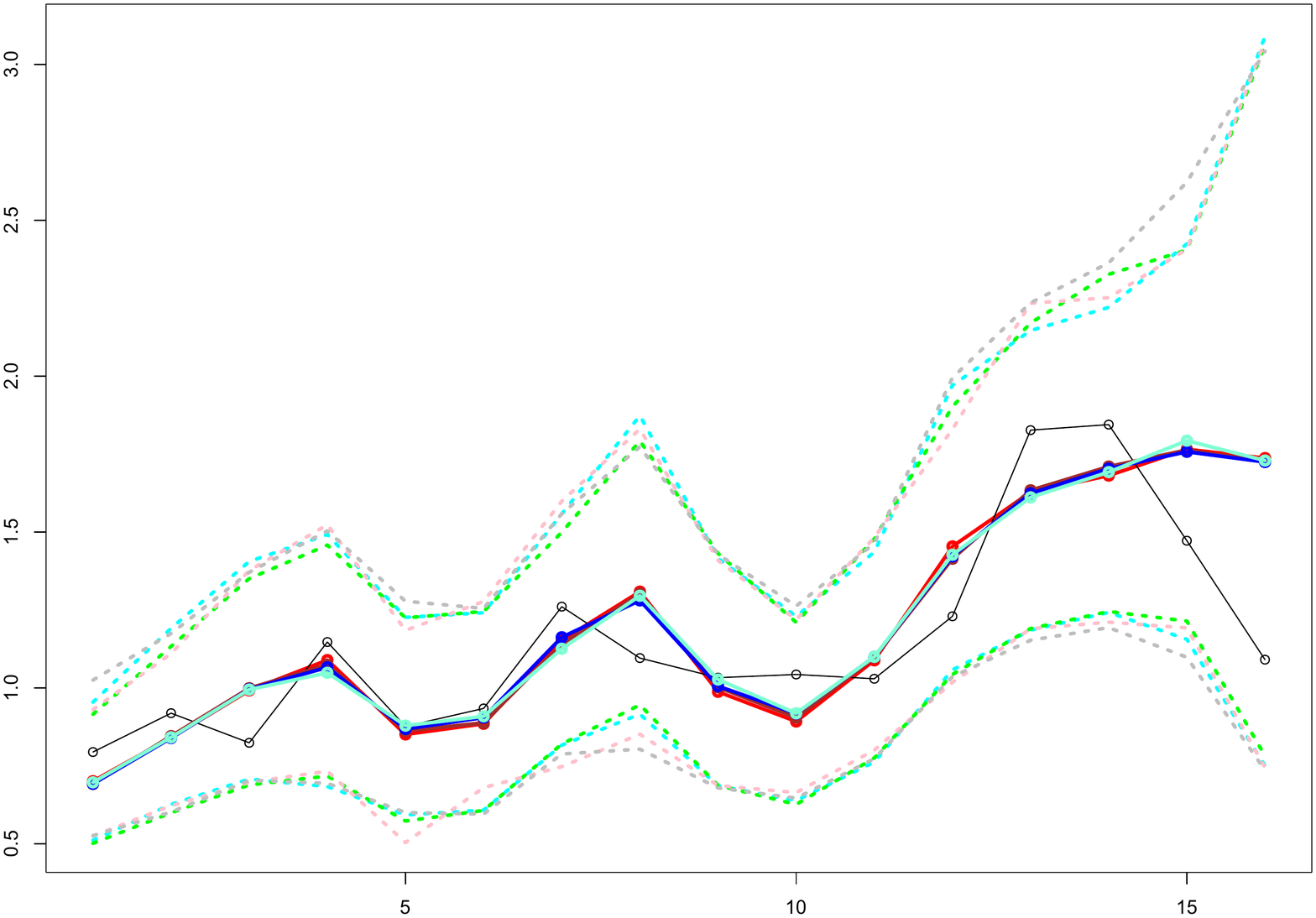}
    \caption{The estimated weights $\{R_n\}_{n=1}^{16}$.}
  \end{subfigure}
  \begin{subfigure}{6cm}
    \centering\includegraphics[width=6cm]{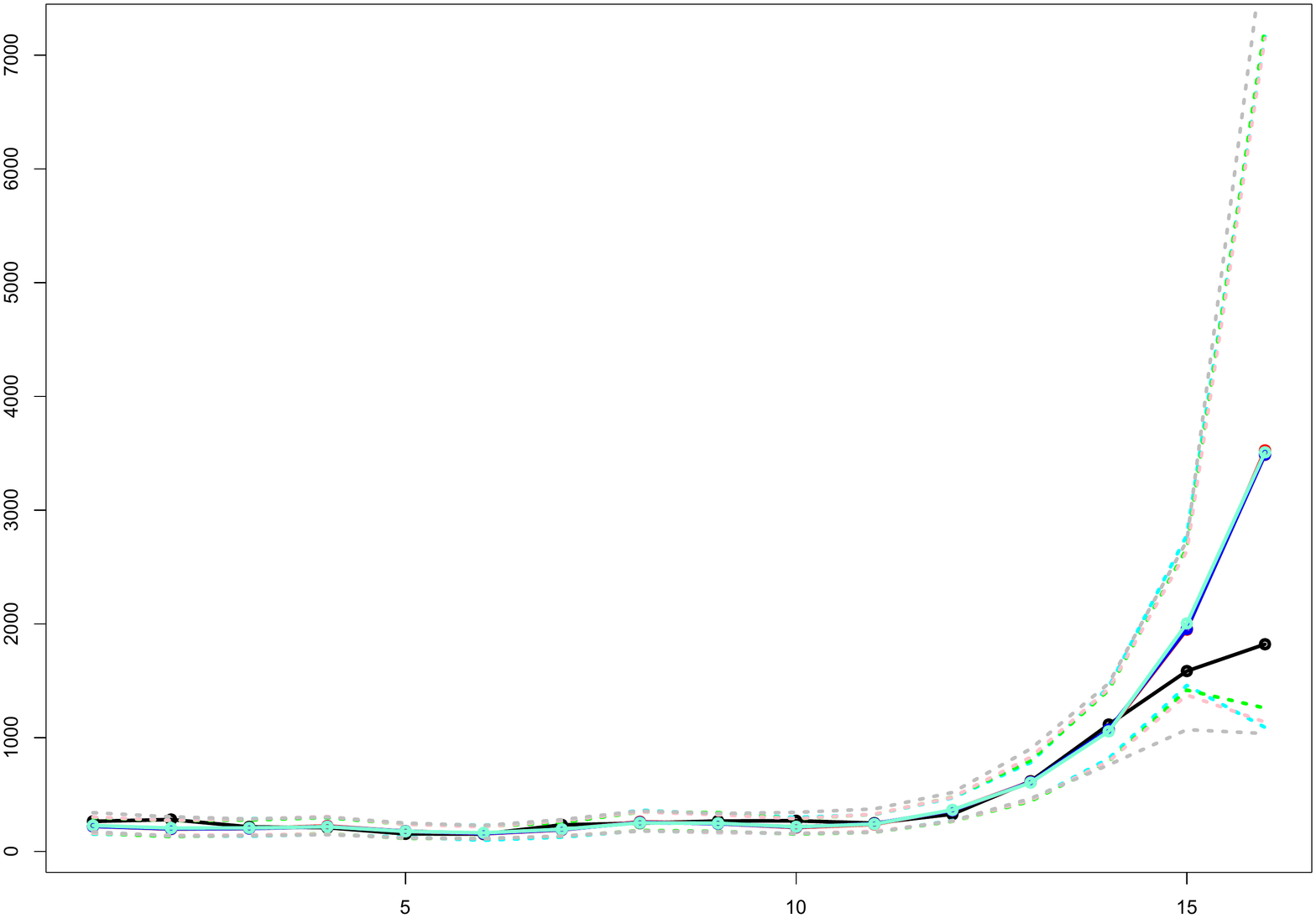}
    \caption{The estimated weekly hidden cases.}
  \end{subfigure}

  \caption{ {\bf The weekly observed cases, the estimated intensity, the estimated reproduction number, the estimated weekly hidden cases using KDPF( median (red line) ; 99\% CI (cyan line)), using APF (median (brown line); 99\% CI (green dashed lines)), using PMMH (median (aquamarine line); 99\% CI (grey dashed lines)), using BF ( median (blue line); 99\% CI (pink dashed lines)) and the true values (black line) in scenario C.} The vertical dotted lines show the beginning of each week in the period we examine.}
  \label{FigSDC}
\end{figure}

Finally, we compare the performance of the KDPF (Algorithm \ref{APAlga}), APF (Algorithm \ref{APFAlg}), bootstrap filter (BF) (Algorithm \ref{BFAlg}) and particle marginal Metropolis-Hastings (PMMH) (Algorithm \ref{PMMHAlg}) \citep{andrieu2010particle} for inferring the latent intensity $\lambda^N(t)$ and the reproduction number $R(t)$ illustrated in a new simulation scenario C (see Figure \ref{FigSDC}). Scenario C concerns a process triggered by 661 infectious and generated similar to scenario A assuming that $\alpha=0.5$, $b=2$, $d=15.11$, $v=0.01$, $R_1=1.57$, $d_{min}=10$, $d_{max}=20$, $v_{min}=0.0001$ and $v_{max}=0.5$. The time-constant parameters $d$ and $v$ are known for BF and APF. We used 10000 iterations of the PMMH sampler with a burn-in of 5000 iterations. We use APF using 50 particles as an SMC sampler. The average acceptance ratio is about 0.1844 resulting in a Markov chain that mixes well. For the KDPF, $\Delta$ was set to 0.99.

We find the Average Absolute Error (AAE) of the computed estimates:
\begin{align*}
AAE(\hat{\lambda}^N) &=\frac{1}{N_t}\sum\limits_{i=1}^{N_t}|\hat{\lambda}^N(x_i)-\lambda^N(x_i)| \\
AAR(\hat{R}) &=\frac{1}{16}\sum\limits_{i=1}^{16}|\hat{R}_i-R_i| 
\end{align*}
 and the Root Mean Square Error (RMSE):
\begin{align*}
RMSE(\hat{\lambda}^N) &=\sqrt{\frac{1}{N_t}\sum\limits_{i=1}^{N_t}\Big(\hat{\lambda}^N(x_i)-\lambda^N(x_i)\Big)^2} \\
RMSE(\hat{R}) &=\sqrt{\frac{1}{16}\sum\limits_{i=1}^{16}(\hat{R}_i-R_i)^2},\\ 
\end{align*}
where $N_t$ is the number of test time points $x_i$ randomly chosen in the time-horizon we consider, $\hat{\lambda}^N(x_i)$ and $\lambda^N(x_i)$ the estimated via posterior median and true intensity at time $x_i$, $\hat{R}_i$ and $R_i$ the estimated via posterior median and true reproduction number in the $i_{th}$ week. 

Table \ref{tableSDA} shows the errors related to KDPF, APF, BF and PMMH for scenario C. The errors associated with KDPF are comparable to those obtained using BF and APF for which the time-constant parameters are known. The performance of KDPF compares well with PMMH, having the advantage that it is a more computationally efficient algorithm than PMMH.

\begin{table}[!ht]
\centering
\caption{\bf Average Absolute Error and Root  Mean Square Error for the reproduction number and the latent intensity in scenario C.}
\begin{tabular}{ |l|l|l|l|l| }
 \hline
 \multicolumn{5}{|l|}{\bf Scenario C}  \\
 \thickhline
 Filter & AAE($\hat{R}$) & RMSE($\hat{R}$) & AAE($\hat{\lambda}^N$) &  RMSE($\hat{\lambda}^N$)  \\
 \hline
 KDPF & 0.16  & 0.22  & 25.19 & 73.04  \\ \hline
 APF & 0.16 & 0.21  & 24.79 &71.56  \\ \hline
  BF & 0.15  & 0.21  &24.7 & 71.43  \\ \hline
  PMMH & 0.16  & 0.22  & 24.87 & 71.55  \\
\hline
\end{tabular}
\label{tableSDA}
\end{table}

 \begin{algorithm}[!h] 
\algsetup{linenosize=\tiny}
\tiny 
\caption{\bf Auxiliary particle filter} 
\label{APFAlg}
\begin{algorithmic}[1]
\STATE{Sample $N$ particles $\{X_{j1}\}_{j=1}^N$, $X_{j1}=(R_{j1},S_{j1}^N)$: \\
\hspace*{0.5em} \textbf{for} $j$ in $1:N$ \textbf{do} \\
\hspace*{1.5em}$R_{j1}\sim \mu(R_1)$ \\
\hspace*{1.5em}$S_{j1}^N\sim P(S_1^N|R_{j1},\mathcal{H}_0)$\\
\hspace*{0.5em}\textbf{end for}
}
\STATE{Find the weights, $\tilde{w}_1=\{\tilde{w}_{j1}\}_{j=1}^N$:\\
\hspace*{0.5em} \textbf{for} $j$ in $1:N$ \textbf{do} \\
\hspace*{1.5em}$\tilde{w}_{j1}=P(Y_1|S_{j1}^N,\beta,\mathcal{H}_0,v)$ \\
\hspace*{0.5em}\textbf{end for}
}
\STATE{Normalize the weights, $w_1=\{w_{j1}\}_{j=1}^N$:\\
\hspace*{0.5em} \textbf{for} $j$ in $1:N$ \textbf{do} \\
\hspace*{1.5em}$w_{j1}=\frac{\tilde{w}_{j1}}{\sum\limits_{j=1}^N\tilde{w}_{j1}}$ \\
\hspace*{0.5em}\textbf{end for}
}
\STATE{Estimate $P(Y_1): \hat{P}(Y_1)=\frac{1}{N}\sum\limits_{j=1}^N\tilde{w}_{j1}$}
\FOR{$n=1,..,k$}
\STATE{ For each particle $j$, we calculate an estimate of $X_{j,n+1}$ called $\tilde{X}_{j,n+1}$ by drawing a sample from $P(X_{n+1}|X_n,\mathcal{H}_0,d)$: \\
\hspace*{0.5em} \textbf{for} $j$ in $1:N$ \textbf{do} \\
\hspace*{1.5em}$\tilde{R}_{j,n+1}\sim P(R_{n+1}|R_{jn},d)$ \\
\hspace*{1.5em}$\tilde{S}_{j,n+1}^N\sim P(S_{n+1}^N|S_{j,1:n}^N,\tilde{R}_{j,n+1},\mathcal{H}_0)$\\
\hspace*{0.5em}\textbf{end for}
}
\STATE{Find the auxiliary weights, $\tilde{g}_{n+1}=\{\tilde{g}_{j,n+1}\}_{j=1}^N$:\\
\hspace*{0.5em} \textbf{for} $j$ in $1:N$ \textbf{do} \\
\hspace*{1.5em}$\tilde{g}_{j,n+1}=g_{jn}w_{jn}P(Y_{n+1}|S_{j,1:n}^N,\tilde{S}_{j,n+1}^N,\beta,\mathcal{H}_0, v)$ \\
\hspace*{0.5em}\textbf{end for}
}
\STATE{Normalize the auxiliary weights, $g_{n+1}=\{g_{j,n+1}\}_{j=1}^N$:\\
\hspace*{0.5em} \textbf{for} $j$ in $1:N$ \textbf{do} \\
\hspace*{1.5em}$g_{j,n+1}=\frac{\tilde{g}_{j,n+1}}{\sum\limits_{j=1}^N\tilde{g}_{j,n+1}}$ \\
\hspace*{0.5em}\textbf{end for}}

\STATE{
Resample and form $N$ equally weighted particles, $\Bar{X}_{n}=\{\Bar{X}_{n}^i\}_{i=1}^N$:\\
\hspace{0.5em}\textbf{for} $j$ in $1:N$ \textbf{do} \\
\hspace*{1em}(i) sample index $i_j$ from a multinomial distribution with probabilities $g_{n+1}$\\
\hspace*{1em}(ii) $\bar{X}_{n}^j=X_{i_j,n}$\\
\hspace*{1em}(iii) $g_{j,n+1}=1$\\
\hspace{0.5em}\textbf{end for}\\
\textbf{end if}

}
\STATE{Using $\bar{X}_{n}$ propagate: \\
\hspace*{0.5em} \textbf{for} $j$ in $1:N$ \textbf{do} \\
\hspace*{1.5em}$R_{j,n+1}\sim P(R_{n+1}|R_{jn},d)$ \\
\hspace*{1.5em}$S_{j,n+1}^N \sim P(S_{n+1}^N|S_{j,1:n}^N,R_{j,n+1},\mathcal{H}_0)$\\
\hspace{1.5em}Set $X_{j,n+1}=(\bar{X}_{n}^j,(R_{j,n+1},S_{j,n+1}^N))$\\
\hspace*{0.5em}\textbf{end for}
}

\STATE{Find the weights, $\tilde{w}_{n+1}=\{\tilde{w}_{j,n+1}\}_{j=1}^N$:\\
\hspace*{0.5em} \textbf{for} $j$ in $1:N$ \textbf{do} \\
\hspace*{1.5em}$\tilde{w}_{j,n+1}=\frac{P(Y_{n+1}|S_{j,1:n+1}^N,\beta,\mathcal{H}_0,v)}{P(Y_{n+1}|S_{i_j,1:n}^N,\tilde{S}_{i_j,n+1}^N,\beta,\mathcal{H}_0,v)}$ \\
\hspace*{0.5em}\textbf{end for}
}

\STATE{Normalize the weights, $w_{n+1}=\{w_{j,n+1}\}_{j=1}^N$:\\
\hspace*{0.5em} \textbf{for} $j$ in $1:N$ \textbf{do} \\
\hspace*{1.5em}$w_{j,n+1}=\frac{\tilde{w}_{j,n+1}}{\sum\limits_{j=1}^N\tilde{w}_{j,n+1}}$ \\
\hspace*{0.5em}\textbf{end for}
}

\STATE{Estimate $P(Y_{n+1}|Y_{1:n}): \hat{P}(Y_{n+1}|Y_{1:n})=\Big(\sum\limits_{i=1}^Nw_{i,n}P(Y_{n+1}|S_{i,1:n}^N,\tilde{S}_{i,n+1}^N,\beta,\mathcal{H}_0, v)\Big)\Big(\frac{1}{N}\sum\limits_{i=1}^N\tilde{w}_{i,n+1}\Big)$. }
\ENDFOR

\STATE{Estimate the marginal likelihood $P(Y_{1:k})$: $\hat{P}(Y_{1:k})=\hat{P}(Y_1)\prod\limits_{n=2}^k\hat{P}(Y_n|Y_{1:n-1})$.} 
\end{algorithmic}
\end{algorithm}

 \begin{algorithm}[!h] 
\algsetup{linenosize=\tiny}

\tiny
\caption{\bf Bootstrap filter} 
\label{BFAlg}
\begin{algorithmic}[1]
\STATE{Sample $N$ particles $\{X_{j1}\}_{j=1}^N$, $X_{j1}=(R_{j1},S_{j1}^N)$: \\
\hspace*{0.5em} \textbf{for} $j$ in $1:N$ \textbf{do} \\
\hspace*{1.5em}$R_{j1}\sim \mu(R_1)$ \\
\hspace*{1.5em}$S_{j1}^N\sim P(S_1^N|R_{j1},\mathcal{H}_0)$\\
\hspace*{0.5em}\textbf{end for}
}
\STATE{Find the weights, $\tilde{w}_1=\{\tilde{w}_{j1}\}_{j=1}^N$:\\
\hspace*{0.5em} \textbf{for} $j$ in $1:N$ \textbf{do} \\
\hspace*{1.5em}$\tilde{w}_{j1}=P(Y_1|S_{j1}^N,\beta,\mathcal{H}_0,v)$ \\
\hspace*{0.5em}\textbf{end for}
}
\STATE{Normalize the weights, $w_1=\{w_{j1}\}_{j=1}^N$:\\
\hspace*{0.5em} \textbf{for} $j$ in $1:N$ \textbf{do} \\
\hspace*{1.5em}$w_{j1}=\frac{\tilde{w}_{j1}}{\sum\limits_{j=1}^N\tilde{w}_{j1}}$ \\
\hspace*{0.5em}\textbf{end for}
}
\STATE{Estimate $P(Y_1): \hat{P}(Y_1)=\frac{1}{N}\sum\limits_{j=1}^N\tilde{w}_{j1}$}

\FOR{$n=2,..,k$}
\STATE{ Resample and form $N$ equally weighted particles, $\Bar{X}_{n-1}=\{\Bar{X}_{n-1}^i\}_{i=1}^N$:\\
\hspace{0.5em}\textbf{for} $j$ in $1:N$ \textbf{do} \\
\hspace*{1em}(i) sample index $i_j$ from a multinomial distribution with probabilities $w_{n-1}$\\
\hspace*{1em}(ii) $\bar{X}_{n-1}^j=X_{i_j,n-1}$\\
\hspace*{1em}(iii) $w_{j,n-1}=1$\\
\hspace{0.5em}\textbf{end for}\\
\textbf{end if}

}
\STATE{Using $\bar{X}_{n-1}$ propagate: \\
\hspace*{0.5em} \textbf{for} $j$ in $1:N$ \textbf{do} \\
\hspace*{1.5em}$R_{j,n}\sim P(R_{n}|R_{j,n-1},d)$ \\
\hspace*{1.5em}$S_{j,n}^N \sim P(S_{n}^N|S_{j,1:n-1}^N,R_{j,n},\mathcal{H}_0)$\\
\hspace{1.5em}Set $X_{j,n}=(\bar{X}_{n-1}^j,(R_{j,n},S_{j,n}^N))$\\
\hspace*{0.5em}\textbf{end for}
}

\STATE{Find the weights, $\tilde{w}_{n}=\{\tilde{w}_{j,n}\}_{j=1}^N$:\\
\hspace*{0.5em} \textbf{for} $j$ in $1:N$ \textbf{do} \\
\hspace*{1.5em}$\tilde{w}_{j,n}=w_{j,n-1}P(Y_{n}|S_{j,1:n}^N,\beta,\mathcal{H}_0,v)$ \\
\hspace*{0.5em}\textbf{end for}
}

\STATE{Normalize the weights, $w_{n}=\{w_{j,n}\}_{j=1}^N$:\\
\hspace*{0.5em} \textbf{for} $j$ in $1:N$ \textbf{do} \\
\hspace*{1.5em}$w_{j,n}=\frac{\tilde{w}_{j,n}}{\sum\limits_{j=1}^N\tilde{w}_{j,n}}$ \\
\hspace*{0.5em}\textbf{end for}
}

\STATE{Estimate $P(Y_{n}|Y_{1:n-1}): \hat{P}(Y_{n}|Y_{1:n-1})=\frac{1}{N}\sum\limits_{j=1}^N\tilde{w}_{j,n}$.}

\ENDFOR
\STATE{Resample and form $N$ equally weighted particles, $\Bar{X}_{k}=\{\Bar{X}_{k}^i\}_{i=1}^N$:\\
\hspace{0.5em}\textbf{for} $j$ in $1:N$ \textbf{do} \\
\hspace*{1em}(i) sample index $i_j$ from a multinomial distribution with probabilities $w_{k}$\\
\hspace*{1em}(ii) $\bar{X}_{k}^j=X_{i_j,k}$\\
\hspace*{1em}(iii) $w_{j,k}=1$\\
\hspace{0.5em}\textbf{end for}\\
\textbf{end if}

}
\STATE{ Estimate the marginal likelihood $P(Y_{1:k})$: $\hat{P}(Y_{1:k})=\hat{P}(Y_1)\prod\limits_{n=2}^k\hat{P}(Y_n|Y_{1:n-1})$.} 
\end{algorithmic}
\end{algorithm}

\begin{algorithm}[!h] 
\algsetup{linenosize=\small}
\small 
\caption{\bf Particle marginal Metropolis-Hastings sampler} 
\label{PMMHAlg}
\begin{algorithmic}[1]
\STATE{Step 1: Initialization,$i=0$, \\
\hspace*{1.5em} (a) $\log d_0 \sim \mathcal{N}(\mu_d,\sigma_d^2)$, $\log v_0 \sim \mathcal{N}(\mu_v,\sigma_v^2)$  \\
\hspace*{1.5em} (b) - run a SMC algorithm targetting $P(x_{1:n}|Y_{1:n},\beta,\mathcal{H}_0,d_0,v_0)$\\
\hspace*{3.5em}-Sample $X_{1:n}(0)\sim \hat{P}(\cdot|Y_{1:n},\beta,\mathcal{H}_0,d_0,v_0)$\\
\hspace*{3.5em}-Let $\hat{P}(Y_{1:n}|d_0,v_0)$ denote the marginal likelihood estimate.
}
\STATE{Step 2: for iteration $i \geq 1$ \\
\hspace*{1.5em} (a) $\log d^* \sim \mathcal{N}(\log d_{i-1},\sigma_{d*})$, $\sigma_{d*}=\frac{|\log d_{i-1}-\log d_{min}|}{4}$,\\ \hspace*{3.5em}$\log v^* \sim \mathcal{N}(\log v_{i-1},\sigma_{v*})$,\ $\sigma_{v*}=\frac{|\log v_{i-1}-\log v_{min}|}{4}$ \\
\hspace*{1.5em} (b) - run a SMC algorithm targetting $P(x_{1:n}|Y_{1:n},\beta,\mathcal{H}_0,d^*,v^*)$\\
\hspace*{3.5em}-Sample $X_{1:n}(i)\sim \hat{P}(\cdot|Y_{1:n},\beta,\mathcal{H}_0,d^*,v^*)$\\
\hspace*{3.5em}-Let $\hat{P}(Y_{1:n}|d^*,v^*)$ denote the marginal likelihood estimate. \\
\hspace*{1.6em}(c) wp $\min\Big(1,\frac{\hat{P}(Y_{1:n}|d^*,v^*)}{\hat{P}(Y_{1:n}|d_{i-1},v_{i-1})}\frac{P(d^*)}{P(d_{i-1})}\frac{P(v^*)}{P(v_{i-1})}\frac{P(d_{i-1}|d^*)}{P(d^*|d_{i-1})}\frac{P(v_{i-1}|v^*)}{P(v^*|v_{i-1})}\Big)$ \\ \hspace*{2em}set $d_i=d^*$, $v_i=v^*$, $X_{1:n}(i)=X^*_{1:n}$, $\hat{P}(Y_{1:n}|d_i,v_i)=\hat{P}(Y_{1:n}|d^*,v^*)$. Otherwise,\\ \hspace*{1em}$d_i=d_{i-1}$, $v_i=v_{i-1}$, $X_{1:n}(i)=X_{1:n}(i-1)$, $\hat{P}(Y_{1:n}|d_i,v_i)=\hat{P}(Y_{1:n}|d_{i-1},v_{i-1})$.
}
\end{algorithmic}
\end{algorithm}

\subsection*{Real Data} \label{RSimulation}
\begin{figure}[!h] 
  \begin{subfigure}{6cm}
    \centering\includegraphics[width=6cm]{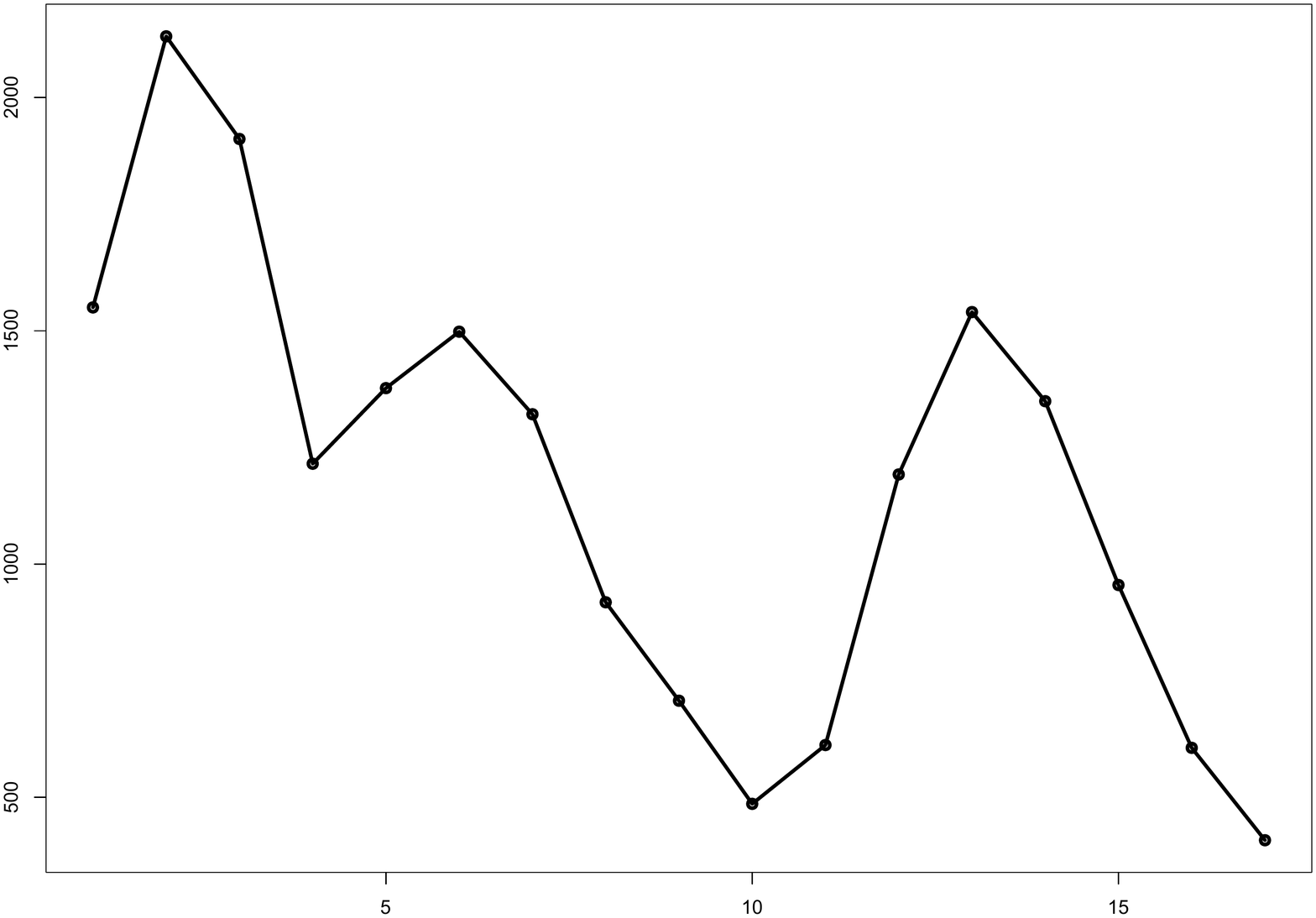}
    \caption{The weekly observed cases in Ashford.} 
  \end{subfigure}
  \begin{subfigure}{6cm}
    \centering\includegraphics[width=6cm]{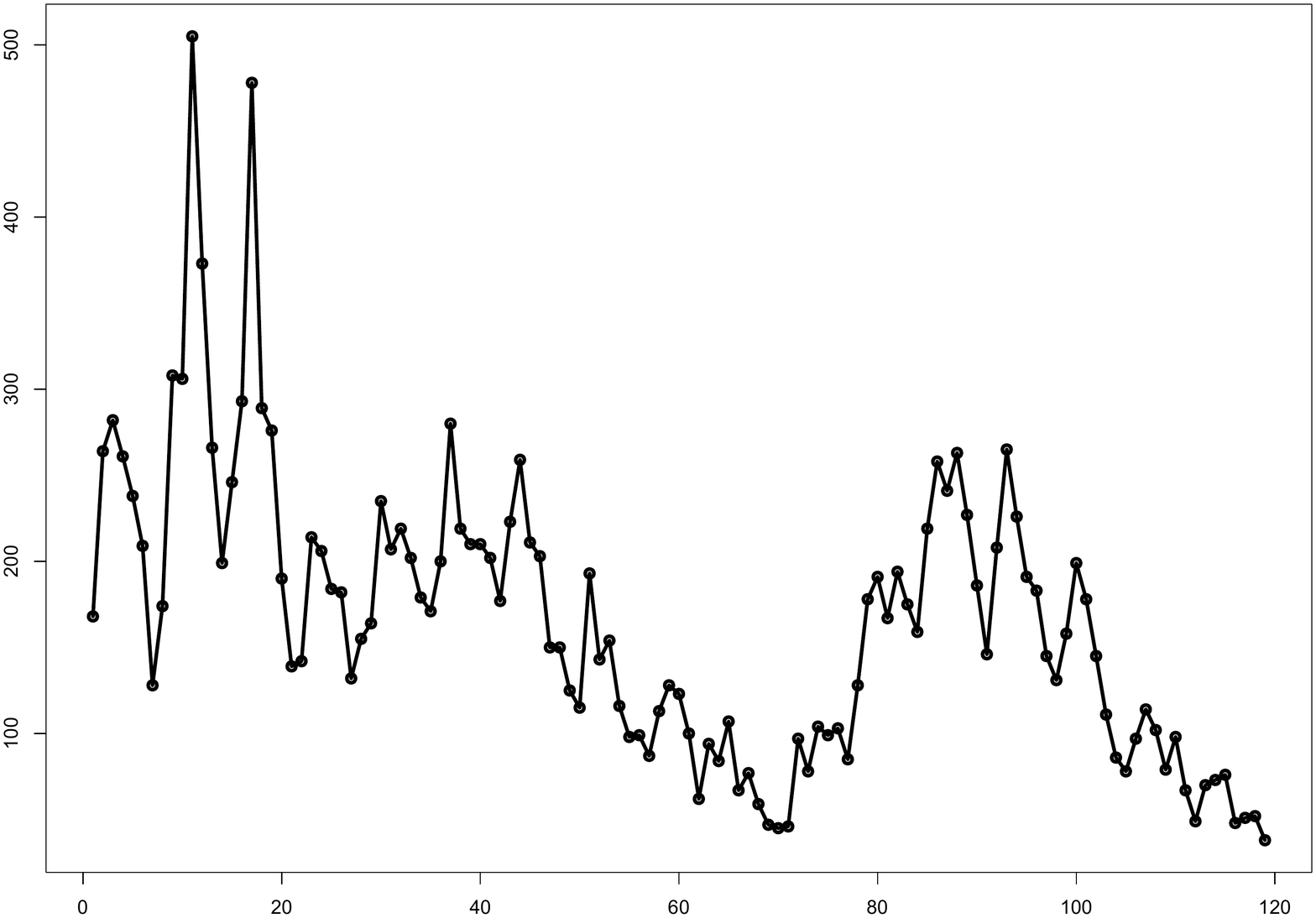}
    \caption{The daily observed cases in Ashford.} 
  \end{subfigure}
   \begin{subfigure}{6cm}
    \centering\includegraphics[width=6cm]{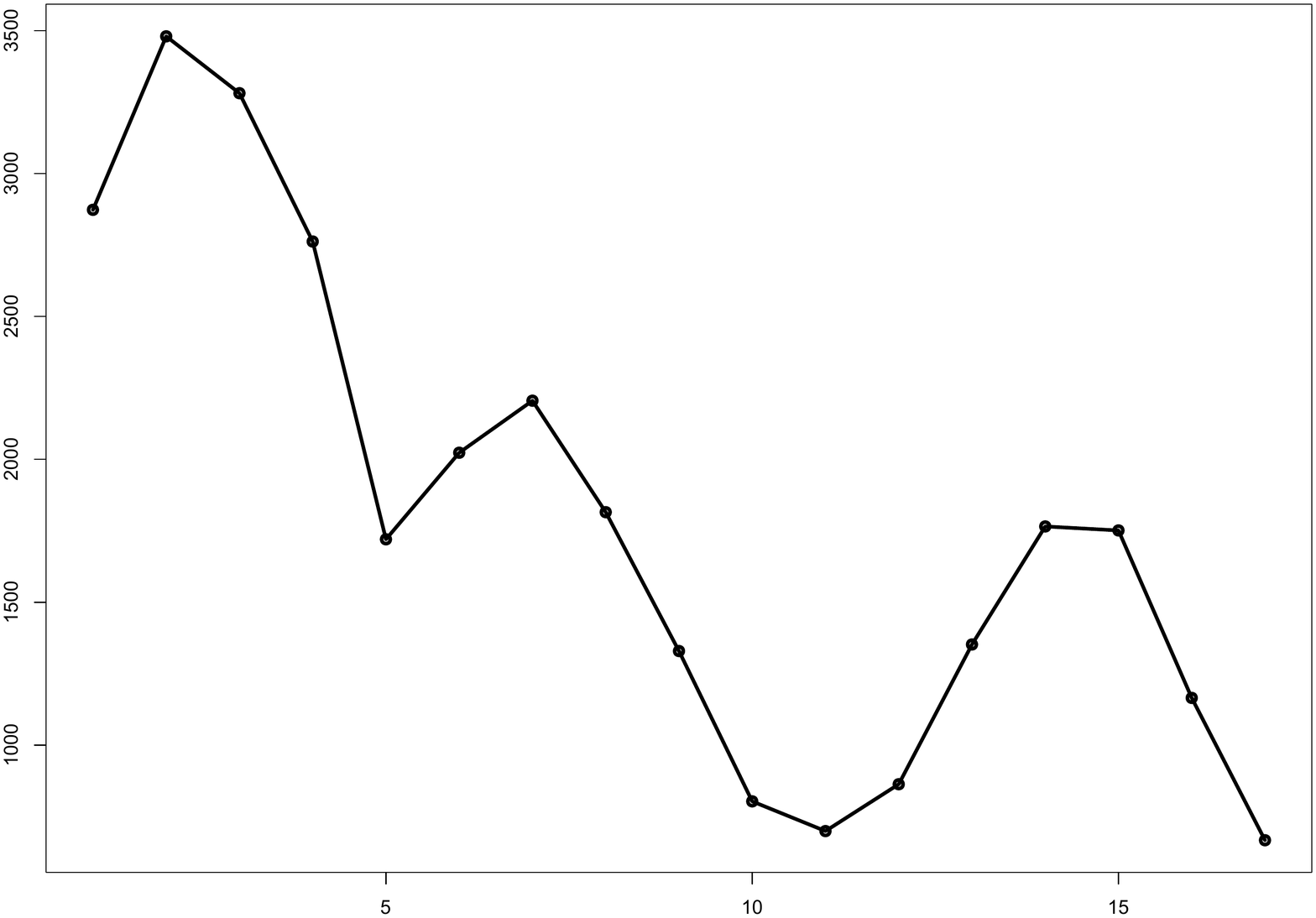}
    \caption{The weekly observed cases in Kingston upon Thames.} 
  \end{subfigure}
  \begin{subfigure}{6cm}
    \centering\includegraphics[width=6cm]{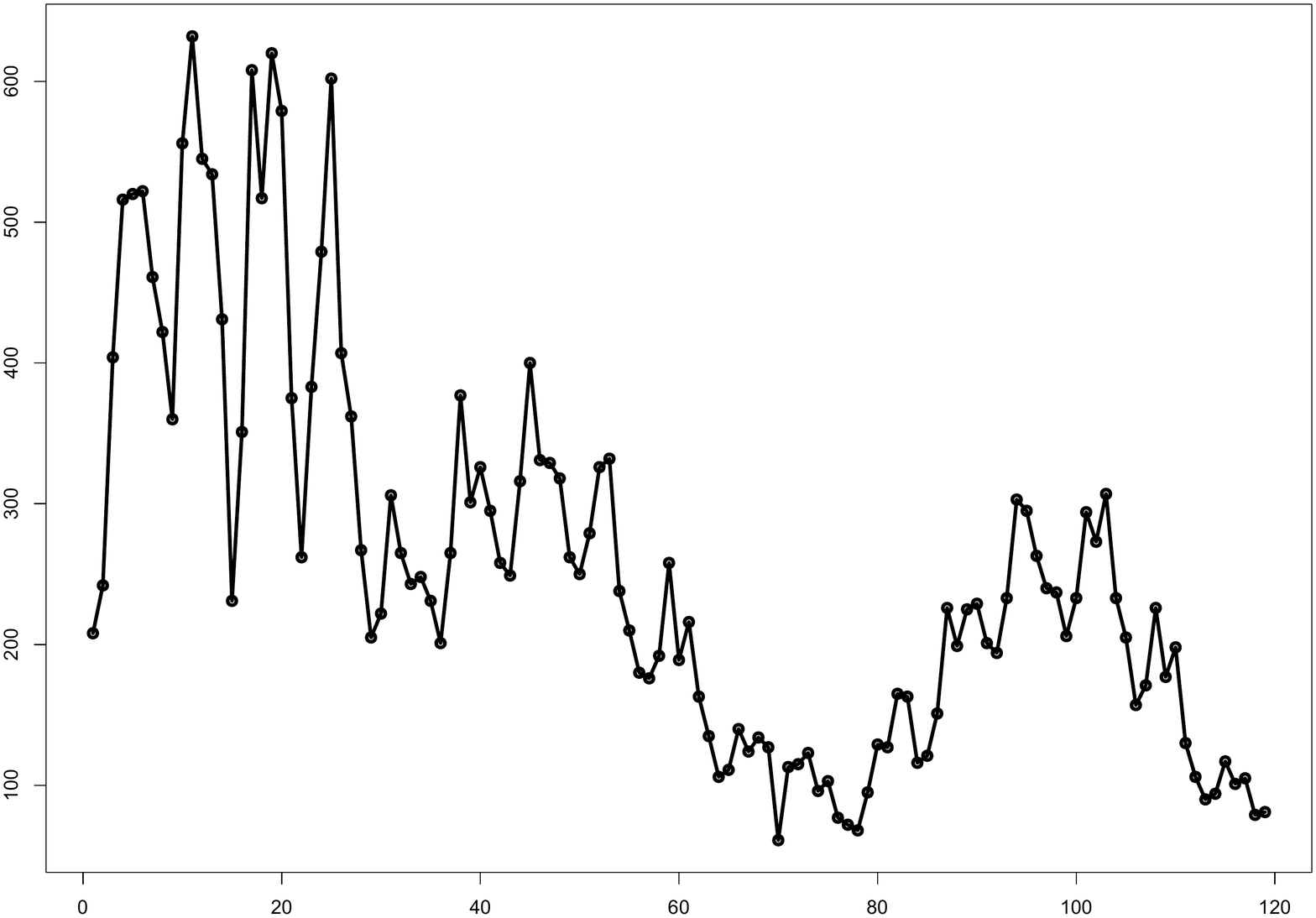}
    \caption{The daily observed cases in Kingston upon Thames.} 
  \end{subfigure}
   \begin{subfigure}{6cm}
    \centering\includegraphics[width=6cm]{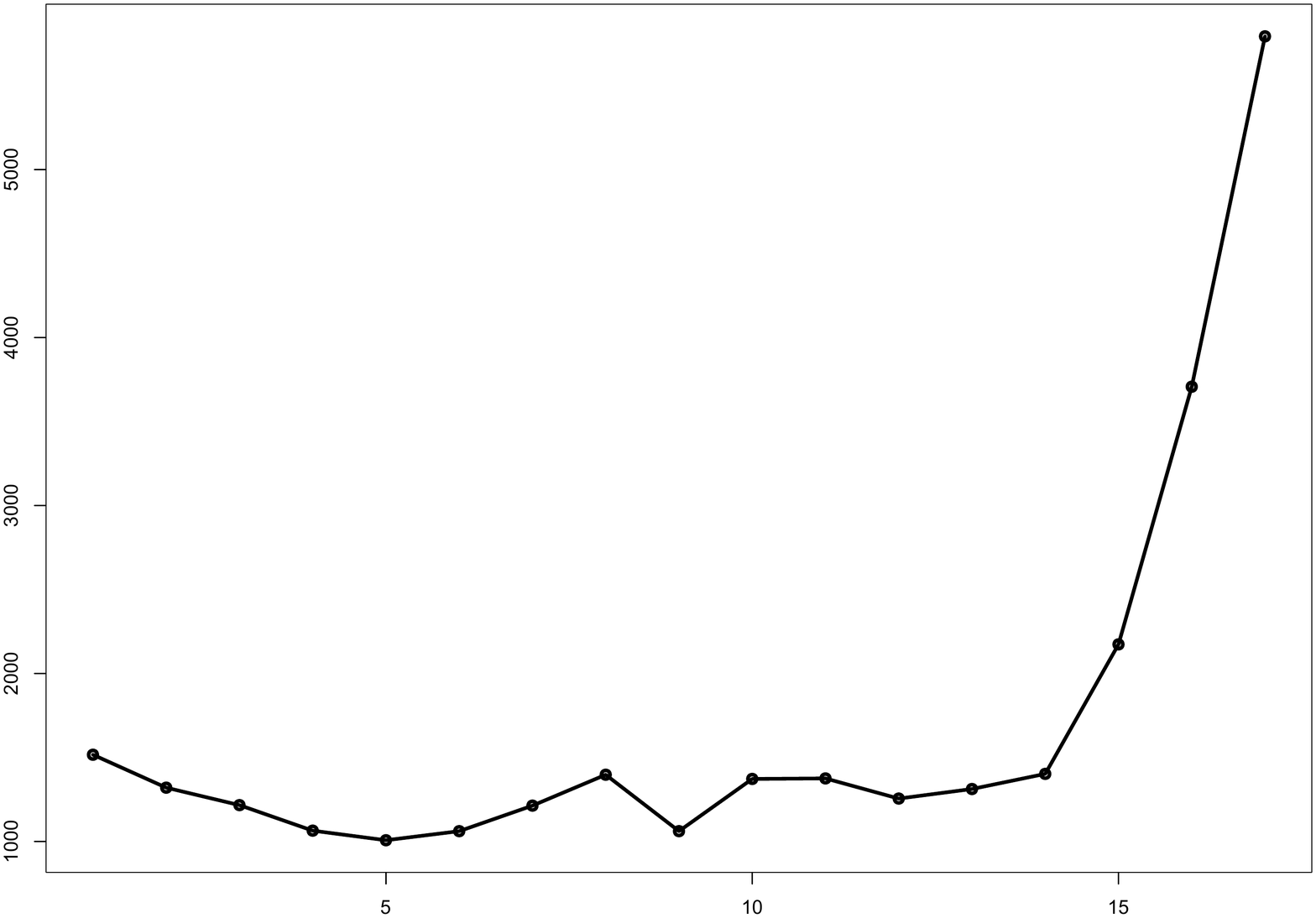}
    \caption{The weekly observed cases in Leicester.} 
  \end{subfigure}
  \begin{subfigure}{8.5cm}
    \centering\includegraphics[width=6cm]{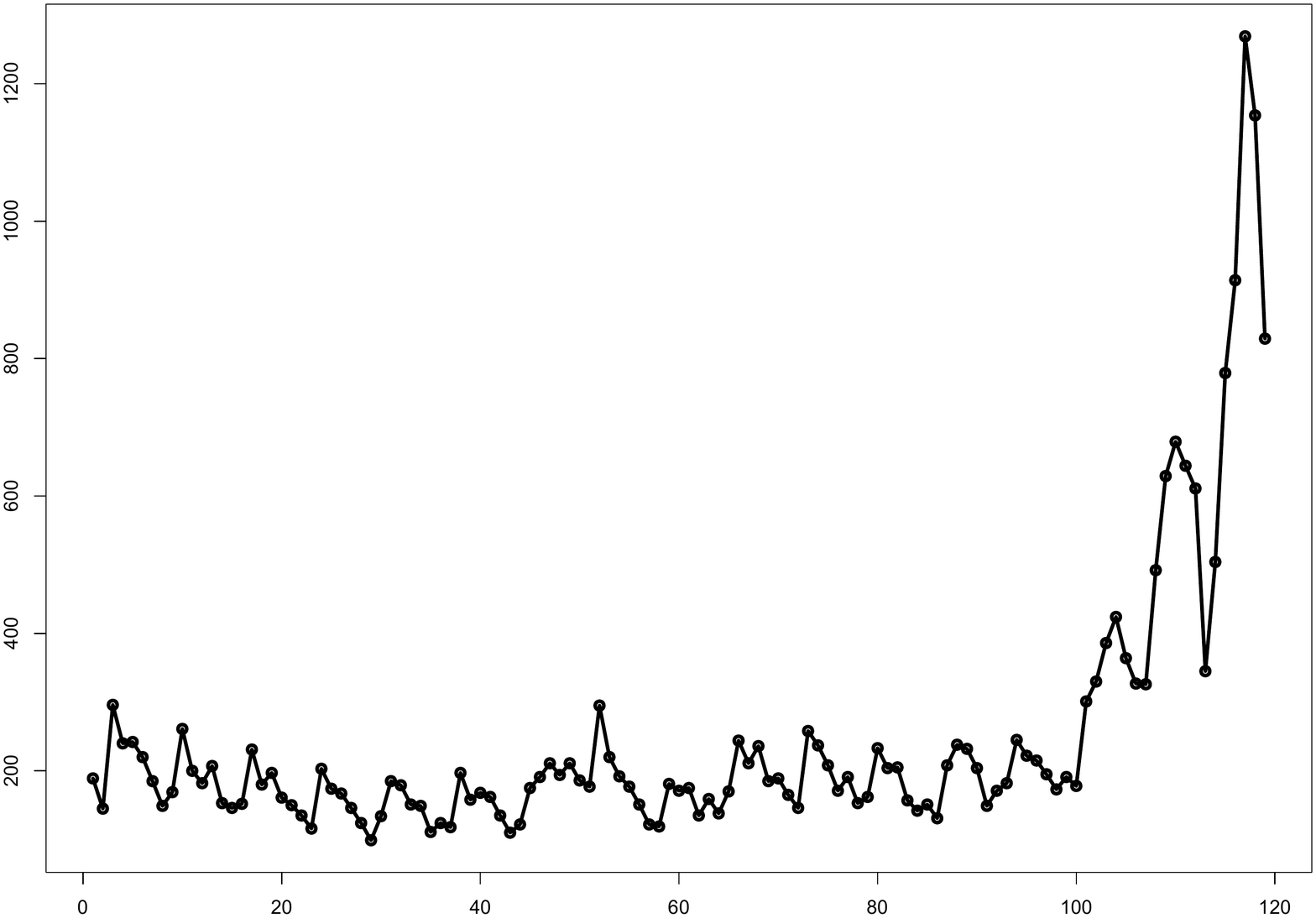}
    \caption{The daily observed cases in Leicester.} 
  \end{subfigure}
  \caption{The daily and weekly observed infections in local authorities.}
  \label{FigData}
\end{figure}

We apply the KDPF (Algorithm~\ref{APAlga}) to real cases in the local authorities: Leicester (4/9/2021 - 24/12/2021), Kingston upon Thames (11/12/2021 - 1/4/2022) and Ashford (19/12/2021 - 9/4/2022) available from the government in the UK~\cite{link_GOVUK}. Figure \ref{FigData} illustrates the daily and weekly observed cases in the local authorities. We deal with 16 hidden states $\{X_n\}_{n=1}^{16}$ and 16 subintervals $\{ \mathcal{T}_n \}_{n=1}^{16}$; each subinterval corresponds to the duration of one week. We infer the latent intensity $\lambda^N(t)$, the reproduction number $R(t)$, and the weekly and daily latent cases via the particle sample derived by drawing samples from the smoothing density with lag equal to 4. We demonstrate that the proposed model can be applied to predict the new observed cases over short time horizons.

We assume that the initial reproduction number during the first week is uniformly distributed  over the interval from 0.5 to 2. Our initialization includes the 90\% Confidence Interval published from the government in the UK: 0.9-1.1 on 4/9/2021 and 11/12/2021, 1-1.2 on 19/12/2021~\cite{link_GOVR}. We also assume $d_{min}=1$, $d_{max}=10$, $v_{min}=0.0001$ and $v_{max}=0.5$.

\begin{figure}[!h] 
   \begin{subfigure}{7cm}
    \centering\includegraphics[width=6cm]{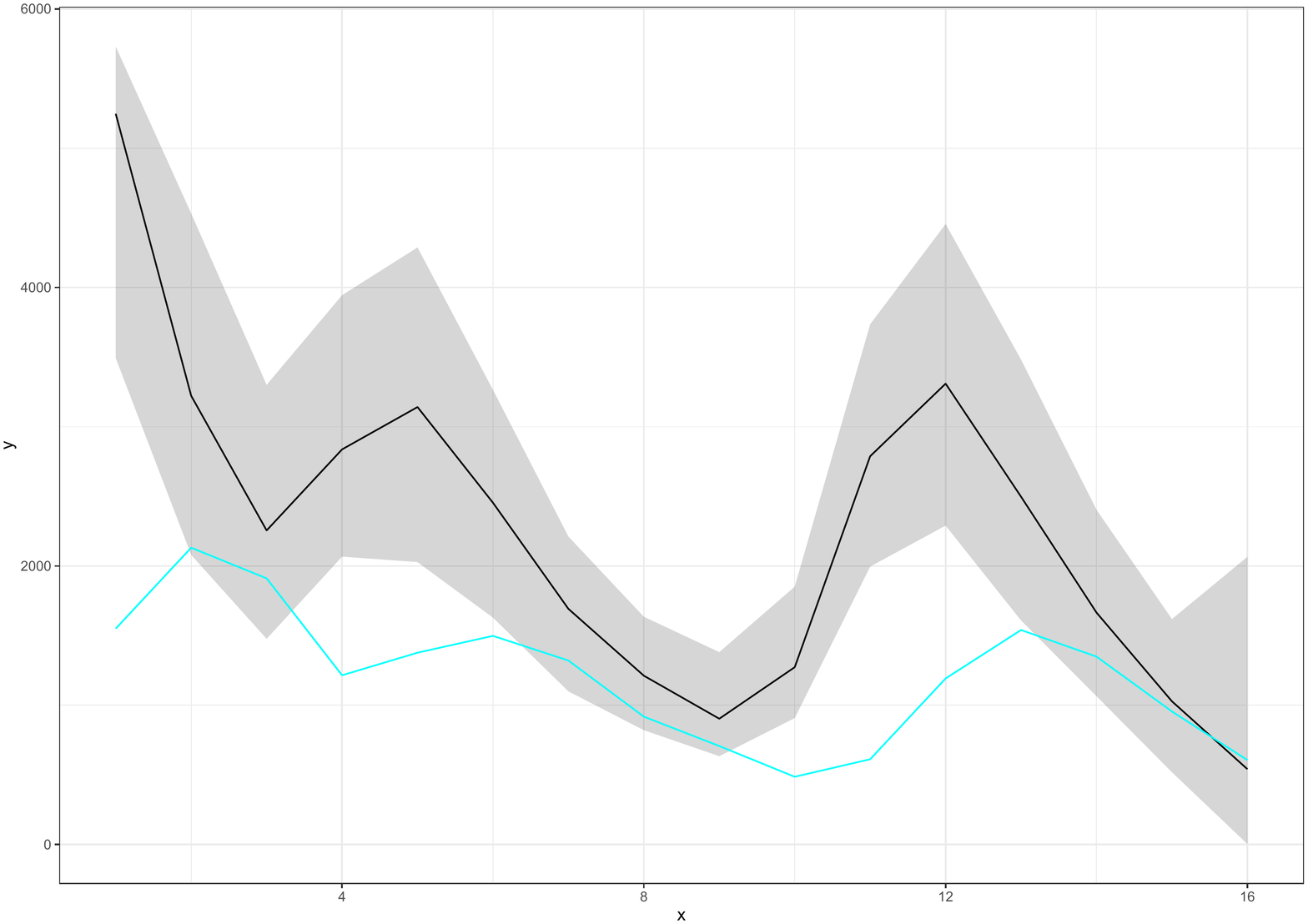}
   \caption{The estimated weekly latent cases (posterior  median (black line) ; 99\% CI (ribbon)), and the weekly observed cases (cyan line).}
  \end{subfigure}
  \begin{subfigure}{7cm}
    \centering\includegraphics[width=6cm]{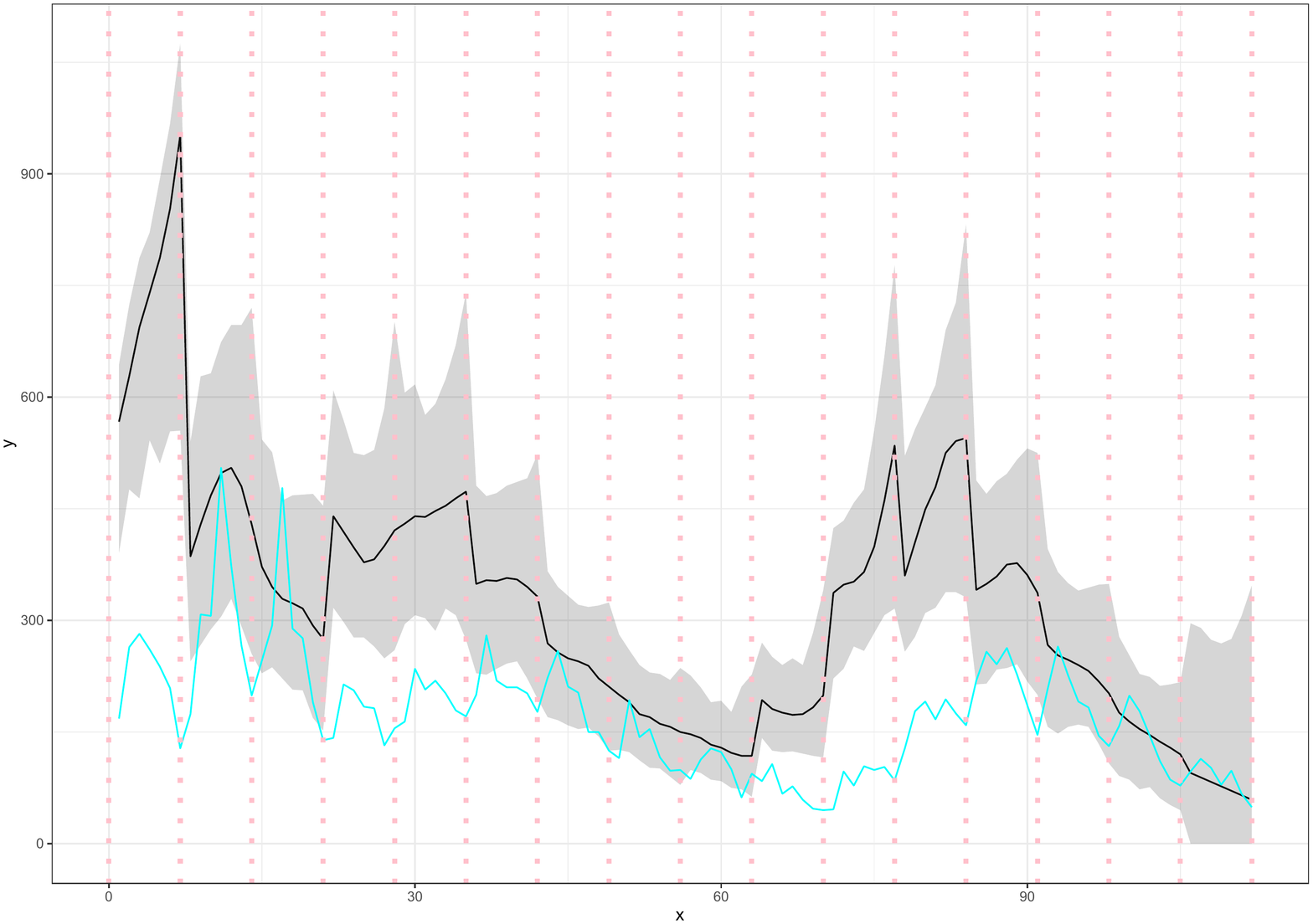}
   \caption{The estimated daily latent cases (posterior median (black line) ; 99\% CI (ribbon)), and the daily observed cases (cyan line).}
  \end{subfigure}
  \begin{subfigure}{7cm}
    \centering\includegraphics[width=6cm]{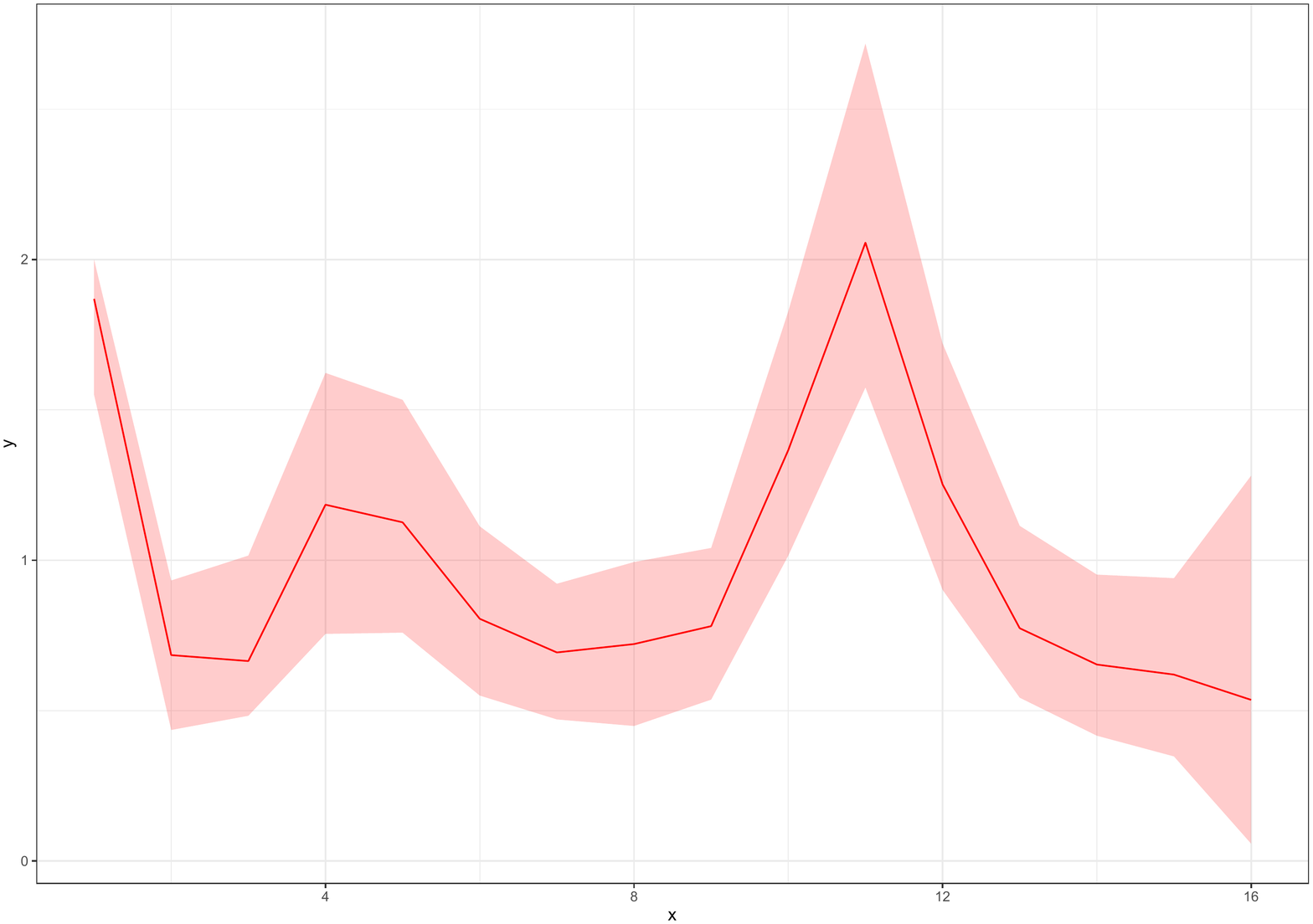}
    \caption{The estimated reproduction number (posterior median (red line); 95\% CI (ribbon)).}
  \end{subfigure}
  \begin{subfigure}{7.0cm}
    \centering\includegraphics[width=6cm]{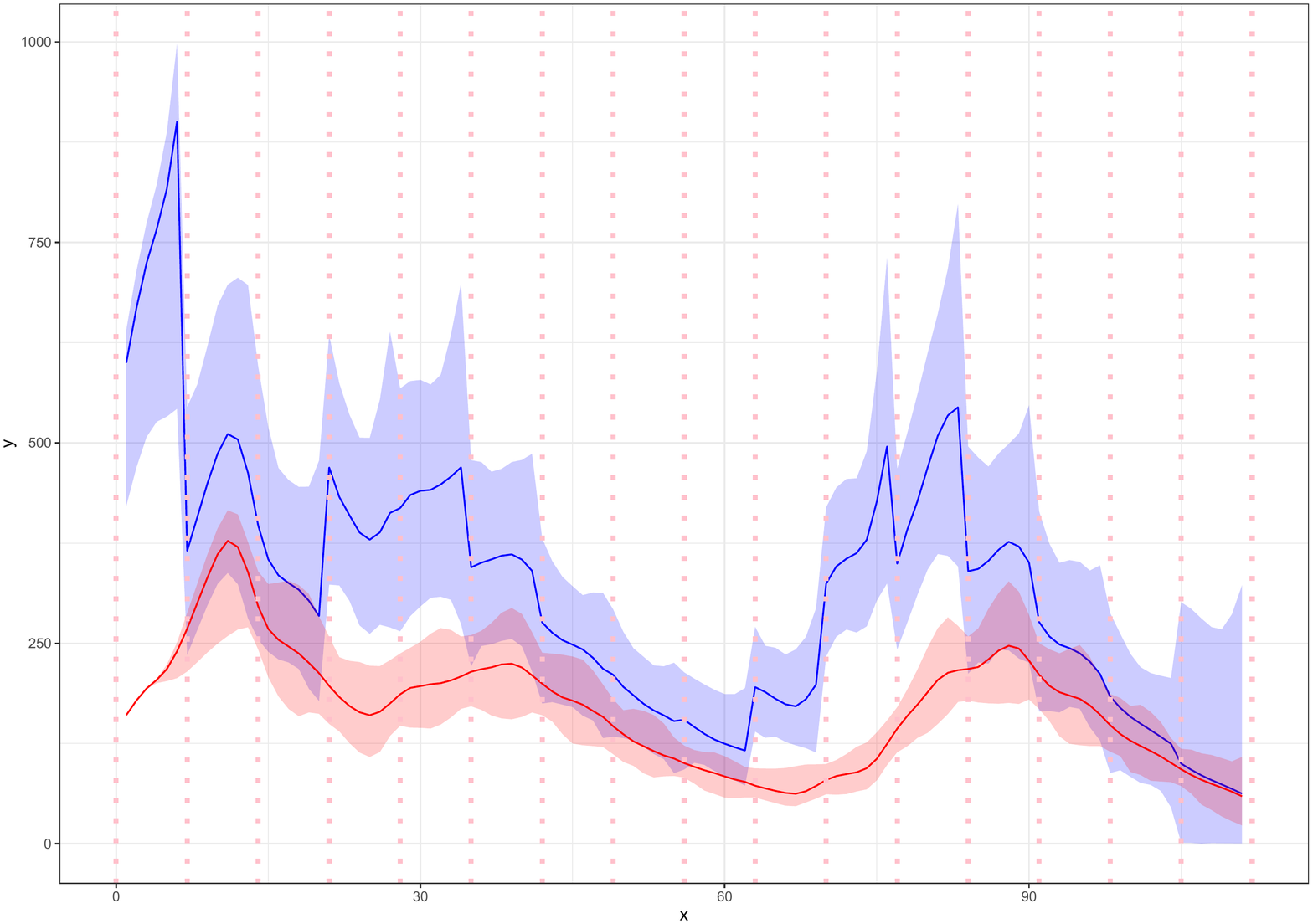}
   \caption{The estimated intensity of latent cases (posterior median (blue line) ; 99\% CI (ribbon)) and the estimated intensity of observed cases (posterior median (red line) ; 99\% CI (ribbon)).} 
  \end{subfigure}
  \begin{subfigure}{7cm}
    \centering\includegraphics[width=6cm]{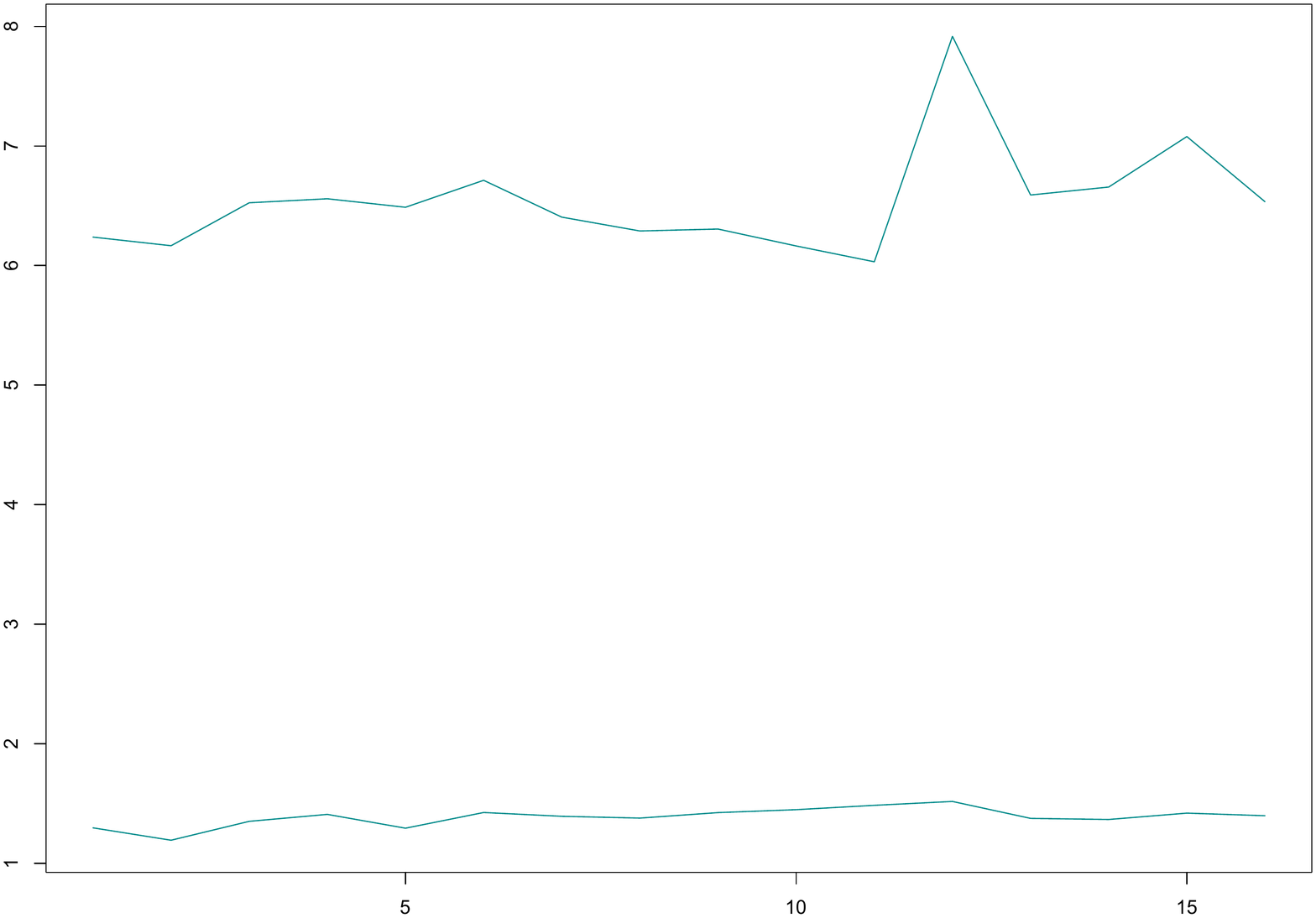}
   \caption{The 99\% CI of d.}
  \end{subfigure}
   \begin{subfigure}{7cm}
    \centering\includegraphics[width=6cm]{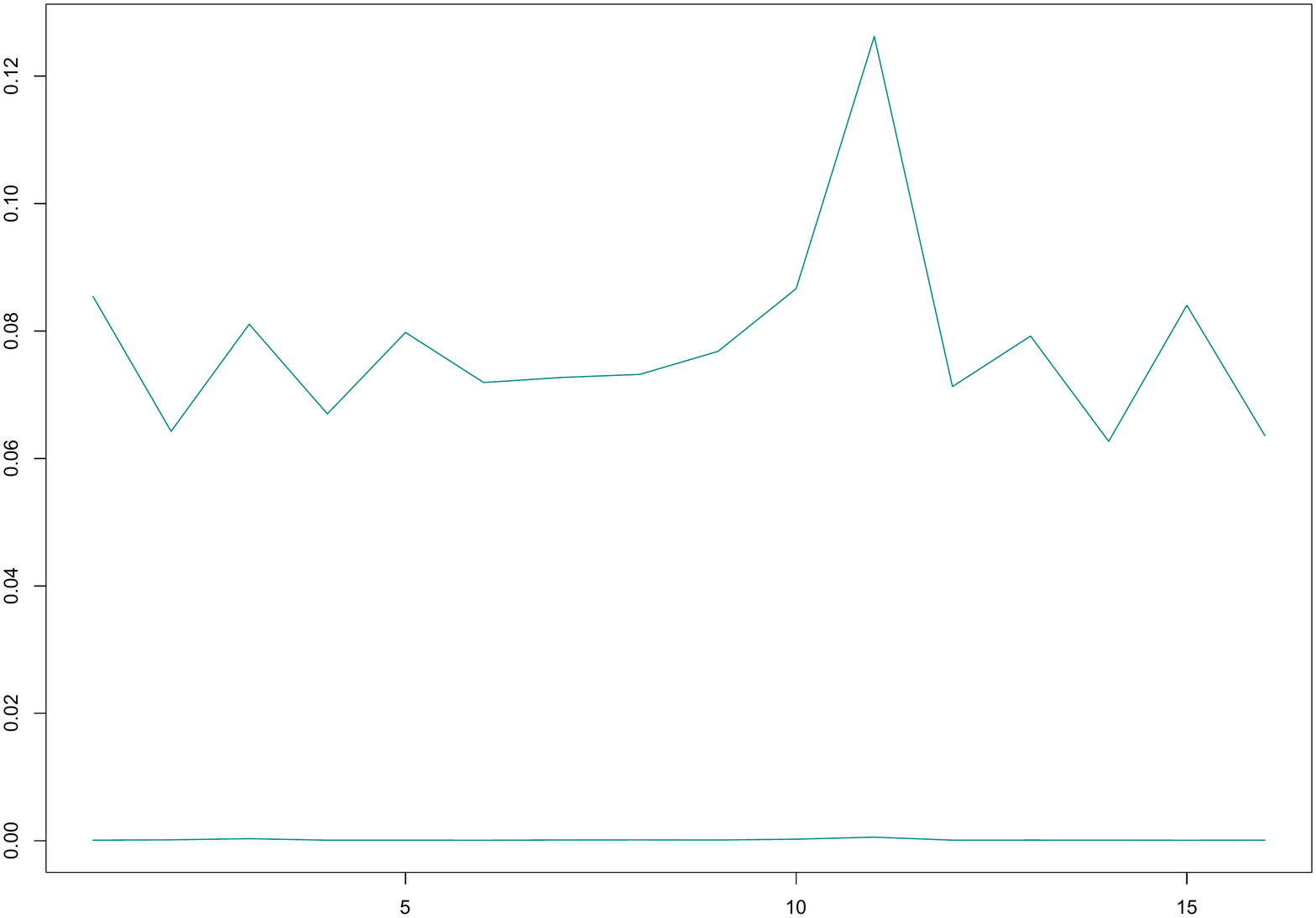}
   \caption{The 99\% CI of v.}
  \end{subfigure}
  \caption{{\bf The weekly and daily latent cases, the reproduction number, the latent and observed intensity and the $99\%$ CIs of time-constant parameters in Ashford.} The time interval between two successive pink vertical dashed lines corresponds to a week.}
  
  \label{FigAshford}
\end{figure}

\begin{figure}[!h] 
   \begin{subfigure}{7cm}
    \centering\includegraphics[width=6cm]{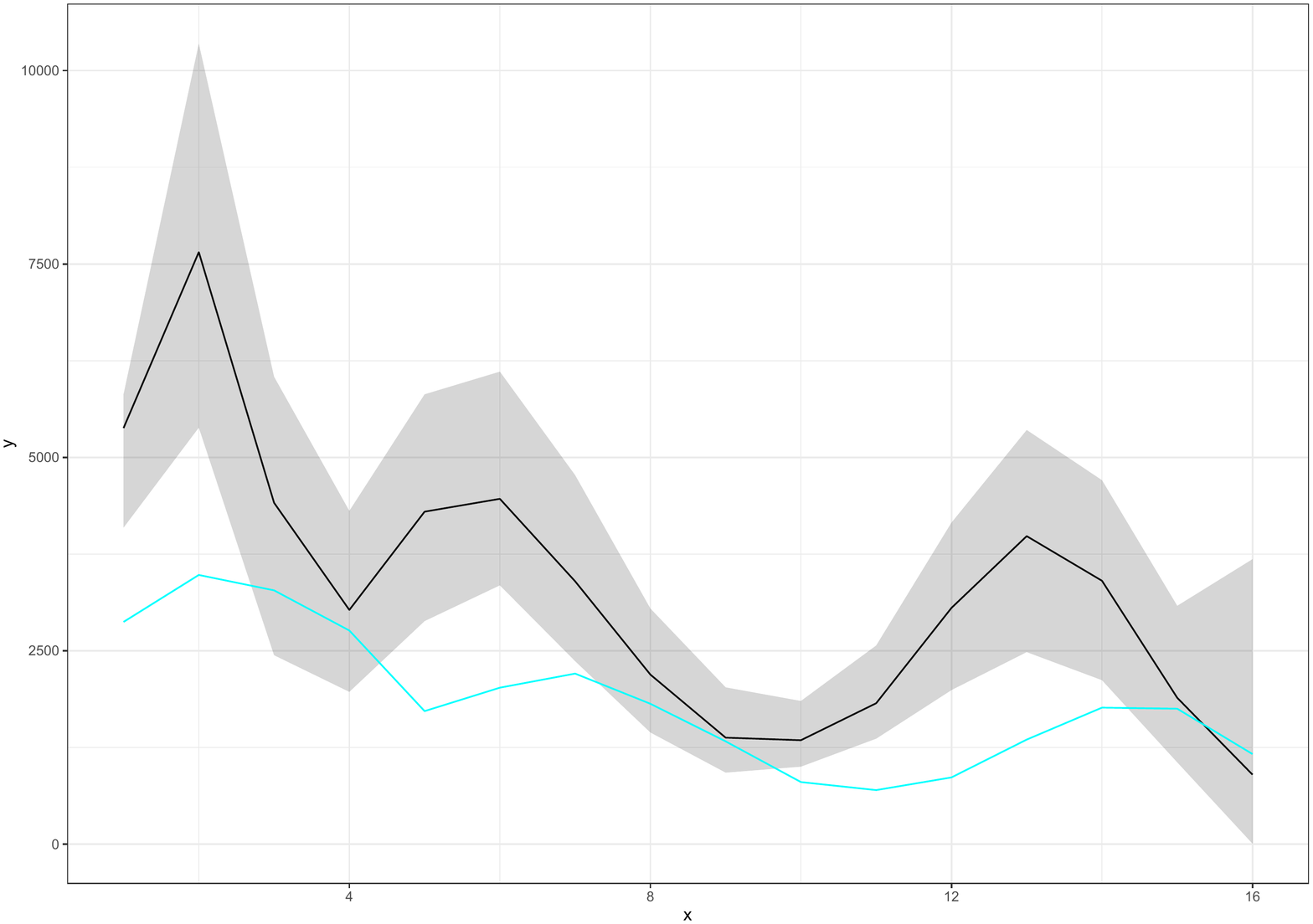}
   \caption{The estimated weekly latent cases (posterior  median (black line) ; 99\% CI (ribbon)), and the weekly observed cases (cyan line).}
  \end{subfigure}
  \begin{subfigure}{7cm}
    \centering\includegraphics[width=6cm]{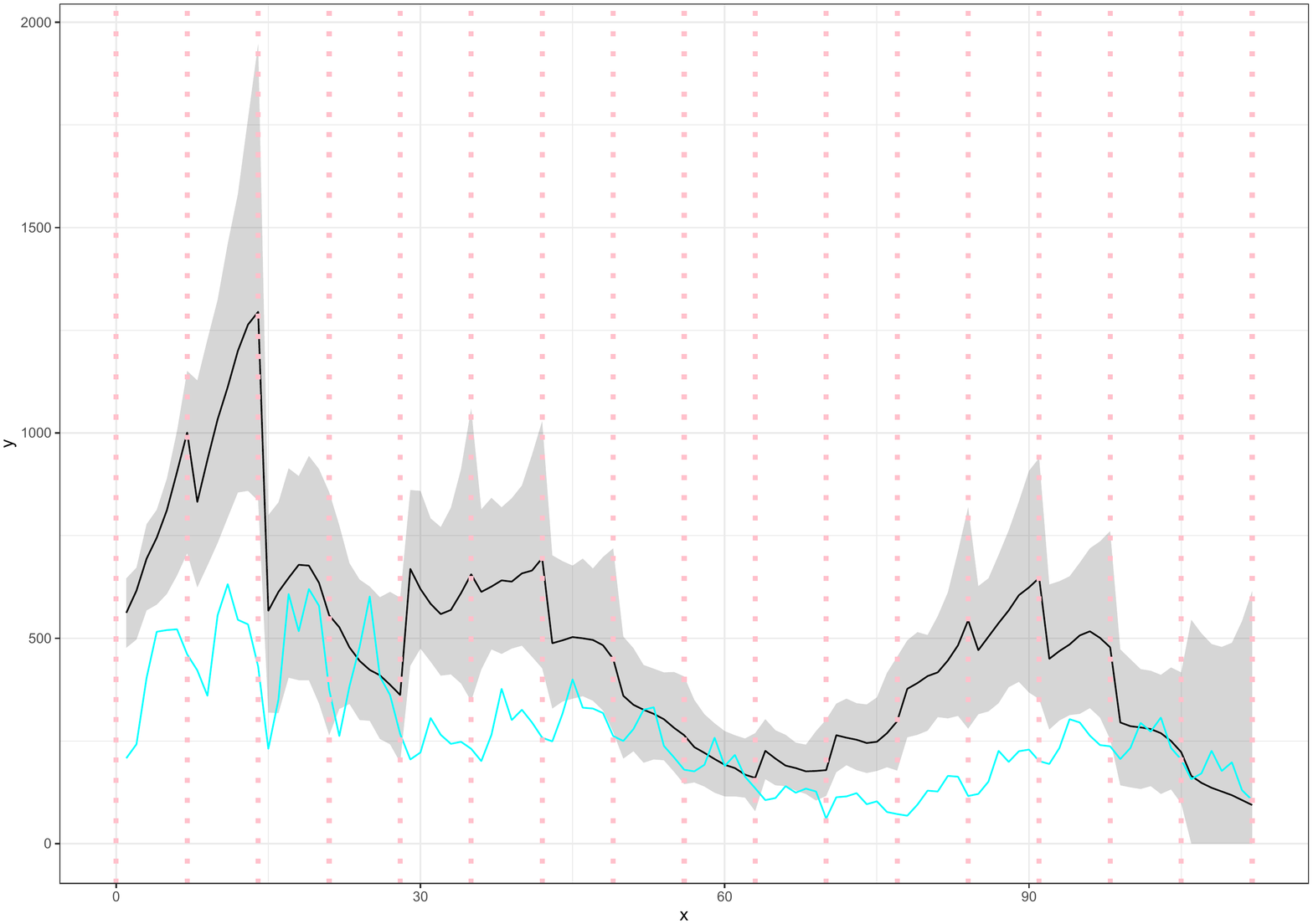}
   \caption{The estimated daily latent cases (posterior median (black line) ; 99\% CI (ribbon)), and the daily observed cases (cyan line).}
  \end{subfigure}
  \begin{subfigure}{7cm}
    \centering\includegraphics[width=6cm]{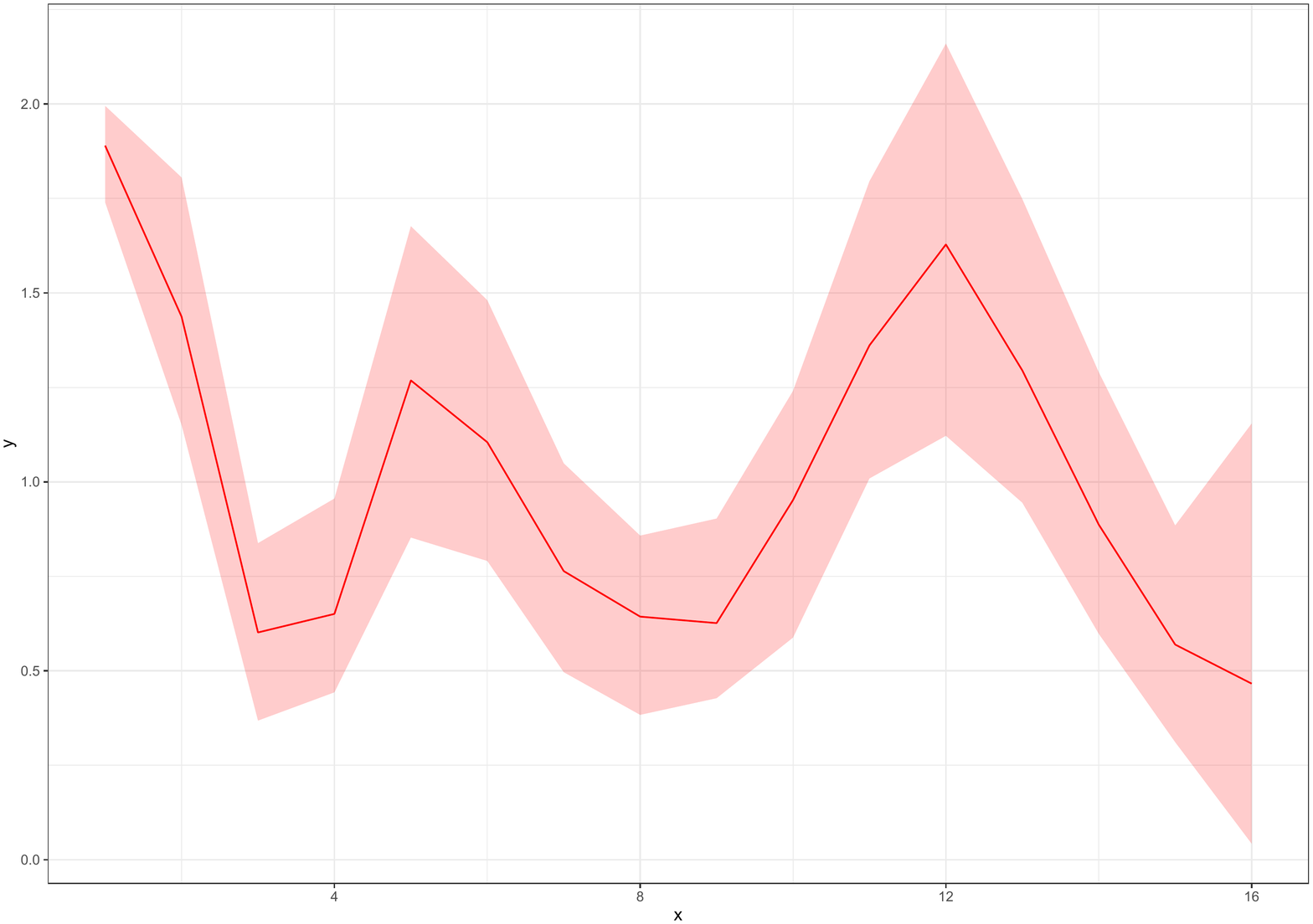}
    \caption{The estimated reproduction number (posterior median (red line); 95\% CI (ribbon)).}
  \end{subfigure}
  \begin{subfigure}{7.0cm}
    \centering\includegraphics[width=6cm]{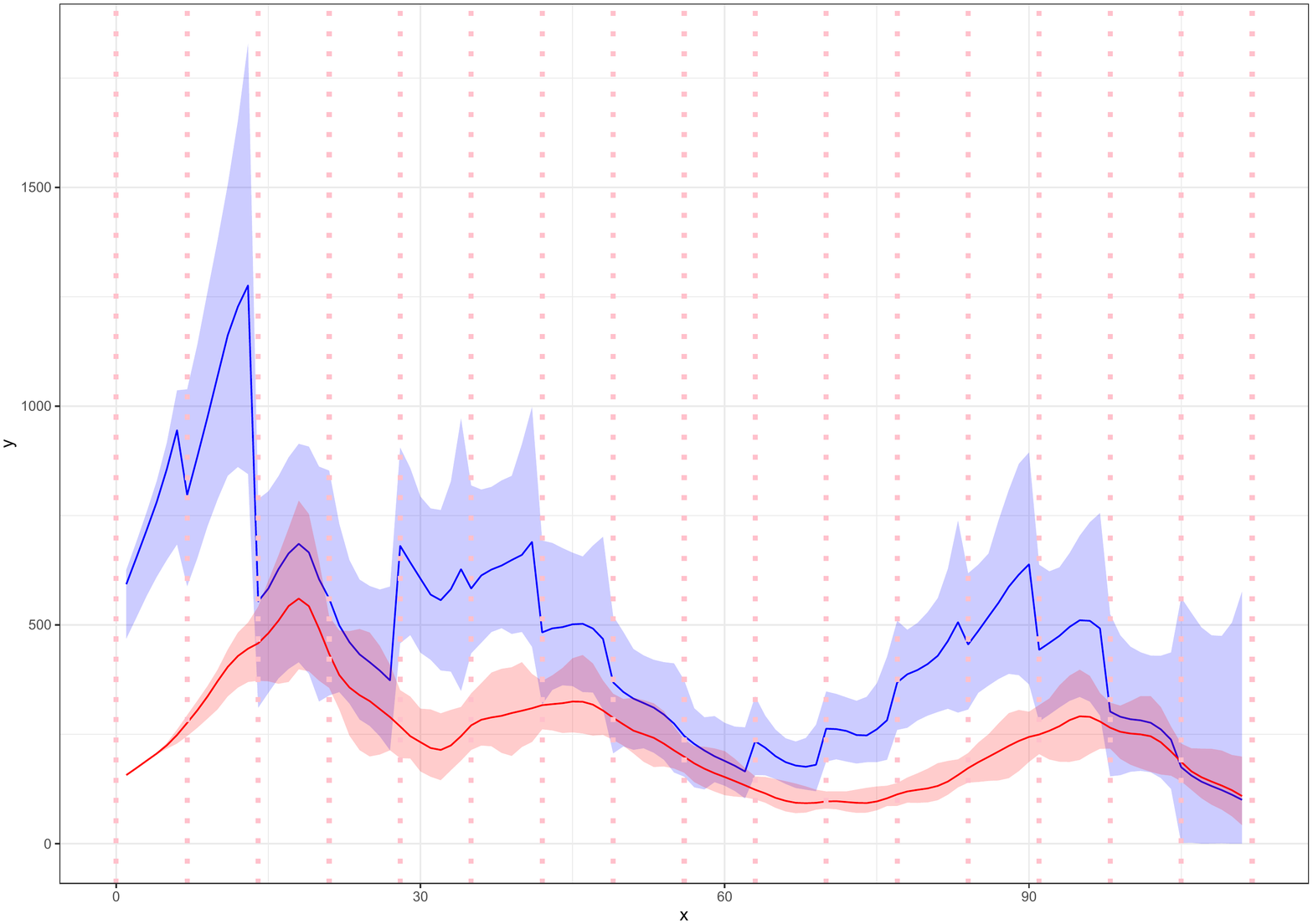}
   \caption{The estimated intensity of latent cases (posterior median (blue line) ; 99\% CI (ribbon)) and the estimated intensity of observed cases (posterior median (red line) ; 99\% CI (ribbon)).} 
  \end{subfigure}
  \begin{subfigure}{7cm}
    \centering\includegraphics[width=6cm]{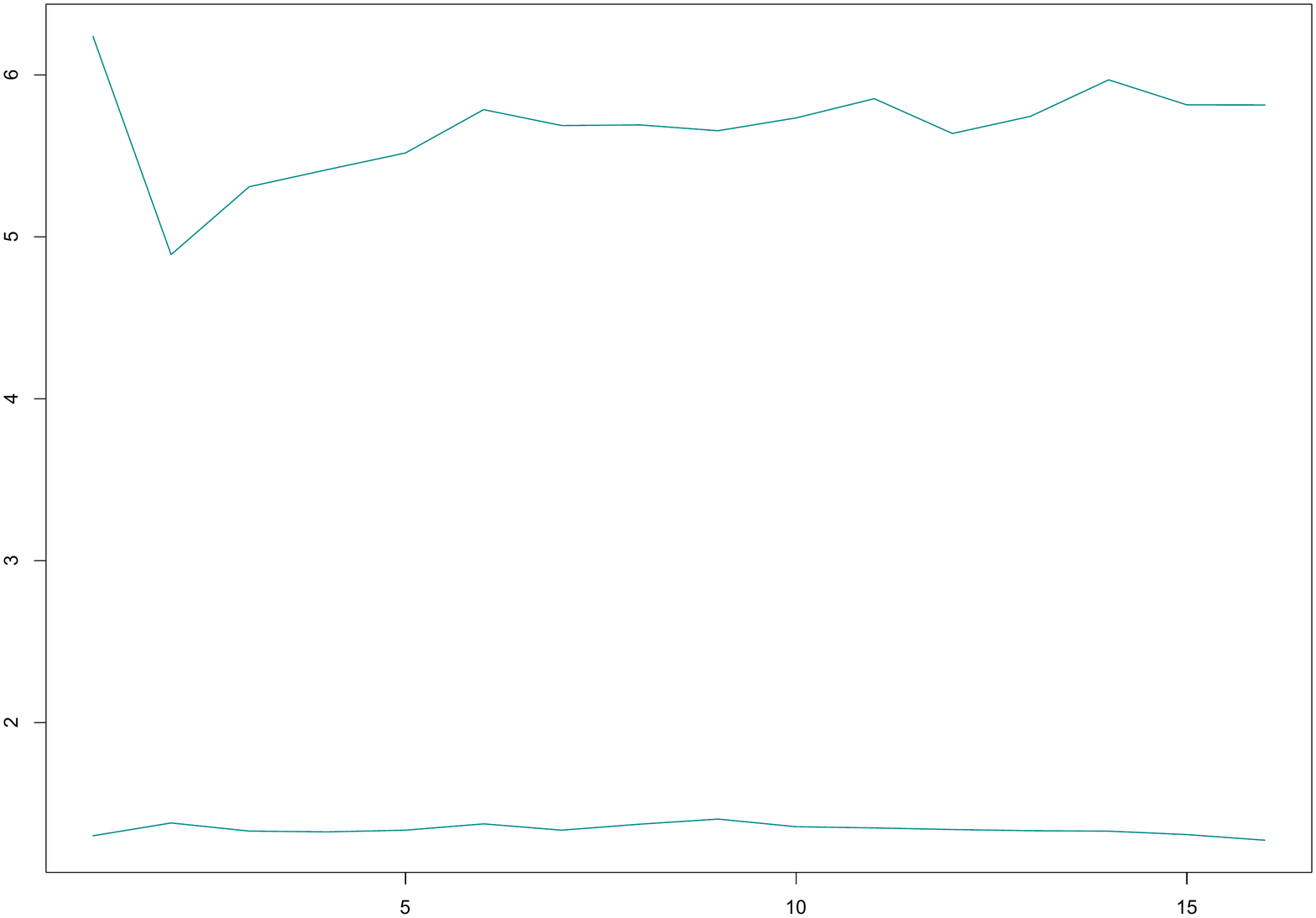}
   \caption{The 99\% CI of d.}
  \end{subfigure}
   \begin{subfigure}{7cm}
    \centering\includegraphics[width=6cm]{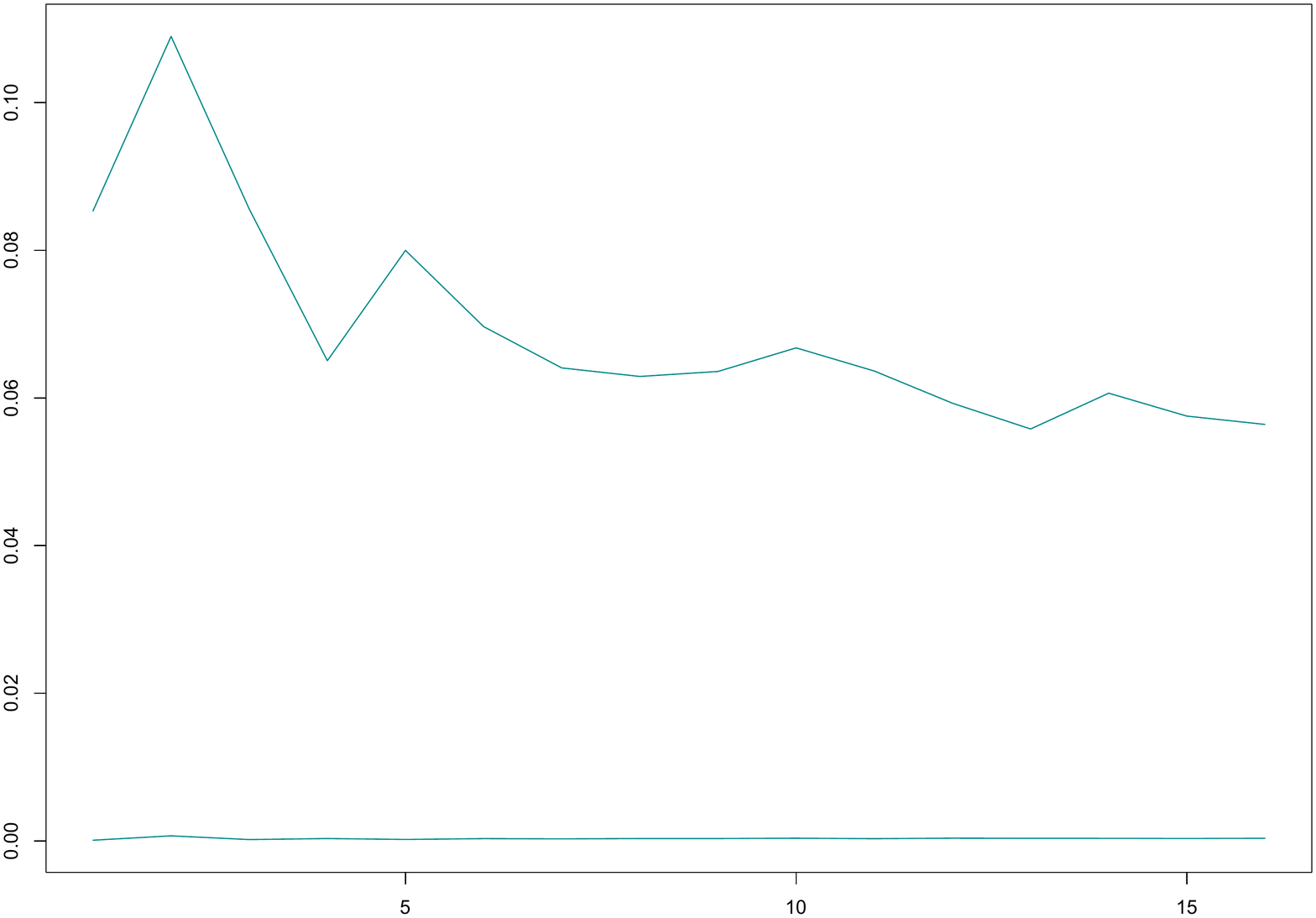}
   \caption{The 99\% CI of v.}
  \end{subfigure}
  \caption{{\bf The weekly and daily latent cases, the reproduction number, the latent and observed intensity and the $99\%$ CIs of time-constant parameters in Kingston upon Thames.} The time interval between two successive pink vertical dashed lines corresponds to a week.}
  \label{FigKings}
\end{figure}

\begin{figure}[!h] 
   \begin{subfigure}{7cm}
    \centering\includegraphics[width=6cm]{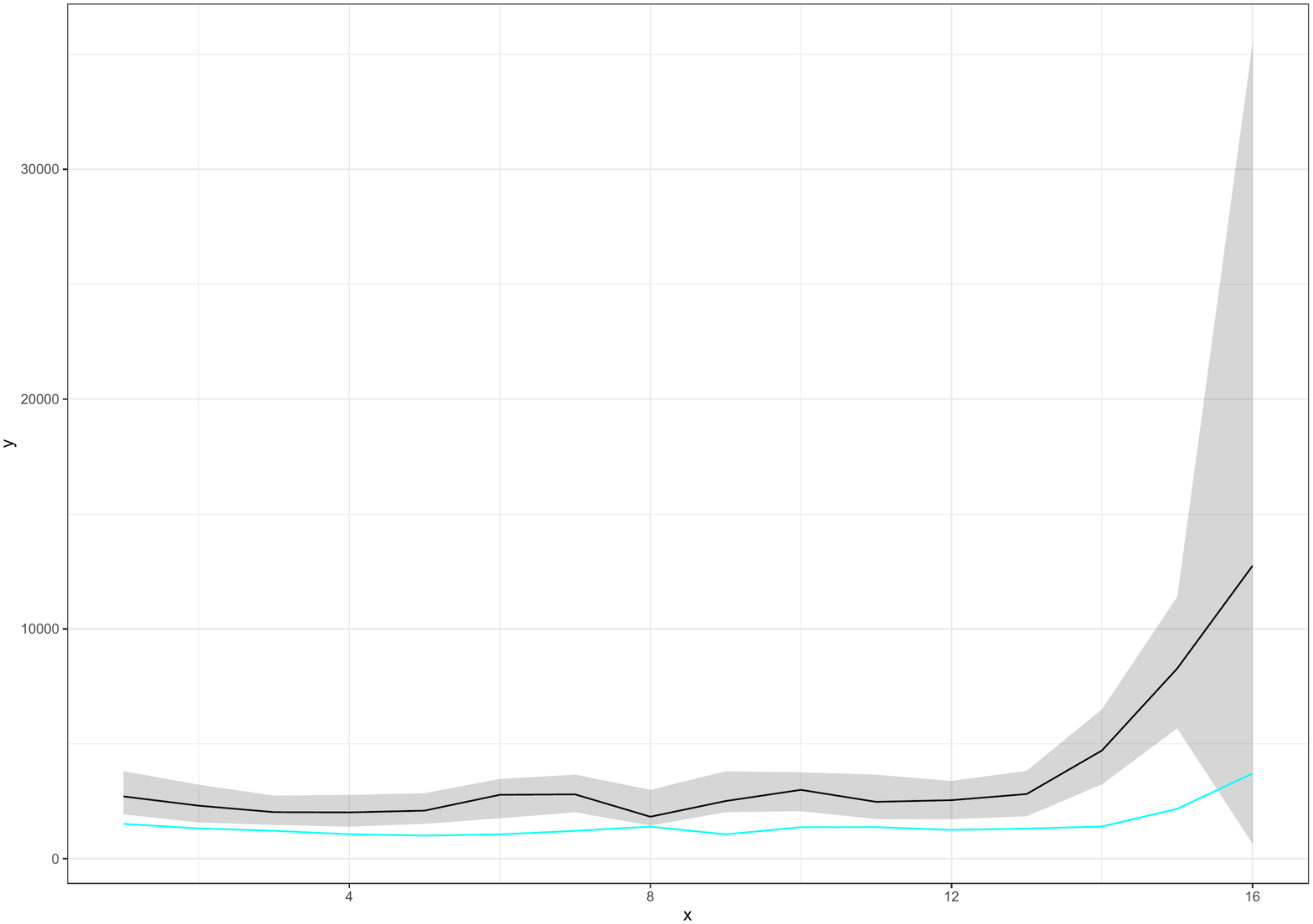}
   \caption{The estimated weekly latent cases (posterior  median (black line) ; 99\% CI (ribbon)), and the weekly observed cases (cyan line).}
  \end{subfigure}
  \begin{subfigure}{7cm}
    \centering\includegraphics[width=6cm]{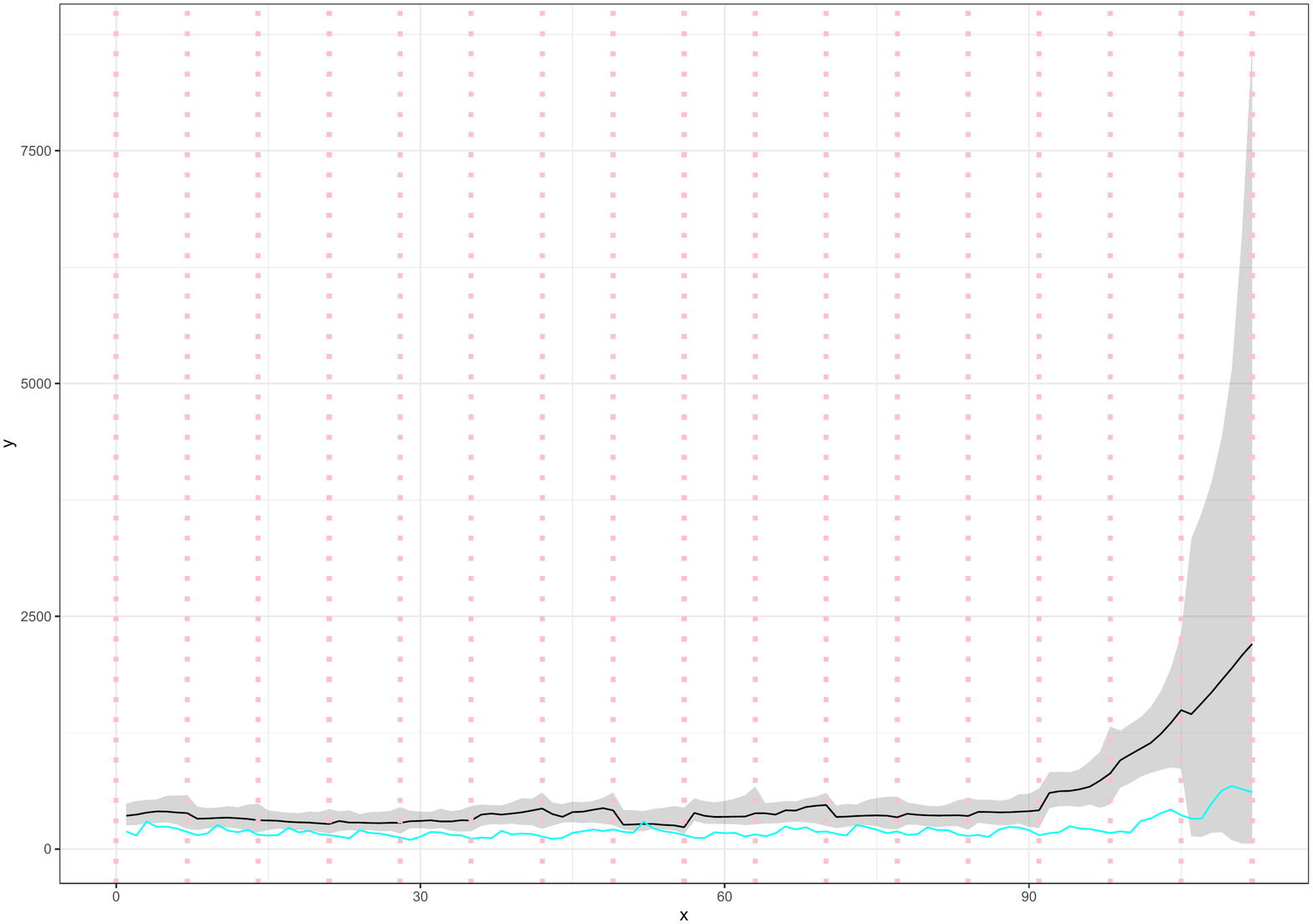}
   \caption{The estimated daily latent cases (posterior median (black line) ; 99\% CI (ribbon)), and the daily observed cases (cyan line).}
  \end{subfigure}
  \begin{subfigure}{7cm}
    \centering\includegraphics[width=6cm]{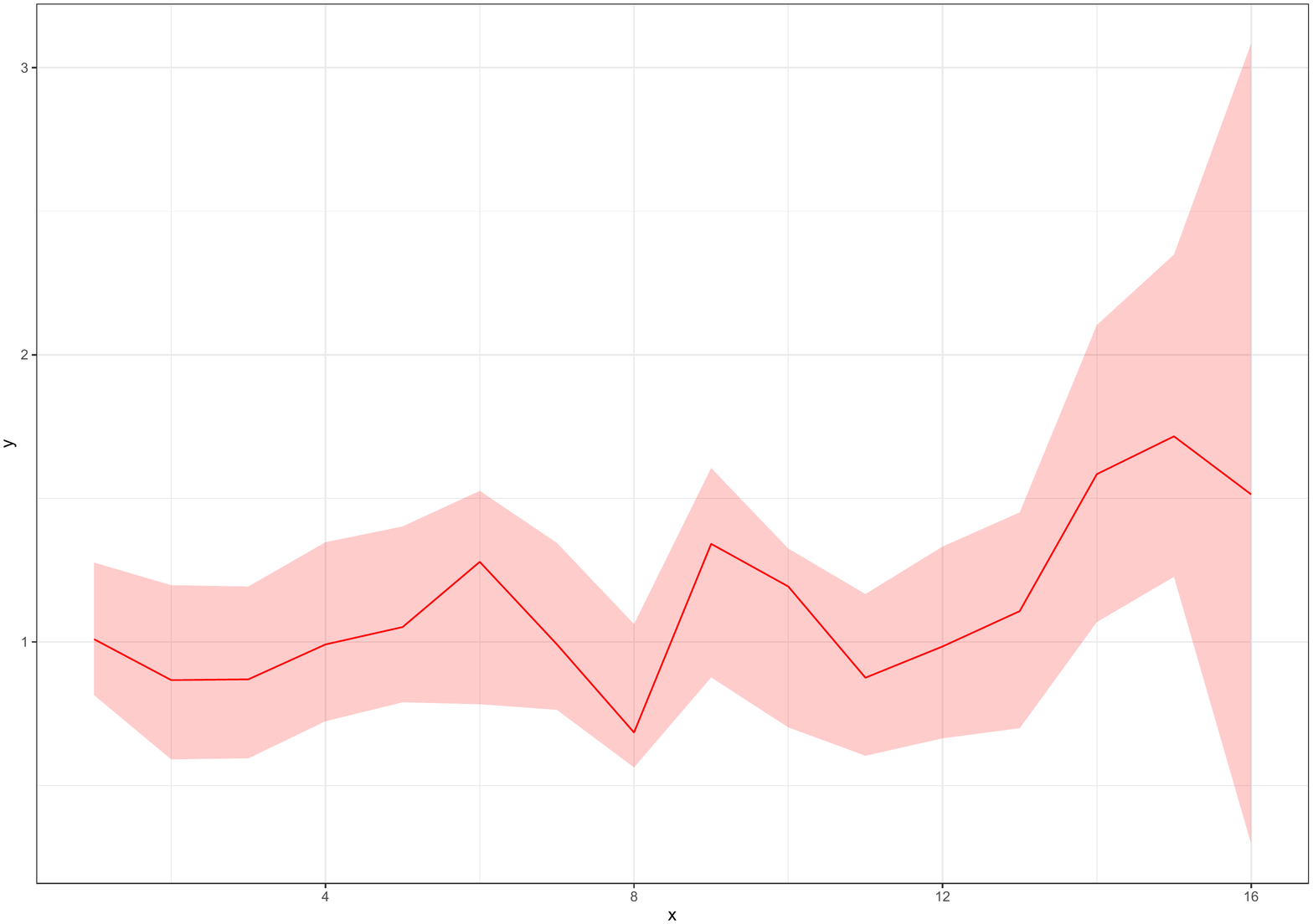}
    \caption{The estimated reproduction number (posterior median (red line); 95\% CI (ribbon)).}
  \end{subfigure}
  \begin{subfigure}{7.0cm}
    \centering\includegraphics[width=6cm]{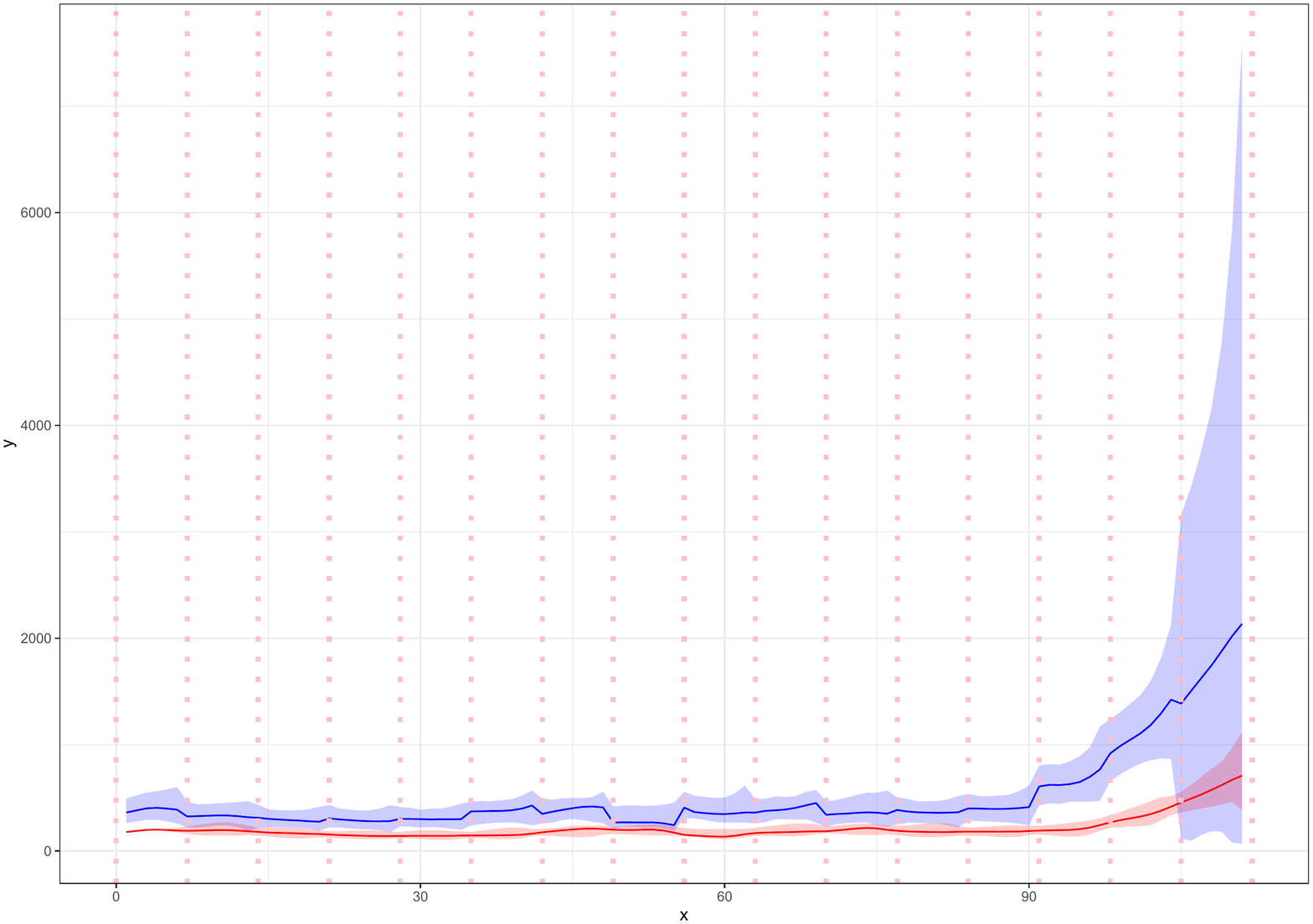}
   \caption{The estimated intensity of latent cases (posterior median (blue line) ; 99\% CI (ribbon)) and the estimated intensity of observed cases (posterior median (red line) ; 99\% CI (ribbon)).} 
  \end{subfigure}
  \begin{subfigure}{7cm}
    \centering\includegraphics[width=6cm]{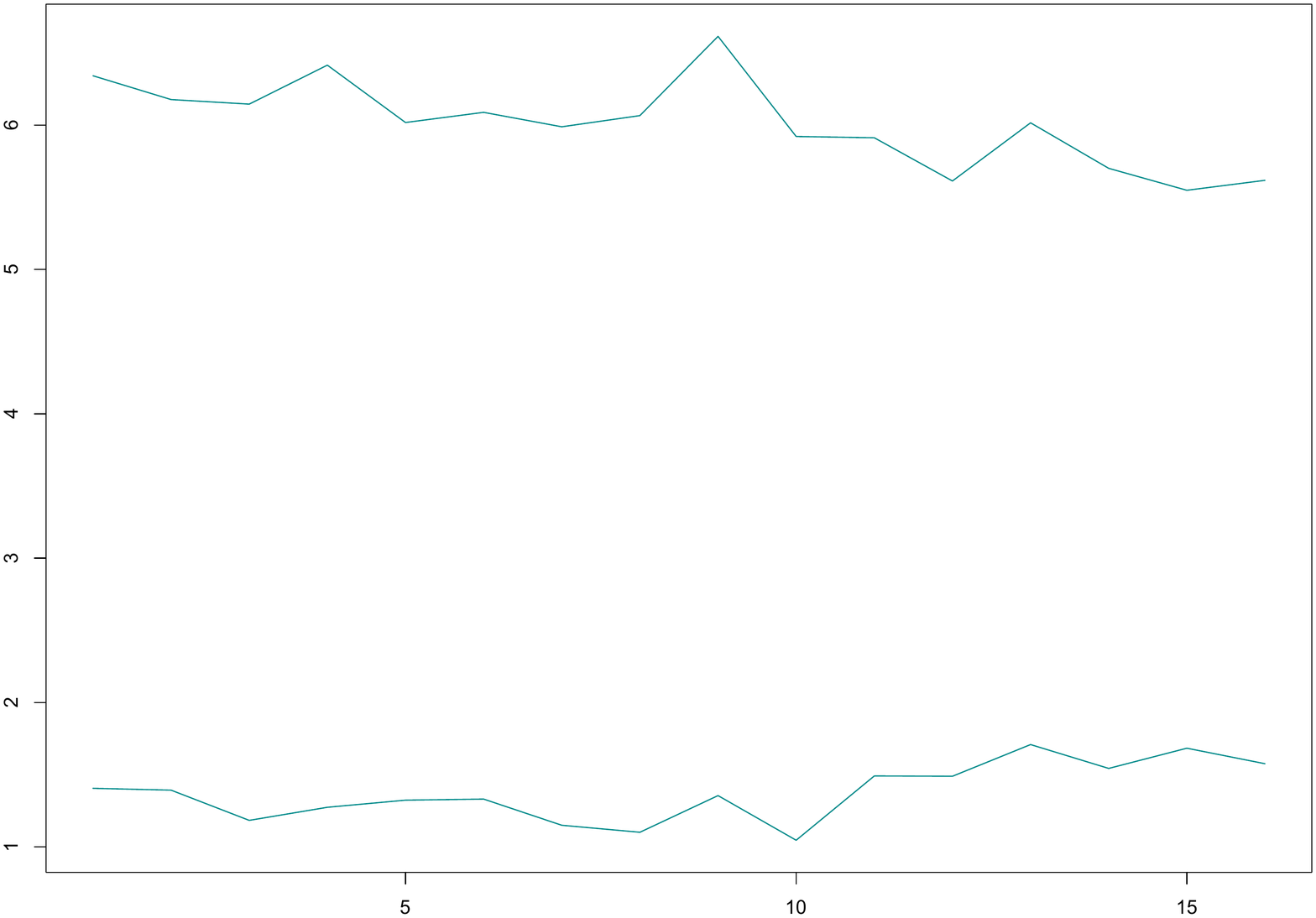}
   \caption{The 99\% CI of d.}
  \end{subfigure}
   \begin{subfigure}{7cm}
    \centering\includegraphics[width=6cm]{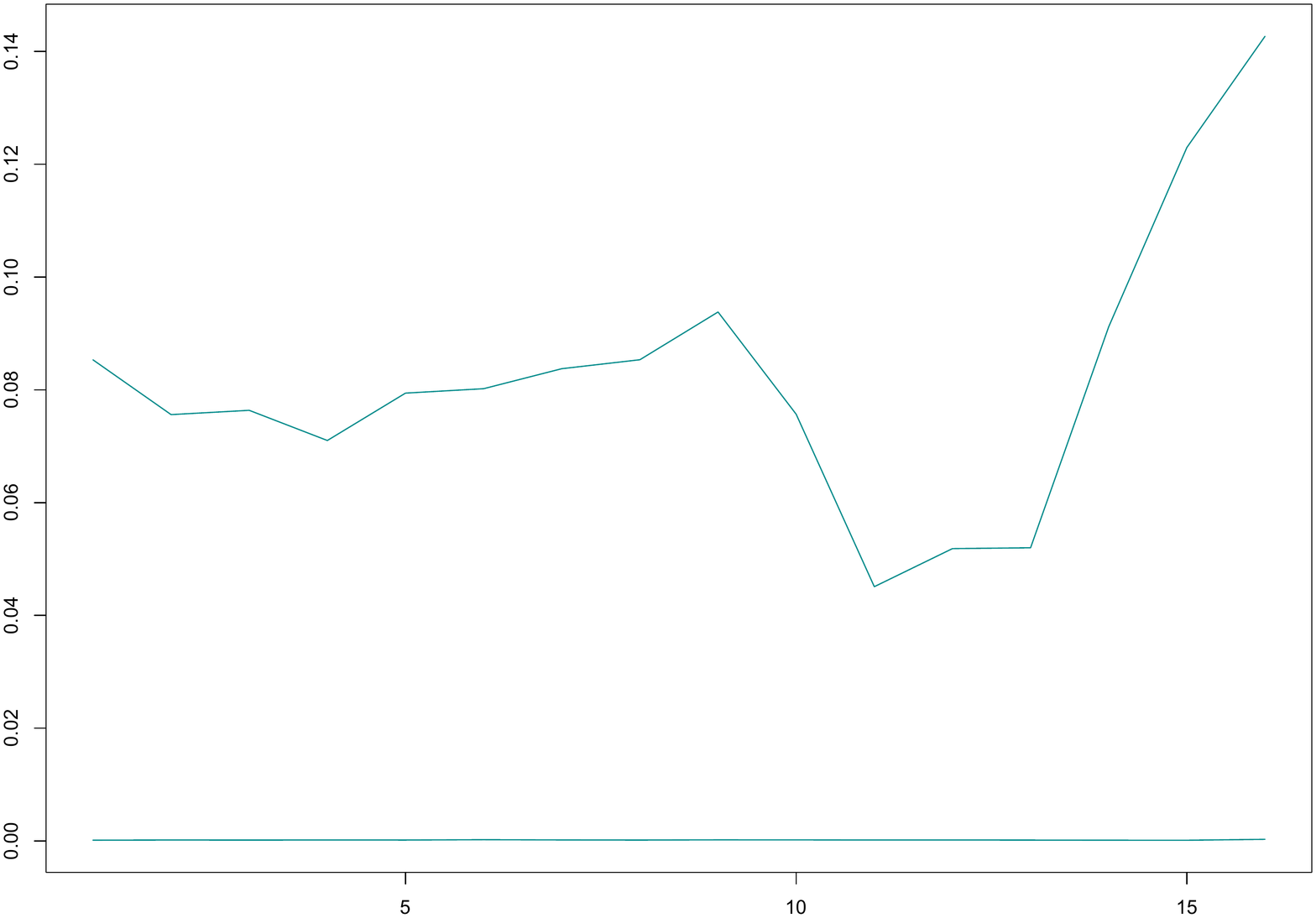}
   \caption{The 99\% CI of v.}
  \end{subfigure}
  \caption{{\bf The weekly and daily latent cases, the reproduction number, the latent and observed intensity and the $99\%$ CIs of time-constant parameters in Leicester.} The time interval between two successive pink vertical dashed lines corresponds to a week.}
  \label{FigLeicester}
\end{figure}

Figures \ref{FigAshford} - \ref{FigLeicester} show the estimated latent and observed intensity, the estimated weekly and daily hidden cases, the estimated reproduction number and the time-constant parameters in the local authorities. We illustrate the intensity of observed cases, approximating via equation \ref{EqIntOC}. We note that the estimated latent intensity and the estimated latent cases are in agreement with the reported cases. According to the analysis, the instantaneous reproduction number $R(t)$ depicts the pandemic's development and capture dynamics. For the COVID-19 pandemic, there is a maximum delay of 21 days between the reported and actual infection times, which provides information regarding the progression of the epidemic. As a result, estimates have become more uncertain towards the end of the horizon.

We compare the proposed algorithm with two methods of estimating the reproduction number. The method suggested by Cori et al.~\cite{cori2013new} (EpiEstim) estimates the reproduction number from incidence time series using a Bayesian framework with a gamma distributed prior imposed on the reproduction number. An alternative method suggested by Koyama et al.~\cite{koyama2021estimating} is a state-space method for estimating the daily reproduction number from a time series of reported infections using a random walk prior to the reproduction number and log-normal distribution as the distribution of the serial interval (SI). We assume that the mean and standard deviation of the SI distribution is at 6.9 days and 5.6 days following Zhao et al.~\cite{zhao2021estimating}. We apply EpiEstim by using the gamma and log-normal distribution as the distribution of SI. Both choices lead to identical results.

\begin{figure}[!h] 
  \begin{subfigure}{7cm}
    \centering\includegraphics[width=6cm]{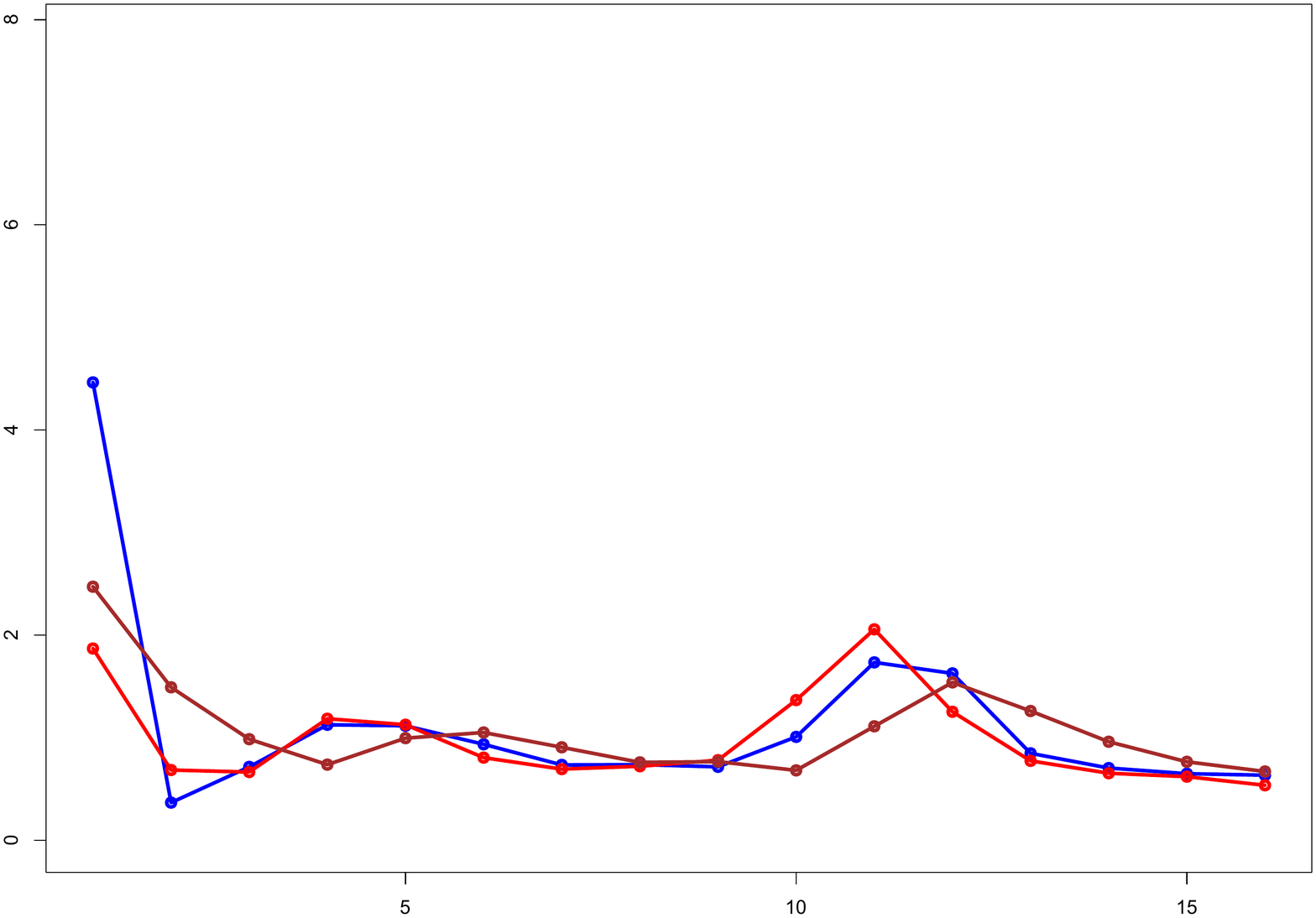}
    \caption{Ashford} 
  \end{subfigure}
  \begin{subfigure}{7cm}
    \centering\includegraphics[width=6cm]{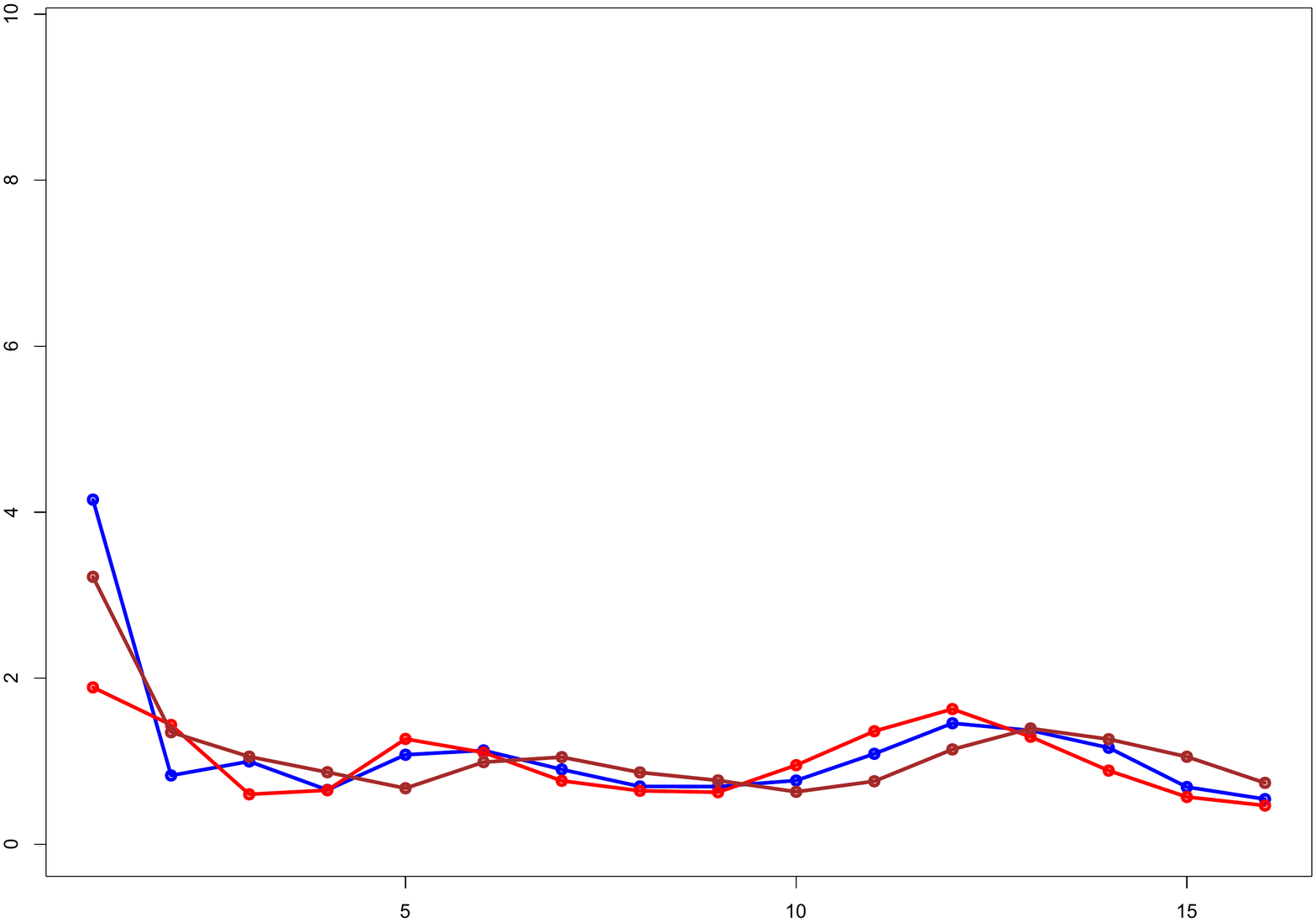}
    \caption{Kingston upon Thames} 
  \end{subfigure}
   \begin{subfigure}{7cm}
    \centering\includegraphics[width=6cm]{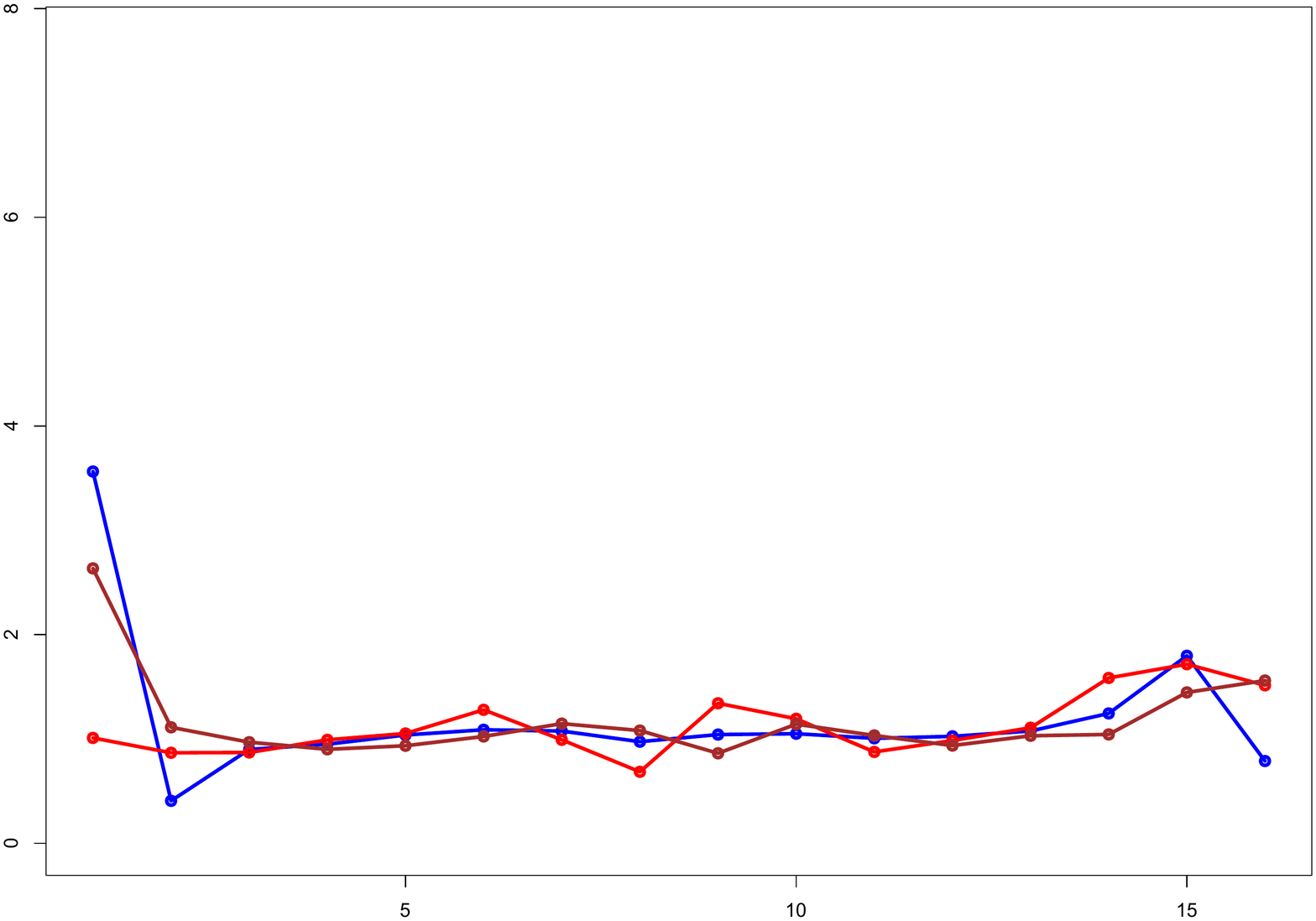}
    \caption{Leicester} 
  \end{subfigure}
  \caption{\bf The weekly average of daily estimates of the reproduction number via posterior median derived by the method of Koyama et al.~\cite{koyama2021estimating} (blue line) and the posterior medians of $R(t)$ given by EpiEstim (brown line) and the proposed algorithm (red line).}
  \label{FigKoyamaR}
\end{figure}

Figure \ref{FigKoyamaR} shows the weekly average of daily estimates of the reproduction number via posterior median derived by the method of Koyama et al.~\cite{koyama2021estimating} and the posterior medians of $R(t)$ given by EpiEstim and the proposed algorithm following the course of the pandemic. The method of Koyama et al.~\cite{koyama2021estimating} and EpiEstim provide similar estimates to those of Algorithm \ref{APAlga}. In the first week, the estimates of Koyama et al.~\cite{koyama2021estimating} and EpiEstim have essential higher values than one of the proposed algorithm due to different initializations. 

\begin{figure}[!h] 
  \begin{subfigure}{7cm}
    \centering\includegraphics[width=6cm]{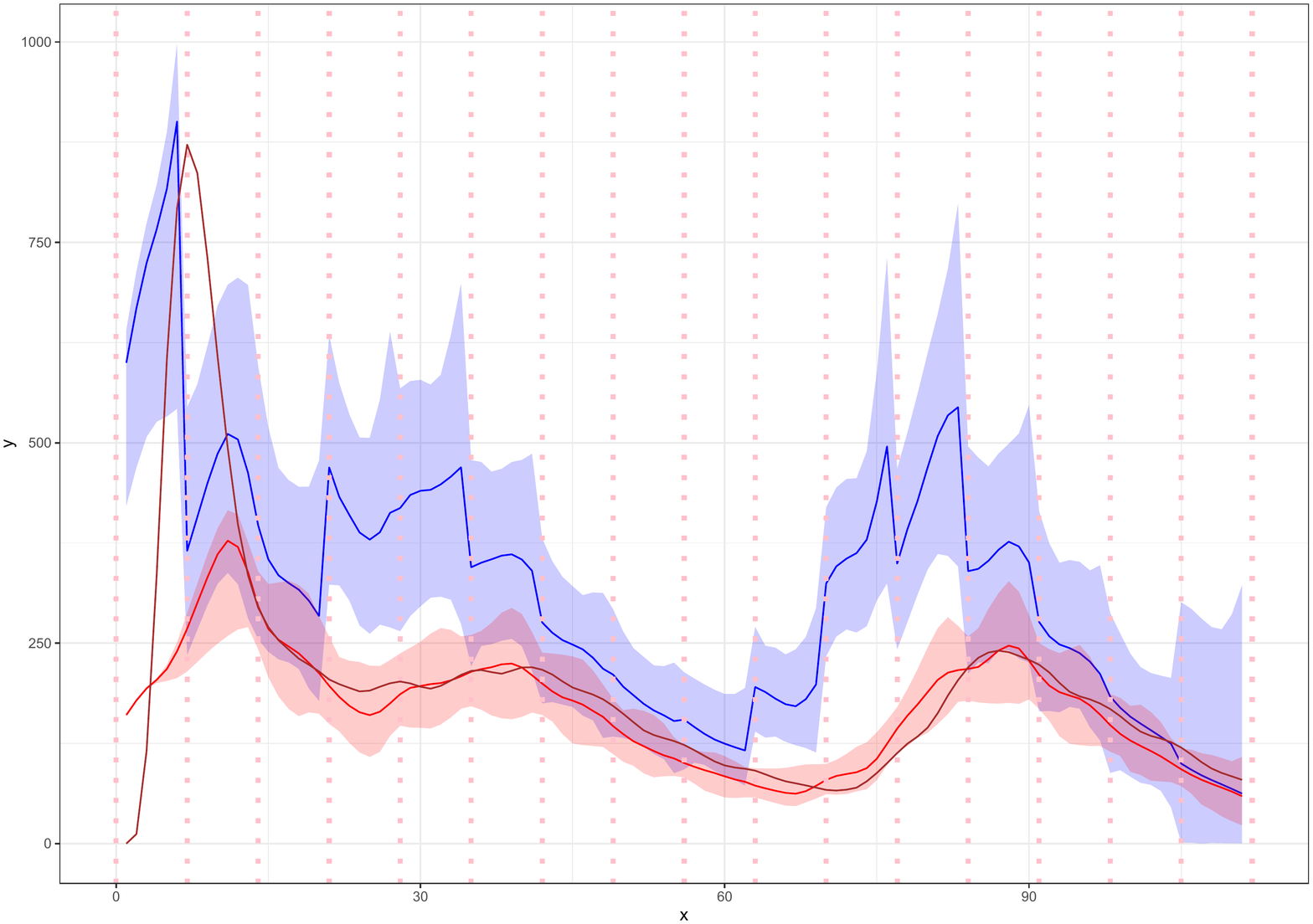}
    \caption{Ashford} 
  \end{subfigure}
  \begin{subfigure}{7cm}
    \centering\includegraphics[width=6cm]{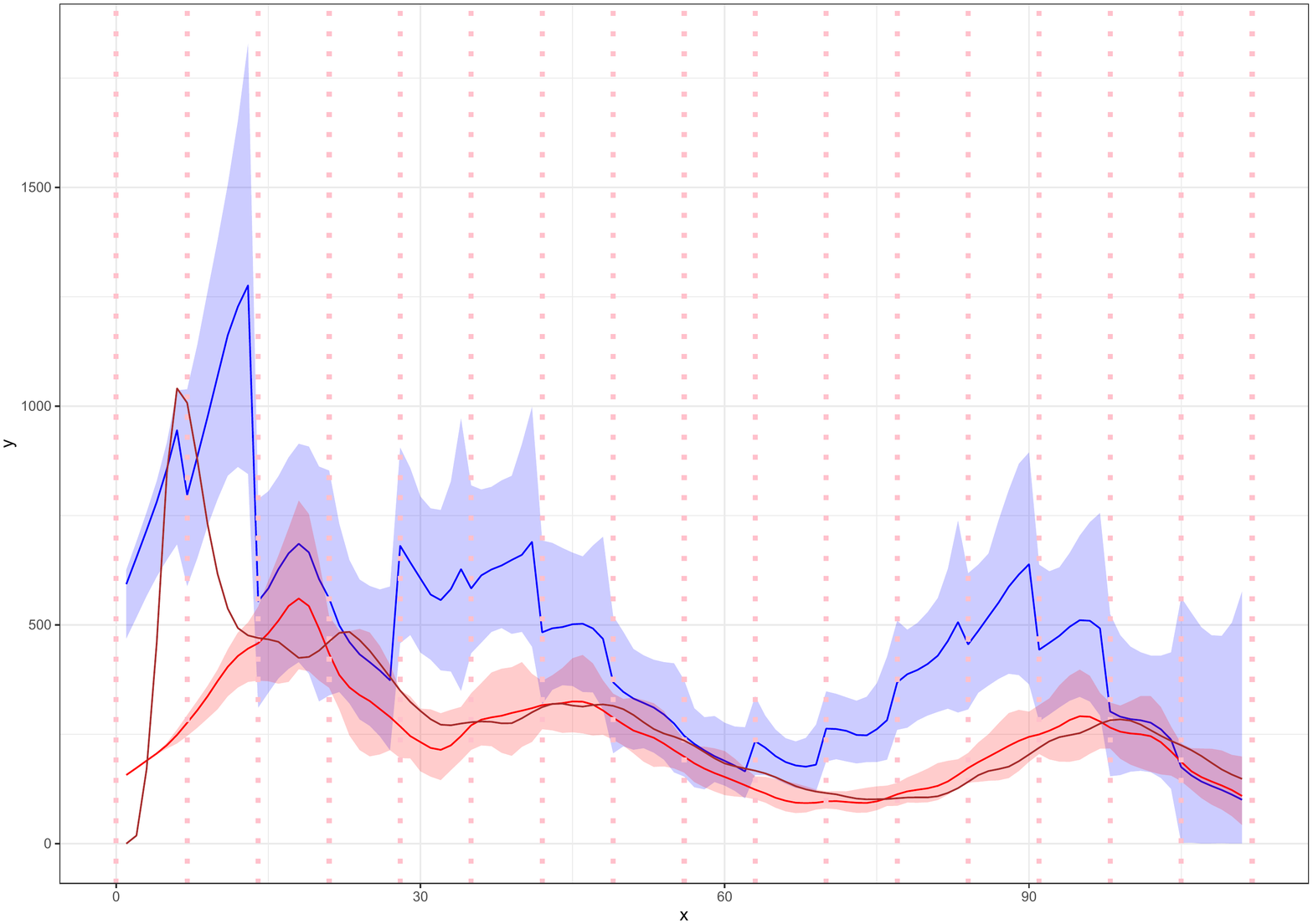}
    \caption{Kingston upon Thames} 
  \end{subfigure}
   \begin{subfigure}{7cm}
    \centering\includegraphics[width=6cm]{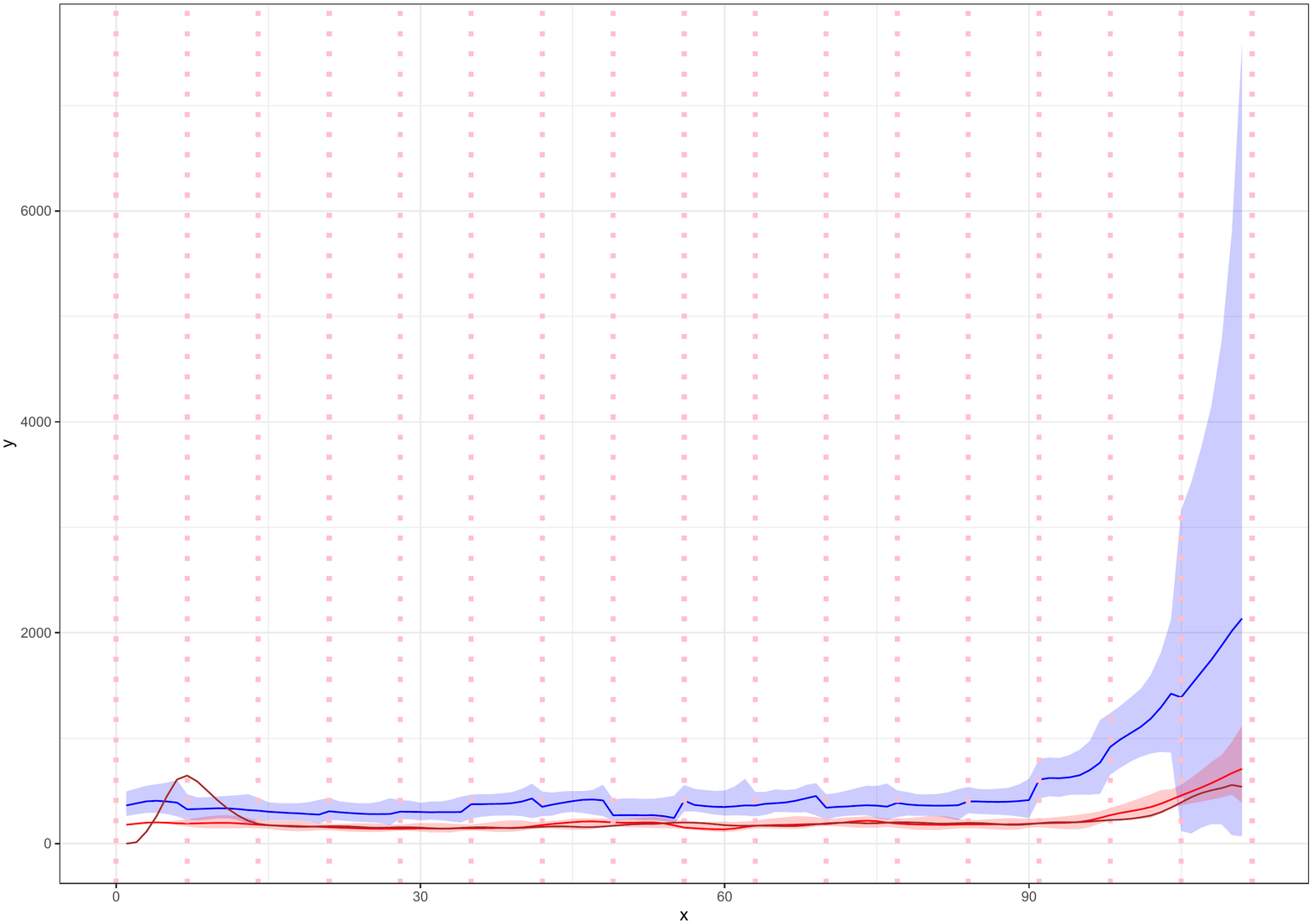}
    \caption{Leicester} 
  \end{subfigure}
  \caption{\bf The estimated daily number of events derived by the method of Koyama et al.~\cite{koyama2021estimating} (brown line), the estimated latent intensity (posterior median (blue line) ; 99\% CI (ribbon)) and the estimated intensity of observed cases (posterior median (red line) ; 99\% CI (ribbon)).}
  \label{FigKoyamaLambda}
\end{figure}

We also compare the estimated rate of latent cases $\lambda^N(t)$ and observed cases $\lambda^M(t)$ with the estimated daily number of events derived by Koyama et al.~\cite{koyama2021estimating}. Figure \ref{FigKoyamaLambda} shows that the expected daily number of events is almost identical to $\lambda^M(t)$ and in agreement with $\lambda^N(t)$ after the end of the 3rd week. The differences in the first three weeks are due to different initializations of the methods.

\paragraph{Future prediction} Using the proposed model, it is also possible to predict the number of new observed cases in the next week ($\mathcal{T}_{17}$) by applying the steps described in Algorithm \ref{AlgPred} (see Table \ref{tablePM}). Table \ref{tableKoyama} shows the estimated number by applying the method introduced by Koyama et al.~\cite{koyama2021estimating} assuming that the reproduction number remains at the value obtained for the last day. Our estimates are similar to those given by Koyama et al.~\cite{koyama2021estimating} in Ashford and Kingston upon Thames. The proposed model gives more accurate results for Leicester, since the current conditions describing the last day are not maintained. Our forecasts are subject to the same level of uncertainty as last week's estimates.
The posterior median and mean, however, can show how the epidemic will develop over the course of the following week.  

\begin{table}[!ht]
\centering
\caption{ \bf The true number of infections in $\mathcal{T}_{17}$, and the posterior median, the posterior mean and the 80\% CIs of the estimated reported infections in $\mathcal{T}_{17}$. }
\begin{tabular}{|l|l|l|l|l|}\hline
 \multicolumn{5}{|l|}{\bf Proposed Method} \\
 \thickhline
 Authority & Posterior Mean & Posterior Median  &  $80\%$ CIs & True Number  \\
 \hline
Ashford & 377 & 341 & (164, 500) & 408 \\ \hline
Kingston & 675 & 609 & (310, 880) & 667  \\ \hline
Leicester & 6519 & 5860 & (2927, 8317) & 5794  \\ 
 \hline
\end{tabular}
\label{tablePM}
\end{table}

\begin{table}[!ht]
\centering
\caption{\bf The estimated number of reported infections in $\mathcal{T}_{17}$. }
\begin{tabular}{ |l|l| }\hline
 \multicolumn{2}{|l|}{\bf Method introduced by Koyama et al.~\cite{koyama2021estimating}} \\
 \thickhline
 Authority & Estimated Number  \\
 \hline
Ashford & 419  \\ \hline
Kingston & 717   \\ \hline
Leicester & 1635   \\
 \hline
\end{tabular}
\label{tableKoyama}
\end{table}

\begin{algorithm}[!h] 
\algsetup{linenosize=\tiny}
\tiny 
\caption{\bf Predicting the new observed cases in near future} 
\label{AlgPred}
\begin{algorithmic}[1]
\STATE{Let $\hat{v}=\frac{1}{N}\sum_{j=1}^Nv_{j,16}$ and $\hat{d}=\frac{1}{N}\sum_{j=1}^Nd_{j,16}$.}
\FOR{$j=1,..,N$}
\STATE{ $R_{j,17}\sim P(R_{17}|R_{j,16}, \hat{d})$} \\
\STATE{$S_{j,17}^N \sim P(S_{17}^N|S_{j,1:16}^N, R_{j,17}, \mathcal{H}_0)$} \\
\STATE{ Calculate the mean of observed cases in the interval denoted by $\mu_{j,17}$.}\\
\STATE{$Y_{j,17}\sim \mbox{NB}(\mu_{j,17},\hat{v})$.}\\

\ENDFOR

\STATE{Use the sample $\{Y_{j,17}\}_{j=1}^N$ to find the posterior mean, the posterior median and the 80\% CI of the estimated observed cases in $\mathcal{T}_{17}$.}

\end{algorithmic}
\end{algorithm}

\section*{Discussion}
 In this paper, we introduce a novel epidemic model using a latent Hawkes process with temporal covariates. Unlike other Hawkes models, we model the infections via a Hawkes process and the aggregated reported cases via a probability distribution $G$ with a mean driven by the underlying Hawkes process. The usual options of $G$ are Negative Binomial and Poisson distribution. We propose a KDPF for inferring the latent cases and the instantaneous reproduction number and for predicting the new observed cases over short time horizons. We demonstrate the performance of the proposed algorithm on COVID-19. 

 The analysis of synthetic data shows that KDPF compares well with PMMH, having the advantage that it is a more computationally efficient algorithm than PMMH. We also demonstrate that our predicted new cases, and our inference for the latent intensity, the daily and weekly hidden cases are consistent with the observed cases in various local authorities in the UK. The simulation analysis shows that the proposed algorithm provides comparable estimates of observed case fluctuations compared with those of Koyama et al.~\cite{koyama2021estimating}. The method of Koyama et al.~\cite{koyama2021estimating} and EpiEstim provide similar estimates of the reproduction number to the proposed algorithm. 

The simulation analysis shows that working with daily reported infections leads to better Effective Sample Sizes using a smaller number of particles, as the data spikes are reduced. According to Cori et al.~\cite{cori2013new}, the estimates of the instantaneous reproduction number are expected to be affected by the selection of the time window size. Large sizes result in more smoothing and reductions in statistical noise, whereas small sizes result in faster detection of transmission changes and more statistical noise. They suggest an appropriate way of choosing the time window size. We have selected a weekly time window to analyse the real data in line with Cori et al.~\cite{cori2013new}. 

Uncovering disease dynamics and tracing how and by whom an infected individual was infected is challenging due to unobservable transmission routes ~\cite{yang2013mixture, kim2019modeling}. Modelling the infections via a Hawkes process allows us to model infection dynamics. 

Isham and Medley~\cite{isham1996models}; Wallinga et al.~\cite{wallinga1999perspective} contend that it is necessary to account for individual heterogeneities while modelling the transmission of an infectious disease. 
Individuals vary in their tendency to interact with others; personal hygiene is a key factor in the propagation of diseases; individuals' community structure and location might be significant in spreading epidemics. The proposed epidemic model can be viewed as a turning point in deriving epidemic models that consider individual heterogeneities and provide insight into underlying dynamics that is the subject of our future work. Future work also considers the inference of ascertainment rate ($\beta$), using various transition kernels for modelling the latent and reported infection cases, as well as more sophisticated ways for initializing the set of infectious triggering the epidemic process, $\mathcal{H}_0$. 

\section*{Acknowledgments}
We thank Professor Young for their constructive comments on this manuscript.


%
%
%

\end{document}